\documentclass[11pt]{article}

\usepackage{jheppub}
\usepackage{silence}
\usepackage{empheq}
\usepackage{bm}
\usepackage[missing=]{gitinfo2}
\usepackage{xcolor}
\definecolor{darkred}{rgb}{0.5,0.15,0.15}
\hypersetup{colorlinks=true,urlcolor=darkred,linkcolor=darkred,citecolor=darkred}
\usepackage{soul}
\usepackage{graphicx}
\usepackage{epsfig}
\usepackage{amsmath}
\usepackage{amssymb}
\usepackage{amsthm}
\usepackage{indentfirst}
\usepackage{xspace}
\usepackage{multirow}
\usepackage{hyperref}
\usepackage{verbatim}
\usepackage{subcaption}
\usepackage{geometry}
\usepackage{tikz-cd}
\usepackage{capt-of}
\usepackage{dynkin-diagrams}

\usepackage{color}

\usepackage{array}
\usepackage{longtable}

\usepackage{tikz}
\usetikzlibrary{positioning}
\usetikzlibrary{calc}  
\usepackage{varioref}

\usetikzlibrary{topaths}
\usetikzlibrary{decorations}
\usetikzlibrary{decorations.pathmorphing}

\numberwithin{equation}{section}

\newcommand{\be}{\begin{equation}}
\newcommand{\ee}{\end{equation}}
\newcommand{\ba}{\begin{aligned}}
\newcommand{\ea}{\end{aligned}}

\newcommand{\cC}{\ensuremath{\mathcal C}}

\newcommand{\cM}{\ensuremath{\mathcal M}}

\newcommand{\cW}{\ensuremath{\mathcal W}}

\newcommand{\ri}{{\rm i}}
\newcommand{\rd}{{\rm d}}

\newcommand{\E}{{\mathrm e}}

\DeclareMathOperator{\re}{{\rm e}}

\usepackage{color}

\newcommand{\FIVc}[9]{F\bigg( \begin{matrix} #2 \\ #1 \end{matrix} \, #3 \, \begin{matrix} #4  \\ \, \end{matrix} \,#5 \, \begin{matrix} #6 \\ #7 \end{matrix} ; #8, #9 \bigg)}
\newcommand{\Fn}[8]{\mathfrak{F}\bigg( \begin{matrix} #2 \\ #1 \end{matrix} \, #3 \, \begin{matrix} #4  \\ \, \end{matrix} \,#5 \, #6\,\begin{matrix} #7 \\  \, \end{matrix} \, #8 \, }
\newcommand{\Fns}[8]{\mathcal{F}\bigg( \begin{matrix} #2 \\ #1 \end{matrix} \, #3 \, \begin{matrix} #4  \\ \, \end{matrix} \,#5 \, #6\,\begin{matrix} #7 \\  \, \end{matrix} \, #8 \, }

\usetikzlibrary{positioning,arrows,patterns}
\usetikzlibrary{decorations.markings}
\usetikzlibrary{calc}
\usetikzlibrary{shapes}
\usetikzlibrary{topaths}
\usetikzlibrary{decorations}
\usetikzlibrary{decorations.pathmorphing,fit}
\usetikzlibrary{bending}
\usetikzlibrary{calc}

\tikzset{dot/.style={circle, fill, inner sep=1.5pt}}

\usepackage{scalerel,stackengine}
\stackMath
\newcommand\reallywidehat[1]{%
\savestack{\tmpbox}{\stretchto{%
  \scaleto{%
    \scalerel*[\widthof{\ensuremath{#1}}]{\kern-.6pt\bigwedge\kern-.6pt}%
    {\rule[-\textheight/2]{1ex}{\textheight}}%WIDTH-LIMITED BIG WEDGE
  }{\textheight}% 
}{0.5ex}}%
\stackon[1pt]{#1}{\tmpbox}%
}
\preprint{CERN-TH-2025-015}
\title{On quivers, spectral networks and black holes}
\author[a]{Paolo Arnaudo}
\author[b,c]{Alba Grassi}
\author[b]{Qianyu Hao}
\affiliation[a]{Mathematical Sciences and STAG Research Centre, \\ University of Southampton, Highfield, Southampton SO17 1BJ, UK}
\affiliation[b]{Section de Math\'ematiques, Universit\'e de Gen\`eve, 1211 Gen\`eve 4, Switzerland}
\affiliation[c]{Theoretical Physics Department, CERN, 1211 Geneva 23, Switzerland }
\abstract{ It was recently found that connection coefficients of the Heun equation can be derived in closed form using crossing symmetry in two-dimensional Liouville theory via the Nekrasov-Shatashvili functions. In this work, we systematize this approach to second-order linear ODEs of Fuchsian type, which arise in the description of  $\mathcal{N}=2$, four-dimensional quiver gauge theories. After presenting the general procedure, we focus  on the specific case of  Fuchsian equations with five regular singularities and present some applications to black hole perturbation theory.  First, we consider a massive scalar perturbation of the Schwarzschild black hole in AdS$_7$. Next, we analyze vector type perturbations of the Reissner-Nordstr\"om-AdS$_5$ black hole. We also discuss the implications of our results in the context of the AdS/CFT correspondence and present explicit results in the large spin limit, where we make connection with the light-cone bootstrap. Furthermore, using the spectral network technology, we identify the region of the moduli space in Seiberg-Witten theory that is relevant for the study of black hole quasinormal modes. Our results suggest that, in some cases, this region corresponds to the strong-coupling regime, highlighting the potential applicability of  the conformal GMN TBA framework to address scenarios where the gravitational dictionary implies that the instanton counting parameters are not parametrically small.}

\begin{document}

\maketitle

\flushbottom

\section{Introduction}

In recent years, many connections have been uncovered between quantum spectral problems on one hand, and supersymmetric gauge theories or topological string theories on the other, see \cite{Marino:2015nla} for a review.
An important result relevant to this work was presented in \cite{ns}, where the authors found that the partition functions of four-dimensional, $\mathcal{N}=2$  gauge theories, evaluated in the Nekrasov--Shatashvili (NS) phase of the $\Omega$ background, serve as fundamental building blocks for constructing solutions to a class of differential equations known as four-dimensional quantum Seiberg--Witten (SW) curves.  Initially, the works \cite{ns,mirmor,mirmor2,Zenkevich2011,Nekrasov:2011bc} primarily focused on the quantization conditions for the energy spectrum, while the study of eigenfunctions was performed in more details later, see for instance \cite{Kozlowski:2010tv,Alday:2010vg,Kanno:2011fw,Jeong:2021rll,Jeong:2023qdr,Jeong:2018qpc,Jeong:2017pai, Alday:2009fs,Drukker:2009id}. In \cite{Grassi:2019coc,Grassi:2021wpw}, a method was developed to obtain a closed-form expression for the Fredholm determinants of such operators by taking a special limit of the TS/ST correspondence \cite{ghm}. More recently in \cite{Bonelli:2022ten}, it was found that these determinants, and the associated connection coefficients, can be derived in an alternative  way using crossing symmetry in two-dimensional Liouville theory\footnote{For differential equations of order higher than two, one needs to use Toda field theory \cite{Wyllard:2009hg}.}, through the AGT correspondence \cite{Alday:2009aq}. In this context, the NS  partition functions compute  the semiclassical conformal blocks.
In \cite{Bonelli:2022ten} the explicit example of Heun equation  was worked out in detail.
In this work, we build on this approach and extend it to second-order linear ODEs of Fuchsian type\footnote{These coefficients can also be derived using the topological string approach, as done in the examples of $SU(N)$ with $N_f = 0$ \cite{Grassi:2019coc} or SU(2) with $N_f = 1$  \cite{Grassi:2021wpw}. However, for the cases of interest here, the method based on crossing symmetry \cite{Bonelli:2022ten} is more straightforward and easier to implement. Let us also mention that yet another way of computing these coefficients can be obtained by extending \cite{Jeong:2018qpc} to open paths.}.  After presenting the general procedure, we focus on the detailed analysis of a Fuchsian equation with five regular singularities. 

One motivation for studying these problems arises from their connection to black hole perturbation theory, which, as discussed in \cite{Aminov:2020yma}, can be systematically explored using the Nekrasov-Shatashvili  functions. This framework has been effectively extended to a variety of gravitational backgrounds and has applications that go well beyond the computation of quasinormal modes (QNMs); see, for example, \cite{Aminov:2020yma,Bonelli:2021uvf,Casals:2021ugr,Bianchi:2021mft,Consoli:2022eey,Bianchi:2022wku,daCunha:2022ewy,Bianchi:2023rlt,Bautista:2023sdf,Bautista:2024agp,Fucito:2024wlg,Zhao:2024rrw,Cipriani:2024ygw,Bianchi:2024mlq,Arnaudo:2024bbd,Bianchi:2024rod,Matone:2024ytm,Arnaudo:2024rhv}. Several applications have also been discussed in the context of the AdS/CFT correspondence, as we discuss in \autoref{sec:applications}.
In the cases studied so far, the relevant differential equation has at most four (regular) singularities.
However, especially in the context of higher-dimensional black holes, the underling differential equation can have an arbitrarily high number of singular points. This makes it interesting to perform a detailed analysis of such scenarios as well. 

This paper is structured as follows. In \autoref{sec2}, we present the detailed procedure for computing the connection coefficients of a second-order linear ODE of Fuchsian type, which corresponds to a four-dimensional  \( \mathcal{N}=2 \) quiver  theory. 
In general, if the equation has more than four singularities, then the relevant  \( \mathcal{N}=2 \)  theory can have multiple representations as a weakly coupled generalized quiver \cite{Gaiotto:2009we}. Here,  we will mostly deal with linear quiver gauge theories.

In \autoref{sec:five}, we apply this analysis to the specific example of a Fuchsian equation with five regular singularities. A new feature for cases with more than four singularities is the appearance of non-comb-like diagrams, an example of which is presented in \autoref{sec:connII}.
 In all cases, we formulate the problem in terms of Frobenius solutions and by using the full NS functions (including classical and one-loop). This enables us to express the ratio of connection coefficients as trigonometric functions with argument  $F^{\rm NS}$. We cross-check our results numerically. 
In \autoref{sec:applications}, we explore some applications. Specifically, in \autoref{sec:AdS7}, we examine a massive scalar perturbation of a Schwarzschild-AdS$_7$ black hole while in \autoref{sec:AdS5}  we study gravitational and electromagnetic (vector type) perturbation a   Reissner–Nordstr\"om (RN) black hole in AdS$_5$.
We present a detailed analysis of  the expansion at large $M/\ell^{(d-2)/2}$ where we make contact with the light-cone bootstrap.

In \autoref{sec:sn}, we initiate the study of spectral networks (SN) within the framework of black hole perturbation theory. Our findings suggest that some QNM spectral problems reside in the strong coupling region of the Seiberg-Witten moduli space. This observation opens the way for exploring alternative techniques, such as the conformal Gaiotto-Moore-Neitzke (GMN) TBA formalism \cite{Gaiotto:2009hg,Gaiotto:2014bza}, to address scenarios like the hydrodynamic limit, where the gravitational dictionary indicates that the instanton counting parameters in the NS functions are not parametrically small. 
Building on these results, it would be important to gain a deeper understanding of the implications of wall-crossing phenomena on the gravity side, as well as to further investigate black hole phase transitions on the spectral networks side. We hope to report on this in the future.

\section*{Acknowledgment}
We would like to thank Giulio Bonelli, Yasuyuki Hatsuda, Lotte Hollands, Cristoforo Iossa, Robin Karlsson, Oleg Lisovyy, Andrew Neitzke, Benjamin Withers, Alessandro Tanzini and Sasha Zhiboedov for useful
discussions. We are particularly grateful to Cristoforo Iossa for his numerous clarifications on \cite{Bonelli:2022ten}, to Robin Karlsson for his careful reading of the draft and insightful feedback, and to Andrew Neitzke for kindly  sharing an updated version of the code for drawing spectral networks. We would also like to thanks an anonymous referee for insightful comments. 
The work of AG and QH is partially supported by the Swiss National Science Foundation Grant No. 185723 and the NCCR SwissMAP.

\section{The general strategy}\label{sec2}
\subsection{Liouville CFT}
\label{Liouville}

Among the 2d conformal field theories (CFT), an important model is the Liouville theory. In this section, we review the background of the 2d Liouville CFT. The vertex algebra of the Liouville CFT is the Virasoro algebra. The central charge $c$ of the Virasoro algebra can be expressed in terms of the background charge $Q$ or the coupling constant $b$ as $c=1+6Q^2$, where $Q=b+1/b$. Liouville theory also depends on a Riemann surface $C$.

The spectrum of the theory is diagonal and continuous. The conformal dimensions of the primary fields are usually denoted by $\Delta$, and they are parametrized by the  momenta $\alpha$ as
\be\Delta=\frac{Q^2}{4}-\alpha^2.\ee

Liouville CFT allows the existence of reducible representations of the Virasoro algebra. 
The invariant submodules are generated by null states, which have the property of being annihilated by all the positive generators of the local conformal transformations $L_n$, $n>0$.
At level $n=2$, 
%t
a reducible representation is generated by the degenerate field $\Phi_{2,1}$ with momentum $\alpha_{2,1}=-b-\frac{1}{2b}$ and conformal weight $\Delta_{2,1}=-\frac{1}{2}-\frac{3}{4}b^2$. The null state  of $\Phi_{2,1}$ is its descendant
\begin{equation}
\left(b^{-2}L_{-1}^2+L_{-2}\right)|\Phi_{2,1}\rangle.
\end{equation}
Since null states decouple from correlation functions, the following equation is satisfied
with all possible insertions of primary fields $V_{\alpha_i}$, $i=0,\dots,n-1$, 
\begin{equation}\label{nullstate}
\langle\prod_{i=0}^{n-1}V_{\alpha_i}(z_i)\left(b^{-2}L_{-1}^2+L_{-2}\right)\Phi_{2,1}(z)\rangle=0,
\end{equation}
where $z_i\in C$. Using the local Ward identities, it is possible to write \eqref{nullstate} as a 
partial differential equation known as BPZ equation \cite{bpz}:
\begin{equation}\label{partialBPZ}
\left(\frac {1}{b^{2}}\frac{\partial^{2}}{\partial z^{2}}+\sum _{i=0}^{n-1}\left(\frac{1}{z-z_{i}}\frac {\partial }{\partial z_{i}}+\frac {\Delta _{i}}{(z-z_{i})^{2}}\right)\right)\langle \prod _{i=0}^{n-1}V_{\alpha_{i}}(z_{i})\Phi_{2,1}(z)\rangle =0.
\end{equation}
We can assume without loss of generality that three of the insertion points, say $z_0, z_1, z_{n-1}$, are located at $z= 0,1, \infty$,  and therefore represent the correlator as
\be \label{n+1correlator} \langle\Delta_{\infty}|V_{\alpha_1}(1)\prod_{i=2}^{n-2}V_{\alpha_i}(z_i)\Phi_{2,1}(z)|\Delta_0\rangle~.\ee
The correlator \eqref{n+1correlator} is calculated using the operator product expansion (OPE). 
When we perform the calculation such that there is an OPE between the degenerate field $\Phi_{2,1}(z)$ and one of the primary fields $V_{\alpha_i}(z_i)$,
\begin{equation}\label{OPEzi}
\Phi_{2,1}(z)V_{\alpha_i}(z_i)=\sum_{\pm}C_{\alpha_{2,1}\alpha_i}^{\alpha_i\pm\frac{b}{2}}(z-z_i)^{\frac{bQ}{2}\mp\alpha_i}V_{\alpha_i\mp\frac{b}{2}}(z_i)+\mathcal{O}\left((z-z_i)^{\frac{bQ}{2}\mp\alpha_i+1}\right),
\end{equation}
inside the correlator, we obtain the \emph{conformal block expansion} of a basis of local solutions of the BPZ equation for $z\sim z_i$. Here
\begin{equation}
    \alpha_{i\theta}=\alpha_i- \theta\frac{b}{2}, \quad \theta=\pm 1~,
\end{equation}
\begin{align}\label{eq:dozz}
 &C_{\alpha_1\alpha_2\alpha_3}
=\frac{\Upsilon'_b(0)\Upsilon_b(Q+2\alpha_1)\Upsilon_b(Q+2\alpha_2)\Upsilon_b(Q+2\alpha_3) \Upsilon_b^{-1}(\frac Q 2+\alpha_1+\alpha_2+\alpha_3)}{\Upsilon_b(\frac Q 2+\alpha_1+\alpha_2-\alpha_3)\Upsilon_b(\frac Q 2+\alpha_1-\alpha_2+\alpha_3)\Upsilon_b(\frac Q 2-\alpha_1+\alpha_2+\alpha_3)},
\end{align}
and 
\begin{equation}
     C^{\alpha_1}_{\alpha_2\alpha_3}=G_{\alpha_1}^{-1}C_{\alpha_1\alpha_2\alpha_3},
\end{equation}
where
\begin{equation}
G_\alpha=\frac{\Upsilon_b(2\alpha+Q)}{\Upsilon_b(2\alpha)}.
\end{equation}
Equation \eqref{eq:dozz}  is known as the DOZZ formula \cite{Dorn:1994xn,ZAMOLODCHIKOV1996577}, we refer to \cite{Teschner:2001rv} for a review and a definition of the $\Upsilon_b$ function.

If we want to study the local solutions of \eqref{partialBPZ} around $ z\sim 0 $, we  use an OPE expansion of the form 
\begin{equation}
\begin{aligned}
\label{conf}&\langle\Delta_{\infty}|V_{\alpha_1}(1)\prod_{i=2}^{n-2}V_{\alpha_i}(z_i)\Phi_{2,1}(z)|\Delta_0\rangle=\\
&=\sum_{\theta=\pm}\int\mathrm{d}\beta_1\dots\mathrm{d}\beta_{n-3}\,C_{\alpha_{2,1}\alpha_0}^{\alpha_{0\theta}}C_{\alpha_{n-2}\alpha_{0\theta}}^{\beta_1}\prod_{i=1}^{n-4}C^{\beta_{i+1}}_{\alpha_{n-i-2}\,\beta_i}C_{\alpha_{\infty}\alpha_1\beta_{n-3}}\times\\
&\quad\quad\quad\times\bigg|\Fn{\alpha_{\infty}}{\alpha_{1}}{\beta_{n-3}}{\alpha_{2}}{\dots}{\beta_1}{\alpha_{n-2}}{\alpha_{0\theta}}\begin{matrix} \alpha_{2,1} \\ \alpha_0\end{matrix};\frac{z_2}{z_1},\frac{z_3}{z_2}\dots,\frac{z_{n-2}}{z_{n-3}},\frac{z}{z_{n-2}}\bigg)\bigg|^2,
\end{aligned}
\end{equation}
where $\Fn{\alpha_{\infty}}{\alpha_{1}}{\beta_{n-3}}{\alpha_{2}}{\dots}{\beta_1}{\alpha_{n-2}}{\alpha_{0\theta}}\begin{matrix} \alpha_{2,1} \\ \alpha_0\end{matrix};\frac{z_2}{z_1},\frac{z_3}{z_2}\dots,\frac{z_{n-2}}{z_{n-3}},\frac{z}{z_{n-2}}\bigg)   $ is the $(n+1)$-point conformal block, and  we assumed $|z_{n-2}|<|z_{n-3}|<\dots<|z_2|<1=z_1$.
Diagrammatically, we represent this conformal block as a ‘‘comb" diagram %
\begin{equation}\label{blocktkk}
\begin{tikzpicture}[baseline={(current bounding box.center)}, node distance=1cm and 1.5cm]
\coordinate[label=above:$\alpha_{1}$] (e1);
\coordinate[below=of e1] (aux1);
\coordinate[left=of aux1,label=left:$\alpha_{\infty}$] (e2);
\coordinate[right=1.5cm of aux1] (aux2);
\coordinate[above=of aux2,label=above:$\alpha_{2}$] (e3);
\coordinate[right=1.5cm of aux2] (aux3);
\coordinate[above=of aux3,label=above:$\alpha_{n-2}$] (e4);
\coordinate[right=1.5cm of aux3] (aux4);
\coordinate[above=of aux4,label=above:$\alpha_{2,1}$] (e5);
\coordinate[right=of aux4,label=right:$\alpha_0$] (e6);

\draw (e1) -- (aux1);
\draw (aux1) -- (e2);
\draw (e3) -- (aux2);
\draw (e4) -- (aux3);
\draw[dashed,red] (aux4) -- (e5);
\draw (aux4) -- (e6);
\draw (aux1) -- node[label=below:$\beta_{n-3}$] {} (aux2);
\draw[dashed] (aux2) -- node[label=below:] {} (aux3);
\draw (aux3) -- node[label=below:$\alpha_{0\theta}$] {} (aux4);
\end{tikzpicture},
\end{equation}
which gives an order for performing OPEs. To be more concrete, each trivalent conponent 
\begin{equation}\label{cca}
\begin{tikzpicture}[baseline={(current bounding box.center)}, node distance=1cm and 1.5cm]
\coordinate[label=above:$\alpha_{k}$] (e1);
\coordinate[below=of e1] (aux1);
\coordinate[left=of aux1,label=left:$\cdots$] (e2);
\coordinate[left=1.5cm of aux1] (aux0);
\coordinate[right=1.5cm of aux1] (aux2);
\coordinate[right=of aux1,label=right:$\cdots$] (e2);

\draw (e1) -- (aux1);
\draw (aux1) --  (e2);
\draw (aux1) -- node[label=below:$\beta_l$] {} (aux0);
\draw (aux1) -- node[label=below:$\beta_{l+1}$] {} (aux2);
\end{tikzpicture}
\end{equation}
represents an OPE
\begin{equation}
V_{\beta_l}(z)V_{\alpha_k}(z_k)=\int\rd{\beta_{l+1}}C_{\beta_l\alpha_k}^{\beta_{l+1}}(z-z_k)^{\Delta_{\beta_{l+1}}-\Delta_{\beta_l}-\Delta_{\alpha_k}}V_{\beta_{l+1}}(z_k)+\mathcal{O}\left((z-z_k)^{\Delta_{\beta_{l+1}}-\Delta_{\beta_l}-\Delta_{\alpha_k}+1}\right).
\end{equation}
An important scaling limit of the Liouville CFT is the \textit{semiclassical limit} given by
\begin{equation}\label{semiclassicallimit}
b\to 0,\quad\quad \beta_i,\alpha_i\to\infty,\quad\quad b\,\alpha_i\equiv a_i, \quad b\,\beta_i\equiv b_i\quad \text{finite},
\end{equation}
under which the divergence of the conformal blocks exponentiates  \cite{1986JETP631061Z} and the $z$-dependence becomes subleading:\footnote{We suppress the dependence of $F\left({z_2\over z_1},\frac{z_3}{z_2}\dots,\frac{z_{n-2}}{z_{n-3}}\right)$ on $a_i$'s and $b_i$'s in our notation since the instanton parameters imply the order of $a_i$'s and $b_i$'s.}
\begin{equation}\label{zamoexp}
\begin{aligned}
&\Fn{\alpha_{\infty}}{\alpha_{1}}{\beta_{n-3}}{\alpha_{2}}{\dots}{\beta_1}{\alpha_{n-2}}{\alpha_{0\theta}}\begin{matrix} \alpha_{2,1} \\ \alpha_0\end{matrix};\frac{z_2}{z_1},\frac{z_3}{z_2}\dots,\frac{z_{n-2}}{z_{n-3}},\frac{z}{z_{n-2}}\bigg)=\\
&=\,z_1^{\Delta_{\infty}-\Delta_{\alpha_1}-\Delta_{\beta_{n-3}}} z_2^{\Delta_{\beta_{n-3}}-\Delta_{\alpha_2}-\Delta_{\beta_{n-4}}}\dots z_{n-2}^{\Delta_{\beta_1}-\Delta_{\alpha_{n-2}}-\Delta_{\alpha_{0\theta}}}z^{\frac{b^2+1}{2}+\theta b\alpha_0}\\
&\times \exp\left(\frac{1}{b^2}\left(F\left(\frac{z_2}{z_1},\frac{z_3}{z_2}\dots,\frac{z_{n-2}}{z_{n-3}}\right)+\mathcal{O}(b^2)\right)\right),
\end{aligned}
\end{equation}
where 
\begin{equation}
F\left(\frac{z_2}{z_1},\frac{z_3}{z_2}\dots,\frac{z_{n-2}}{z_{n-3}}\right)=F\left(\begin{matrix} a_{1} \\ a_\infty\end{matrix} \begin{matrix} b_{n-3} \\ \end{matrix} \dots \begin{matrix} b_{1} \\ \end{matrix} \begin{matrix} a_{n-2} \\ a_0\end{matrix};\frac{z_2}{z_1},\frac{z_3}{z_2}\dots,\frac{z_{n-2}}{z_{n-3}}\right)
\end{equation}
is the semiclassical $n$-point conformal block which is defined as 
%. 
\begin{equation}\label{definitionF}
\begin{aligned}
&\mathfrak{F}\left(\begin{matrix} \alpha_{1} \\ \alpha_{\infty}\end{matrix}\,\beta_{n-3}\,\dots\,\beta_1\begin{matrix} \alpha_{n-2} \\ \alpha_0\end{matrix};{z_2\over z_1},\frac{z_3}{z_2}\dots,\frac{z_{n-2}}{z_{n-3}}\right)=\\
&=\,z_1^{\Delta_{\infty}-\Delta_{\alpha_1}-\Delta_{\beta_{n-3}}}z_2^{\Delta_{\beta_{n-3}}-\Delta_{\alpha_2}-\Delta_{\beta_{n-4}}}\dots z_{n-2}^{\Delta_{\beta_1}-\Delta_{\alpha_{n-2}}-\Delta_{\alpha_{0}}}\re^{\frac{1}{b^2}\left(F\left({z_2\over z_1},\frac{z_3}{z_2}\dots,\frac{z_{n-2}}{z_{n-3}}\right)+\mathcal{O}(b^2)\right)}.
\end{aligned}
\end{equation}
Since the divergent parts for \eqref{zamoexp} and \eqref{definitionF} are the same, following the procedure outlined in \cite{Bonelli:2022ten} we cure the divergences of \eqref{zamoexp} by dividing it by \eqref{definitionF} and we denote the resulting finite semiclassical block by \footnote{We remark that the notation $a_{0\theta}$ inside the semiclassical conformal blocks does not denote a shift in the momentum anymore, but it specifies the two choices of the intermediate channel labeled by $\theta$.
}
\begin{equation}\label{semiclassical6pointaround0}
\begin{aligned}
&\Fns{a_{\infty}}{a_{1}}{b_{n-3}}{a_{2}}{\dots}{b_1}{a_{n-2}}{a_{0\theta}}\begin{matrix} a_{2,1} \\ a_0\end{matrix};\frac{z_2}{z_1},\frac{z_3}{z_2}\dots,\frac{z_{n-2}}{z_{n-3}},\frac{z}{z_{n-2}}\bigg)=\\
&=\lim_{b\to 0}\frac{\Fn{\alpha_{\infty}}{\alpha_{1}}{\beta_{n-3}}{\alpha_{2}}{\dots}{\beta_1}{\alpha_{n-2}}{\alpha_{0\theta}}\begin{matrix} \alpha_{2,1} \\ \alpha_0\end{matrix};{z_2\over z_1},\frac{z_3}{z_2}\dots,\frac{z_{n-2}}{z_{n-3}},\frac{z}{z_{n-2}}\bigg)}{\mathfrak{F}\left(\begin{matrix} \alpha_{1} \\ \alpha_{\infty}\end{matrix}\,\beta_{n-3}\,\dots\,\beta_1\begin{matrix} \alpha_{n-2} \\ \alpha_0\end{matrix};{z_2\over z_1},\frac{z_3}{z_2}\dots,\frac{z_{n-2}}{z_{n-3}}\right)}=\\
&=z_{n-2}^{-\theta a_0}\,z^{\frac{1}{2}+\theta a_0}\,\exp\left(-\frac{\theta}{2}\partial_{a_0}F\left({z_2\over z_1},\frac{z_3}{z_2}\dots,\frac{z_{n-2}}{z_{n-3}}\right)\right)\cdot\left(1+\mathcal{O}\left(z\right)\right).
\end{aligned}
\end{equation}
In the semiclassical limit, the BPZ equation \eqref{partialBPZ}  becomes a second-order ordinary differential equation with regular singularities at $z=z_i$, $i=0,\dots,n-1$
\begin{equation}\label{diffeqnnormal}
\begin{aligned}
\left(\frac{\mathrm{d}^2}{\mathrm{d}z^2}+\sum_{i=0}^{n-2}\frac{\frac{1}{4}-a_{i}^2}{(z-z_i)^2}+\frac{(\frac{1}{4}-a_{\infty}^2)-\sum_{i=0}^{n-2}\left(\frac{1}{4}-a_{i}^2\right)}{z(z-1)}+\sum_{i=2}^{n-2}\frac{(z_i-1)\,u_i}{z(z-1)(z-z_i)}\right)\psi_{0,\theta}(z)=0,
\end{aligned}
\end{equation}
where 
\be\label{martone} u_i=\lim_{b\to 0}b^2\,z_i\,\partial_{z_i}\,\log\,\mathfrak{F}\left(\begin{matrix} \alpha_{1} \\ \alpha_{\infty}\end{matrix}\,\beta_{n-3}\,\dots\,\beta_1\begin{matrix} \alpha_{n-2} \\ \alpha_0\end{matrix};{z_2\over z_1},\frac{z_3}{z_2}\dots,\frac{z_{n-2}}{z_{n-3}}\right)~,\ee
and
\be \label{conf0}\ba 
\Fns{a_{\infty}}{a_{1}}{b_{n-3}}{a_{2}}{\dots}{b_1}{a_{n-2}}{a_{0\theta}}\begin{matrix} a_{2,1} \\ a_0\end{matrix};{z_2\over z_1},\frac{z_3}{z_2},\dots,\frac{z_{n-2}}{z_{n-3}},\frac{z}{z_{n-2}}\bigg)~&\propto~\psi_{0,\theta}(x),\ea\ee
where $ \psi_{0,\theta}(x)=(z)^{{1\over 2}+\theta a_0}(1+\mathcal{O}(z))$, $\theta=\pm $ denote the Frobenius solutions around $z=0$. 

Let us stress that the forms of \eqref{martone} and \eqref{conf0} are affected by the specific choice and order of the punctures \( z_i \), as well as the point at which the degenerate field \( \Phi_{2,1}(z) \) is inserted because they specify how the OPEs are taken. Our choices lead to a basis of solutions centered around \( z=0 \). However, as discussed in \autoref{sec:crossing}, solutions around different singular points can also be achieved by alternative choices of comb diagrams.

As we  discuss in \autoref{sec:gauge}, the semiclassical BPZ equations are quantum Seiberg-Witten curves and can be solved by using the Nekrasov-Shatashvili approach \cite{ns}.
In this context,
\eqref{martone} is  known as Matone relation \cite{Matone:1995rx,Flume:2004rp,lmn},
 and the local solutions to \eqref{diffeqnnormal} correspond to partition functions of specific 4d supersymmetric gauge theories coupled to 2d defects. 

\subsection{Connection coefficients from crossing symmetry}\label{sec:crossing}

Around each singular point $z_i$ of  equation \eqref{diffeqnnormal} we can find a basis of Frobenius solutions $ \psi_{z_i,\pm}(z)$ such that 
\begin{equation}\label{basislocalsolutions}
    \psi_{z_i,\pm}(z)\sim(z-z_i)^{\frac 1 2 \pm a_i}\left(1+\mathcal{O}\left(z-z_i\right)\right)\,.
\end{equation}
as well as
\begin{equation}
    \psi_{\infty,\pm}=z^{\frac{1}{2}\mp a_{\infty}}\left[1+\mathcal{O}\left(\frac{1}{z}\right)\right].
\end{equation}
The radius of convergence of \eqref{basislocalsolutions} is determined by the distance to the next closest singularity.
However, it is possible to analytically continue $\psi_{z_i,\pm}(z)$ to a different puncture $z_j$ and express $\psi_{z_j,\pm}(z)$  as
\be\label{tdiagram} \psi_{z_j,\pm}(z)=\cC^{z_i,+}_{z_j,\pm}\psi_{z_i,+}(z)+\cC^{z_i,-}_{z_j,\pm}\psi_{z_i,-}(z),\ee
where the coefficients $\cC^{z_j,\pm}_{z_i,\pm}$ are known as the connection coefficients. 

As pointed out in \cite{Bonelli:2022ten}, an efficient approach to derive closed-form expressions for these coefficients involves the use of crossing symmetry in the underlying Liouville CFT. In \cite{Bonelli:2022ten}, the case of the Heun equation (four regular singular points), was thoroughly analyzed. In this paper, we extend these results to scenarios in which the ODE has more than four singularities, with special attention to the case of five singular points.

To study the connection coefficients in \eqref{tdiagram}, we use the Frobenius expansions of $\psi_{z_i,\theta}(z)$ in \eqref{basislocalsolutions} corresponding to the semiclassical limit of a particular conformal block of the form 
$\mathfrak{F}\left(\begin{matrix}  \\ \alpha_{j} \end{matrix}\cdots \begin{matrix} \alpha_{1} \\ \hspace{0.1cm} \end{matrix}\cdots \alpha_{i\theta}\begin{matrix} \alpha_{2,1} \\ \alpha_{i} \end{matrix} ; \cdots\right)$, where the OPE of $\Phi(2,1)(z)$ and $V_{\alpha_i}(z_i)$ is given in \eqref{OPEzi}. 
Such block can be represented  schematically  as the following comb diagram

\begin{equation}\label{blockt}
\begin{tikzpicture}[baseline={(current bounding box.center)}, node distance=1cm and 1.5cm]
\coordinate[label=above:$\alpha_{\cdots}$] (e1);
\coordinate[below=of e1] (aux1);
\coordinate[left=of aux1,label=left:$\alpha_{j}$] (e2);
\coordinate[right=1.5cm of aux1] (aux2);
\coordinate[above=of aux2,label=above:$\alpha_{1}$] (e3);
\coordinate[right=1.5cm of aux2] (aux3);
\coordinate[above=of aux3,label=above:$\alpha_{\cdots}$] (e4);
\coordinate[right=1.5cm of aux3] (aux4);
\coordinate[above=of aux4,label=above:$\alpha_{2,1}$] (e5);
\coordinate[right=of aux4,label=right:$\alpha_i$] (e6);

\draw (e1) -- (aux1);
\draw (aux1) -- (e2);
\draw (e3) -- (aux2);
\draw (e4) -- (aux3);
\draw[dashed,red] (aux4) -- (e5);
\draw (aux4) -- (e6);
\draw (aux1) -- node[label=below:$\beta_2$] {} (aux2);
\draw[dashed] (aux2) -- node[label=below:] {} (aux3);
\draw (aux3) -- node[label=below:$\alpha_{i\theta}$] {} (aux4);
\end{tikzpicture}.
\end{equation}
This configuration can be obtained by imposing the M\"{o}bius transform for the $PGL(2,\mathbb{C})$ symmetry starting from \eqref{blocktkk} up to a normalization which comes from the transformation of the primary fields under $x\rightarrow x'$ as 
\be\label{jacobiank} V_{\alpha}(x)~\to~ V'_{\alpha}(x')=\left(\frac{\rd x'}{\rd x}\right)^{-\Delta_{\alpha}}V_{\alpha}(x)~.\ee
In this way one has a direct map between Frobenius solutions and conformal blocks. Similarly we use the Frobenius expansions of $\psi_{z_j,\theta}(z)$ corresponding to the semiclassical limit of a particular conformal block of the form 
\begin{equation}
\begin{tikzpicture}[baseline={(current bounding box.center)}, node distance=1cm and 1.5cm]
\coordinate[label=above:$\alpha_{\cdots}$] (e1);
\coordinate[below=of e1] (aux1);
\coordinate[left=of aux1,label=left:$\alpha_{i}$] (e2);
\coordinate[right=1.5cm of aux1] (aux2);
\coordinate[above=of aux2,label=above:$\alpha_{1}$] (e3);
\coordinate[right=1.5cm of aux2] (aux3);
\coordinate[above=of aux3,label=above:$\alpha_{\cdots}$] (e4);
\coordinate[right=1.5cm of aux3] (aux4);
\coordinate[above=of aux4,label=above:$\alpha_{2,1}$] (e5);
\coordinate[right=of aux4,label=right:$\alpha_j$] (e6);

\draw (e1) -- (aux1);
\draw (aux1) -- (e2);
\draw (e3) -- (aux2);
\draw (e4) -- (aux3);
\draw[dashed,red] (aux4) -- (e5);
\draw (aux4) -- (e6);
\draw (aux1) -- node[label=below:$\beta_2$] {} (aux2);
\draw[dashed] (aux2) -- node[label=below:] {} (aux3);
\draw (aux3) -- node[label=below:$\alpha_{j\theta}$] {} (aux4);
\end{tikzpicture}.
\end{equation}
We remark that when there are six or more insertions, a new structure of diagram can arise, which cannot be represented as a comb-like one. The first example of this kind arises when there are six punctures, which are arranged in three pairs of nearby insertions. In the topological string and gauge theory literature, these are associated to non-linear (or Sicilian) quivers (see \cite{Benini:2009mz, Hollands:2011zc, Coman:2019eex} for a discussion). We will see an example of this kind (in the presence of a degenerate insertion) in \autoref{sec:connII}.

Then, we want to relate the solutions around $z_i$, i.e.~$\psi_{z_i,\pm}(z)$, to the solutions around $z_j$, i.e.~ $\psi_{z_j,\pm}(z)$. 
In the language of CFT, this implies that our task is to examine the effects of exchanging the positions of $\alpha_{2,1}$  and the punctures in between $z_i$ and $z_j$. This effect can be obtained by the product of the factors corresponding to the swapping of $\alpha_{2,1}$ and $\alpha_k$. Each single step from 
\begin{equation}\label{ccb}
\begin{tikzpicture}[baseline={(current bounding box.center)}, node distance=1cm and 1.5cm]
\coordinate[label=above:$\alpha_{k}$] (e1);
\coordinate[below=of e1] (aux1);
\coordinate[left=of aux1,label=left:$\cdots$] (e2);
\coordinate[left=1.5cm of aux1] (aux0);
\coordinate[right=1.5cm of aux1] (aux2);
\coordinate[above=of aux2,label=above:$\alpha_{2,1}$] (e3);
\coordinate[right=1.5cm of aux2, label=right: $\cdots$] (aux3);

\draw (e1) -- (aux1);
\draw (aux1) --  (e2);
\draw (aux1) -- node[label=below:$\beta_l$] {} (aux0);
\draw[dashed] (e3) -- (aux2);
\draw (aux1) -- node[label=below:$\beta_{m\theta}$] {} (aux2);
\draw (aux2) -- node[label=below:$\beta_m$] {} (aux3);
\end{tikzpicture},
\end{equation}
to
\begin{equation}
\begin{tikzpicture}[baseline={(current bounding box.center)}, node distance=1cm and 1.5cm]
\coordinate[label=above:$\alpha_{2,1}$] (e1);
\coordinate[below=of e1] (aux1);
\coordinate[left=of aux1,label=left:$\cdots$] (e2);
\coordinate[left=1.5cm of aux1] (aux0);
\coordinate[right=1.5cm of aux1] (aux2);
\coordinate[above=of aux2,label=above:$\alpha_{k}$] (e3);
\coordinate[right=1.5cm of aux2, label=right: $\cdots$] (aux3);

\draw[dashed] (e1) -- (aux1);
\draw (aux1) --  (e2);
\draw (aux1) -- node[label=below:$\beta_l$] {} (aux0);
\draw (e3) -- (aux2);
\draw (aux1) -- node[label=below:$\beta_{l\theta'}$] {} (aux2);
\draw (aux2) -- node[label=below:$\beta_m$] {} (aux3);
\end{tikzpicture}
\end{equation}
is a local effect analogous to the one of hypergeometric functions. To be more precise, the comb diagrams \eqref{cca} and \eqref{ccb} are related by crossing symmetry as\footnote{Up to an overall factor coming from \eqref{jacobiank} which we will fix in concrete examples below. 
}
\begin{equation}\begin{aligned}
&\int \prod_i\rd\beta_i\ \sum_{\theta=\pm}C^{\beta_{m\theta}}_{\alpha_{2,1}\beta_m}C_{\alpha_{k}\beta_{m\theta}}^{\beta_l}\,\bigg|\mathfrak{F}\left( \cdots\beta_l\begin{matrix} \alpha_{k} \\ \, \end{matrix}  \,\beta_{m\theta} \, \begin{matrix} \alpha_{2,1} \\ \, \end{matrix}\beta_m\cdots ; \cdots \right)\bigg|^2\\
\label{crosssympre}&=
\int \prod_i\rd\beta_i\ \sum_{\theta'=\pm}C^{\beta_{l\theta'}}_{\alpha_k\beta_m}C_{\alpha_{2,1}\beta_{l\theta'}}^{\beta_l}\,\bigg|\mathfrak{F}\left( \cdots\beta_l\begin{matrix} \alpha_{2,1} \\ \, \end{matrix}  \,\beta_{l\theta'} \, \begin{matrix} \alpha_{k} \\ \, \end{matrix}\beta_m\cdots ; \cdots \right)\bigg|^2,\end{aligned}\end{equation}
which is satisfied if 
\begin{equation}\begin{aligned} \label{crosssym}
\sum_{\theta=\pm}C^{\beta_{m\theta}}_{\alpha_{2,1}\beta_m}C_{\alpha_{k}\beta_{m\theta}}^{\beta_l}\,\bigg|\mathfrak{F}\left( \cdots\beta_l\begin{matrix} \alpha_{k} \\ \, \end{matrix}  \,\beta_{m\theta} \, \begin{matrix} \alpha_{2,1} \\ \, \end{matrix}\beta_m\cdots ; \cdots \right)\bigg|^2=\\
\sum_{\theta'=\pm}C^{\beta_{l\theta'}}_{\alpha_k\beta_m}C_{\alpha_{2,1}\beta_{l\theta'}}^{\beta_l}\,\bigg|\mathfrak{F}\left( \cdots\beta_l\begin{matrix} \alpha_{2,1} \\ \, \end{matrix}  \,\beta_{l\theta'} \, \begin{matrix} \alpha_{k} \\ \, \end{matrix}\beta_m\cdots ; \cdots \right)\bigg|^2,\end{aligned}\end{equation}
and we also refer to \eqref{crosssym} as crossing symmetry.
Parallel to \cite{Bonelli:2022ten}, we can check that \eqref{crosssym}  is satisfied by 
\begin{equation}
\label{conn_state}\mathfrak{F}\left( \cdots\beta_l\begin{matrix} \alpha_{k} \\ \, \end{matrix}  \,\beta_{m\theta} \, \begin{matrix} \alpha_{2,1} \\ \, \end{matrix}\beta_m\cdots ; \cdots \right)=\sum_{\theta'=\pm}\E^{\ri\wp_{\theta'}}\cM_{\theta\theta'}(b\,\beta_m,b\,\beta_l;b\,\alpha_k)\mathfrak{F}\left( \cdots\beta_l\begin{matrix} \alpha_{2,1} \\ \, \end{matrix}  \,\beta_{l\theta'} \, \begin{matrix} \alpha_{k} \\ \, \end{matrix}\beta_m\cdots ; \cdots \right)
\end{equation}
if 
\begin{equation}
\label{defM}\mathcal{M}_{\theta \theta'}(a_1,a_2;a_3) = \frac{\Gamma(-2\theta'a_2)\Gamma(1+2\theta a_1)}{\Gamma\left(\frac{1}{2}+\theta a_1-\theta' a_2 + a_3\right) \Gamma\left(\frac{1}{2}+\theta a_1-\theta' a_2 - a_3\right)},\quad \theta,\theta'=\pm,
\end{equation}
where $\E^{\ri\wp_{\pm}}$ is a phase factor which can be fixed by looking at the leading power in the $z$-expansion of the OPE. We will fix this phase in concrete examples below.
The above formulae, as well as the ones we will compute below, hold for generic values of the indicial parameters. 
In principle, one should assume $2a_i \notin \mathbb{Z}$ in order to avoid logarithmic singularities. However, it has been verified in numerous examples in \cite{Jia:2024zes} that, if ones takes the limit $2a_i \to \mathbb{Z}$ carefully, the formulas above continue to hold.

As an example, let us take the special case where we connect the points at $0$ and $\infty$ and we suppose that the singularity structure can be represented by a comb-like diagram. By applying  the connection formula repeatedly
we find (up to phases)
\begin{equation}\label{eq:connex}
\begin{aligned}
&\Fn{\alpha_{\infty}}{\alpha_{{1}}}{\beta_{n-3}}{\alpha_{{2}}}{\dots}{\beta_1}{\alpha_{n-2}}{\alpha_{0\theta}}\begin{matrix} \alpha_{2,1} \\ \alpha_0\end{matrix};\frac{z_{2}}{z_{1}},\frac{z_{3}}{z_{2}},\dots,\frac{z}{z_{n-2}}\bigg)=\\
&=\sum_{\theta_2,\dots,\theta_{n-1}=\pm}\mathcal{M}_{\theta\theta_2}\left(b\alpha_0,b\beta_1;b\alpha_{n-2}\right)\mathcal{M}_{(-\theta_2)\theta_3}\left(b\beta_1,b\beta_2;b\alpha_{n-3}\right)\dots\mathcal{M}_{(-\theta_{n-3})\theta_{n-2}}\left(b\beta_{n-4},b\beta_{n-3};b\alpha_{{2}}\right)\times\\
&\times\mathcal{M}_{(-\theta_{n-2})\theta_{n-1}}\left(b\beta_{n-3},b\alpha_{\infty};b\alpha_{{1}}\right) (-1)^{\Delta_0+\Delta_\infty}\, \\
&\times\prod_{\substack{k=1 }}^{n-2} (-z_k^{-2})^{\Delta_{z_k}}(-z^{-2})^{\Delta_{2,1}}\,\Fn{\alpha_{0}}{\alpha_{n-2}}{\beta_{1\theta_2}}{\alpha_{{n-3}}}{\dots}{\beta_{(n-3)\theta_{n-2}}}{\alpha_{{1}}}{\alpha_{\infty\theta_{n-1}}}\begin{matrix} \alpha_{2,1} \\ \alpha_{\infty}\end{matrix};\frac{z_{n-2}}{z_{n-3}},\dots,\frac{z_{2}}{z_1},\frac{z_{1}}{z}\bigg).
\end{aligned}
\end{equation}
In the semiclassical limit $b\to 0$, using \eqref{zamoexp}, we relate the latter conformal block to the one without the shifts in the intermediate momenta: 
\begin{equation}\label{momentumshift}
\begin{aligned}
&\Fn{\alpha_{0}}{\alpha_{n-2}}{\beta_{1\theta_2}}{\alpha_{{n-3}}}{\dots}{\beta_{(n-3)\theta_{n-2}}}{\alpha_{{1}}}{\alpha_{\infty\theta_{n-1}}}\begin{matrix} \alpha_{2,1} \\ \alpha_{\infty}\end{matrix};\frac{z_{n-2}}{z_{n-3}},\dots,\frac{z_{2}}{z_1},\frac{z_{1}}{z}\bigg)\\
&=\left(\frac{z_{2}}{z_{1}}\right)^{\theta_{n-2}b_{n-3}}\dots\left(\frac{z_{n-2}}{z_{n-3}}\right)^{\theta_2b_{1}}\exp\left(-\sum_{k=1}^{n-3}\frac{\theta_{k+1}}{2}\partial_{b_k}F\left(\frac{z_{n-2}}{z_{n-3}},\dots,\frac{z_{2}}{z_1}\right)+\mathcal{O}(b)\right)\times\\
&\ \times\Fn{\alpha_{0}}{\alpha_{n-2}}{\beta_{1}}{\alpha_{{n-3}}}{\dots}{\beta_{n-3}}{\alpha_{{1}}}{\alpha_{\infty\theta_{n-1}}}\begin{matrix} \alpha_{2,1} \\ \alpha_{\infty}\end{matrix};\frac{z_{n-2}}{z_{n-3}},\dots,\frac{z_{2}}{z_1},\frac{z_{1}}{z}\bigg).
\end{aligned}
\end{equation}
Therefore, if we take  the semiclassical limit
\begin{equation}
b\to 0,\quad \beta_i, \alpha_i\to\infty,\quad b\alpha_i\equiv a_i\quad b\beta_i\equiv b_i\ \text{finite},
\end{equation}
of the connection formula \eqref{eq:connex}, we find (up to phases)
\begin{equation}\label{Mfactors}
\begin{aligned}
&\Fns{a_{\infty}}{a_{{1}}}{b_{n-3}}{a_{{2}}}{\dots}{b_1}{a_{n-2}}{a_{0\theta}}\begin{matrix} a_{2,1} \\ a_0\end{matrix};\frac{z_{2}}{z_{1}},\frac{z_{3}}{z_{2}},\dots,\frac{z}{z_{n-2}}\bigg)=\\
&=\sum_{\theta_2,\dots,\theta_{n-1}=\pm}\mathcal{M}_{\theta_1\theta_2}\left(a_0,b_1;a_{n-2}\right)\mathcal{M}_{(-\theta_2)\theta_3}\left(b_1,b_2;a_{n-3}\right)\dots\mathcal{M}_{(-\theta_{n-3})\theta_{n-2}}\left(b_{n-4},b_{n-3};a_{{2}}\right)\times\\
&\ \times\mathcal{M}_{(-\theta_{n-2})\theta_{n-1}}\left(b_{n-3},a_{\infty};a_{1}\right) \left(\frac{z_{2}}{z_{1}}\right)^{\theta_{n-2}b_{n-3}}\dots\left(\frac{z_{n-2}}{z_{n-3}}\right)^{\theta_2b_{1}}\re^{-\sum_{k=1}^{n-3}\frac{\theta_{k+1}}{2}\partial_{b_k}F\left(\frac{z_{n-2}}{z_{n-3}},\dots,\frac{z_{2}}{z_1}\right)}\times\\
&\ \times \ri\, z\,\Fns{a_{0}}{a_{n-2}}{b_{1}}{a_{{n-3}}}{\dots}{b_{n-3}}{a_{{1}}}{a_{\infty\theta_{n-1}}}\begin{matrix} \alpha_{2,1} \\ a_{\infty}\end{matrix};\frac{z_{n-2}}{z_{n-3}},\dots,\frac{z_{2}}{z_1},\frac{z_{1}}{z}\bigg).
\end{aligned}
\end{equation}
To fully solve the connection problem we also want to find the proportionality factor between the blocks and the Frobenius solutions. From \eqref{semiclassical6pointaround0} we have 
 \be  \Fns{a_{\infty}}{a_{{1}}}{b_{n-3}}{a_{{2}}}{\dots}{b_1}{a_{n-2}}{a_{0\theta}}\begin{matrix} a_{2,1} \\ a_0\end{matrix};\frac{z_{2}}{z_{1}},\frac{z_{3}}{z_{2}},\dots,\frac{z}{z_{n-2}}\bigg) = z_{n-2}^{-\theta a_0}\re^{-{\theta\over 2} \partial_{a_0} F\left(\frac{z_{2}}{z_{1}},\frac{z_{3}}{z_{2}},\dots,\frac{z_{n-2}}{z_{n-3}}\right)}\psi_{0,\theta}
\ee
\be \,z\,\Fns{a_{0}}{a_{n-2}}{b_{1}}{a_{{n-3}}}{\dots}{b_{n-3}}{a_{{1}}}{a_{\infty\theta}}\begin{matrix} \alpha_{2,1} \\ a_{\infty}\end{matrix};\frac{z_{n-2}}{z_{n-3}},\dots,\frac{z_{2}}{z_1},\frac{z_{1}}{z}\bigg) =z_{1}^{-\theta a_{\infty}}\re^{-{\theta\over 2} \partial_{a_{\infty}} F\left(\frac{z_{2}}{z_{1}},\frac{z_{3}}{z_{2}},\dots,\frac{z_{n-2}}{z_{n-3}}\right)} \psi_{\infty,\theta}~.\ee

In this general discussion, we only consider comb-like  configurations of the punctures. As already mentioned, when dealing with differential equations with 5 or more singularities, the situation can be more complicated because of a possibly more involved decomposition of the singularity structure. The analysis of the connection steps remains unchanged, and the final result is still a concatenation of hypergeometric-like connections, but the expansion parameters of the conformal blocks may differ.

\subsection{Solving BPZ equations via gauge theory}\label{sec:gauge}

The connection between classical integrable systems and Seiberg-Witten theory has been known for some time, see \cite{Mbook} for a review and references. However, the extension of this relation to the quantum level was not immediately obvious. Nekrasov and Shatashvili addressed this in \cite{ns}, where they found that transitioning to quantum integrable systems involves activating one of the two parameters in the $\Omega$-background \cite{Moore:1997dj,Lossev:1997bz,no2}. Among various aspects,  this result had a notable implication: it provided a systematic framework for solving the spectral theory of a class of differential equations known as four-dimensional quantum Seiberg-Witten curves. The key of this approach lies in the NS functions, which serve as  building blocks to construct such solutions.

Since the BPZ equations in the semiclassical limit are specific examples of four-dimensional quantum Seiberg-Witten curves, the work of Nekrasov and Shatashvili naturally finds interpretation within Liouville CFT via the AGT correspondence \cite{Alday:2009aq}.  Although viewing the problem from the perspective of Liouville CFT is not strictly necessary, it can sometimes offer an alternative and efficient approach to performing computations.

 For the purpose of this work, we focus on four-dimensional $\mathcal{N}=2$ superconformal linear quiver gauge theories with gauge group $SU(2)^{n-3}$ in the $\Omega$ background which have class $\mathcal{S}$ descriptions \cite{Gaiotto:2009we}. 
A theory of class $\mathcal{S}$ is specified by the following data: a 6d $\mathcal{N}=(2,0)$ theory specified by a Lie algebra $\mathfrak{g}$ of ADE type; a punctured Riemann surface $C$, where the punctures are at ${z_i}$'s; a codimension-2 defect for each puncture ${z_i}$.
With the above data given, a class $\mathcal{S}$ theory is defined as the 4d theory obtained by compactifying the 6d $\mathcal{N}=(2,0)$ theory with defects on the punctured Riemann surface $C$. For $\mathfrak{g}=\text{A}_1$, the Seiberg-Witten curve has the form 
\begin{equation}
\label{SWcurve}
    \Sigma=\{\lambda^2-P(z)\rd z^2=0\}\subset T^*C,
\end{equation}
where $z\in C$ and $P(z)\rd z^2$ is a meromorphic quadratic differential.

Note that, in general, if we have  more than five punctures, there are different realizations of the $\mathcal{N}=2$ theory as a weakly coupled generalized quiver \cite{Gaiotto:2009we}. Here,  we will consider the case of linear quiver gauge theory.

An $SU(2)^{n-3}$ linear quiver gauge theory can be represented by (see for instance \cite{Jeong:2017mfh} and reference therein) \begin{center}
\begin{tikzpicture}
[
node distance = 8mm,
b/.style={rectangle, draw=black, minimum size=25},
c/.style={circle, draw=black, minimum size=25}
]
\node[b] (n1) at (0,0) {$2$};
\node[c] (n2) [right=of n1] {2};
\node(nd) [right = 1.75cm of n1] {$\cdots$};
\node[c] (n4) [right=of n2] {2};
\node[b] (n5) [right=of n4] {2};

\draw[] (n1)   -- (n2);
\draw[] (n4)   -- (n5);

    \node at (2.6,-0.8) {$\underbrace{\quad\quad\quad\quad\quad\quad\quad}_{n-3}$};
\end{tikzpicture},
\end{center}
where the $n-3$ circles represent the $n-3$ $SU(2)$ gauge groups. 
There are $n-3$ vector multiplets of the $SU(2)^{n-3}$ gauge groups.

Besides the vector multiplets  of the $SU(2)^{n-3}$ gauge group, such a quiver theory contains the following matter superfields
\begin{itemize}
\item $n-4$ bifundamental hypermultiplets for $SU(2)_{j-1}\times SU(2)_{j}$ whose masses are given by the parameters $a_{j}$, $j=2,\dots,n-3$. 
\item 2 fundamental hypermultiplets for $SU(2)_{n-3}$, represented by the square boxes on the right, with masses\footnote{There are 16 possible dictionaries between the masses of the (anti-)fundamental hypermultiplets and the semiclassical momenta, which can be found by considering all the possible sign flips. }
\begin{equation}
m_1=a_0-a_{n-2},\quad m_2=a_0+a_{n-2},
\end{equation}
\item 2 antifundamental hypermultiplets, represented by the square boxes on the left, for $SU(2)_1$, with masses 
\begin{equation}
m_3=a_{\infty}-a_{1},\quad m_4=a_{\infty}+a_{1}.
\end{equation}
\end{itemize}
For each gauge factor $SU(2)_i$, the corresponding Yang-Mills coupling constant $g_i$ and the theta-angles $\theta_i$, $i=1,\dots,n-3$, are combined into the complexified gauge coupling
\begin{equation}
\tau_i=\frac{\theta_i}{2\pi}+\ri\,\frac{4\pi}{g_i^2}.
\end{equation}

The vacuum expectation value of the adjoint scalar in the vector multiplet for each gauge factor $SU(2)$ is 
$\vec{b}_i=\mathrm{diag}(b_{i},-b_{i})$.

We can turn on the $\Omega$ background deformation of the theory which introduces two more parameters $\epsilon_1$ and $\epsilon_2$ corresponding to the rotations on two perpendicular planes. In this case, the partition function of the 4d supersymmetric gauge theory can be obtained by the localization method \cite{Moore:1997dj,Lossev:1997bz,Nekrasov:2002qd,no2}. The result is known as the Nekrasov partition function, which is defined as a series in the exponential of the gauge coupling constants $\tau_i$, while each coefficient is a meromorphic function of $b_i$'s and $m_i$'s.  An
example can be found in \eqref{eq:nekfunction}.

We can further insert a surface defect into the theory \cite{Alday:2010vg,Gaiotto:2014ina,Kanno:2011fw,Gaiotto:2009hg,Jeong:2021rll,Jeong:2023qdr,Jeong:2018qpc,Jeong:2017pai,Piatek:2017fyn,Alday:2009fs,Drukker:2009id}. Since we restrict ourselves to the theories of class $\mathcal{S}$, there exists a canonical surface defect whose moduli space is the same Riemman surface $C$. Without loss of generality, we assume that the surface defect corresponds to $z\in C$ such that $0<z<z_1<\cdots z_n<\infty$. The Nekrasov partition function with the insertion of a surface defect has one more expansion parameter, $\frac{z}{z_1}$. An example can be found in \eqref{blockaround0}. 

Recall that we have two $\Omega$ background parameters $\epsilon_1$ and $\epsilon_2$. There is an interesting limit where, if we assume that $\epsilon_1$ corresponds to the rotation along the canonical surface defect,  we take $\epsilon_2\rightarrow 0$ while keeping $\epsilon_1$ finite. This is known as the NS limit. In this particular deformation, the Seiberg-Witten curve is quantized into an oper
\begin{equation}
    \nabla=\partial_z+\epsilon_1^{-1}
\begin{pmatrix} 0 & -P(z)\\ 1 & 0\end{pmatrix}.\end{equation}
Setting $\epsilon_1=1$, such an oper can be recasted into a 2nd order ODE 
\begin{equation}\label{eq:ode}
\begin{aligned}
\left(\frac{\mathrm{d}}{\mathrm{d}z^2}+P(z)\right)\psi(z)=0,
\end{aligned}
\end{equation}
whose spectral theory is fully determined by the NS functions.

The above gauge theory is related to conformal field theory via the famous AGT correspondence, see \cite{LeFloch:2020uop} for a review. The AGT correspondence relates  Liouville and Toda CFTs to  4d $\mathcal{N}=2$ theories which have class $\mathcal{S}$ realizations \cite{Alday:2009aq, Alday:2009fs,Wyllard:2009hg} where conformal blocks are identified with Nekrasov partition functions. For $SU(2)^{n-3}$ quiver theories, the masses $a_i$'s are mapped to the momenta of regular singularities via $b\alpha_i\equiv a_i$, which are denoted by the same $\alpha_i$'s for the reason of this identification. And the scalar v.e.v $b_i$ are mapped to the momenta on internal legs through $b\beta_i\equiv b_i$. The gauge coupling constants are related to the positions of the punctures as
\begin{equation}
 \exp(2\pi\ri\tau_i)=\frac{z_i}{z_{i+1}} ,\end{equation}
where we assume $0<z_1<\cdots z_n<\infty$. 

In the dictionary relating the two sides, the parametrization $b$ of the central charge $c$ can be related to the $\Omega$ background parameters $\epsilon_i$ by
\begin{equation}
b=\sqrt{\frac{\epsilon_2}{\epsilon_1}}.
\end{equation}
In the convention of this paper, we take $\epsilon_1=1$ and $\epsilon_2=b^2$. The semiclassical limit of the 2d CFT is therefore mapped to the NS limit of the supersymmetric quiver gauge theories. The 2nd order ODE \eqref{eq:ode} we mentioned above gives exactly the semiclassical limit of the corresponding BPZ equation. The Nekrasov partition function, incorporating a surface defect in the 4d/2d coupled gauge theory system, is mapped to the conformal blocks arising in the decomposition of the $(n+1)$-point correlation function~\eqref{n+1correlator}, which involves \( n \) primary fields and a degenerate field \( \Phi_{2,1}(z) \).

\section{ Example: Fuchsian equation with five singularities} \label{sec:five}

In this section we focus on the case of a second order  differential equations with five regular singular points at $z=\{0,1,t,q,\infty\}$: 
\begin{equation}\label{diffeq5normal}
\begin{aligned}
\left(\frac{\mathrm{d}}{\mathrm{d}z^2}+\sum_{i=0,1,t,q}\frac{\frac{1}{4}-a_{i}^2}{(z-z_i)^2}+\frac{\frac{1}{4}-a_{\infty}^2-\sum_{i=0,1,t,q}(\frac{1}{4}-a_{i}^2)}{z(z-1)}+\frac{(t-1)\,u_t}{z(z-1)(z-t)}+\frac{\left(q-1\right)\,u_{q}}{z(z-1)\left(z-q\right)}\right)\psi(z)=0.
\end{aligned}
\end{equation}
We examine the regime for the positions of the singularities \( t \) and \( q \) illustrated in \autoref{fig:5point}, specifically \( 0 < |t| < 1 < |q| < \infty \). 
    \begin{figure}[h]
\centering
\begin{tikzpicture}[scale=2]
\def\Rs{0.75}; 
\def\tl{0.5}
\def\skw{0.4}

\draw [line width=\skw mm, domain=20:340] plot ({\Rs*cos(\x)}, {\Rs*sin(\x)}); %top half
\draw [line width=\skw mm,domain=0:\tl] plot ({\Rs*cos(20)+\x}, {\Rs*sin(20)}); 
\draw [line width=\skw mm,domain=0:\tl] plot ({\Rs*cos(340)+\x}, {\Rs*sin(340)}); 
\draw [line width=\skw mm,domain=20:160] plot ({\Rs*cos(\x)+\Rs+\Rs*cos(20)+\tl-0.05}, {\Rs*sin(\x)});
\draw [line width=\skw mm,domain=200:340] plot ({\Rs*cos(\x)+\Rs+\Rs*cos(20)+\tl-0.05}, {\Rs*sin(\x)});
\draw [line width=\skw mm,domain=0:\tl] plot ({\Rs*cos(20)*3+\tl+\x}, {\Rs*sin(20)}); 
\draw [line width=\skw mm,domain=0:\tl] plot ({\Rs*cos(340)*3+\tl+\x}, {\Rs*sin(340)});
\draw [line width=\skw mm,domain=0:160] plot ({\Rs*cos(\x)+\Rs+\Rs*cos(20)*3+\tl*2-0.05}, {\Rs*sin(\x)});
\draw [line width=\skw mm,domain=200:360] plot ({\Rs*cos(\x)+\Rs+\Rs*cos(20)*3+\tl*2-0.05}, {\Rs*sin(\x)}); 

\node[dot,label=below:{\color{black}\( \infty \)}] at (-0.1,-0.3) {};
\node[dot,label=right:{\color{black}\(q \)}] at (0.1,0.3) {};
\node[dot,label=right:{\color{black}\(1\)}] at ({\Rs+\Rs*cos(20)+\tl-0.05},0.3) {};
\node[dot,label=right:{\color{black}\(0\)}] at ({+\Rs+\Rs*cos(20)*3+\tl*2-0.05-0.1},0.3) {};
\node[dot,label=below:{\color{black}\(t\)}] at ({+\Rs+\Rs*cos(20)*3+\tl*2-0.05+0.1},-0.3) {};
\end{tikzpicture}
\caption{Sphere decomposition.%, example I
}
\label{fig:5point}
\end{figure}

\subsection{The gauge theory}\label{sec:gaugefive}

The Riemann surface where $z$ of the differential equation \eqref{diffeq5normal}  lives can be described as a sphere with 5 punctures.
This curve admits a weakly-coupled description as a $\mathcal{N}=2$ superconformal linear quiver gauge theory with gauge group $SU(2)\times SU(2)$, see \autoref{sec:gauge}. In this section the masses of the fundamental hypermultiplets are defoted my by $m_{1,2}=a_t \pm a_0$ and $m_{3,4}=a_q \pm a_{\infty}$, while $a_1$ represents the mass of the matter in the bifundamental. There are two parameters, that we denote with $b_1$ and $b_2$, which represent the v.e.v.~of the scalars in the $\mathcal{N}=2$ vector multiplets. These  are related to the complex moduli $u_t$ and $u_q$ appearing in the potential of \eqref{diffeq5normal} by the Matone relations \eqref{martone}, whose form depends on the decomposition of the five-punctured sphere.

In the case where the punctures are distributed as on   \autoref{fig:5point}, we have $z_1=0, z_2=t, z_3=1, z_4=q, z_5=\infty $. Using this  in \eqref{martone} gives 
\begin{align}
u_{t}&=\lim_{b\to 0}b^2\,t\,\partial_{t}\,\log\,\mathfrak{F}\left(\begin{matrix} \alpha_{q} \\ \alpha_{\infty}\end{matrix}\,\beta_{2}\,
\begin{matrix} \alpha_{1} \\ ~\end{matrix}\, \beta_1\,
\begin{matrix}  \alpha_{t} \\ \alpha_0\end{matrix};\frac{1}{q}, t\right)=-\frac{1}{4}- b_1^2 + a_0^2 + a_t^2+t\frac{\partial F\big({1\over q},t\big)}{\partial t},\label{Matonet}\\
u_{q}&=\lim_{b\to 0}b^2\,q\,\partial_{q}\,\log\,\mathfrak{F}\left(\begin{matrix} \alpha_{q} \\ \alpha_{\infty}\end{matrix}\,\beta_{2}\,
\begin{matrix} \alpha_{1} \\ ~\end{matrix}\, \beta_1\,
\begin{matrix}  \alpha_{t} \\ \alpha_0\end{matrix};\frac{1}{q}, t\right)=-\frac{1}{4}+b_2^2-a_{\infty}^2+a_{q}^2+q\frac{\partial F\big({1\over q},t\big)}{\partial q},\label{Matoneq}
\end{align}
where $F$ is the Nekrasov-Shatashvili  function which we add more details about in \autoref{appendixA}. 
The instanton expansions of the v.e.v. parameters in this case are of the following forms:
\begin{equation}\label{matone7d}
\begin{aligned}
b_1&=\sum_{i,j\ge 0}b_1^{(i,j)}\,t^i\,\frac{1}{q^j},\quad &\text{with}\quad b_1^{(i,j)}=0\quad\text{for}\quad i<j,\\
b_2&=\sum_{i,j\ge 0}b_2^{(i,j)}\,t^i\,\frac{1}{q^j},\quad &\text{with}\quad b_2^{(i,j)}=0\quad\text{for}\quad j<i.
\end{aligned}
\end{equation}
We will say that the $K$th instanton expansion of $b_i$, $i=1,2$, is determined if all the coefficients $b_i^{(m,n)}$ with $m+n=K$ are known.
The Matone relations can be inverted order by order in the two instanton parameters. In particular, for each $K>0$, the $K$th instanton expansion of $b_1$ requires the $(K-1)$th instanton expansion of $b_2$, and vice versa.

The first few  coefficients in the expansions \eqref{matone7d} are given by 
\begin{equation}
\begin{aligned}
b_1^{(0,0)}&=\sqrt{-\frac{1}{4}-u_{t}+a_{0}^2+a_{t}^2},\\
b_1^{(1,0)}&=\frac{\left(1-4 a_{t}^2+2 u_{t}\right) \left(4 a_{0}^2+4 a_{1}^2-4 a_{\infty}^2+4 a_{t}^2+4 a_{q}^2-4 u_{t}-4 u_{q}-3\right)}{8 \sqrt{4 a_{0}^2+4 a_{t}^2-4 u_{t}-1} \left(2 a_{0}^2+2 a_{t}^2-2 u_{t}-1\right)},\\
b_2^{(0,0)}&=\sqrt{\frac{1}{4}+u_{q}+a_{\infty}^2-a_{q}^2},\\
b_2^{(0,1)}&=-\frac{u_{q} \left(-4 a_{0}^2+4 a_{1}^2+4 a_{\infty}^2-4 a_{t}^2-4 a_{q}^2+4 u_{t}+4 u_{q}+1\right)}{8 \left(a_{\infty}^2-a_{q}^2+u_{q}\right) \sqrt{4 a_{\infty}^2-4 a_{q}^2+4 u_{q}+1}}.
\end{aligned}
\end{equation}
To be more precise, the inversion of the Matone relation would lead to two solutions for each v.e.v. parameter, differing for an overall sign; in the above results we chose the branch with the plus sign in front of the canonical branch of  square root.
We computed the inverted Matone relations up to 4 instantons and we noticed the triangular structure of the expansion of the v.e.v. parameters as in \autoref{matone7d}. 
We do not have a proof of the triangular structure of these expansions. However, we found that it holds at all analytically computed orders (up to $i+j=5$), and that it is in good agreement with the numerical results presented in later sections. It would be interesting to gain a better understanding of this structure from a CFT perspective.

In the forthcoming sections we apply the strategy outlined in \autoref{sec:crossing} to some specific examples.

\subsection{Connection problem I}\label{conn1}
We consider the configuration in \autoref{fig:5point}. As a warm-up, we  compute the connection coefficients between $0$ and $t$. This case closely resembles that of the Heun equation as discussed in \cite{Jia:2024zes,Lisovyy:2022flm, Liu:2024eut}.
We define the Frobenius solutions around these points as
\begin{equation}\label{frobenius0andt}
\begin{aligned}
\psi_{t,\pm}(z)&=(z-t)^{\frac{1}{2}\pm a_t}\left(1+\sum_{j\ge 1}c^{t,-}_{j}(z-t)^j\right),\\
\psi_{0,\pm}(z)&=z^{\frac{1}{2}\pm a_0}\left(1+\sum_{j\ge 1}c^{0,\pm}_{j}z^j\right).
\end{aligned}
\end{equation}
Using the gauge theoretic expression of the conformal blocks \eqref{semist} given in \autoref{appendix5+1} we find 
\be \psi_{0,\theta}(z)=t^{\theta a_{0} }\E^{\frac1 2\theta\partial_{a_0}F\left(\frac1 q , t\right)}\mathcal{F}\left(\begin{matrix}a_q\\ a_{\infty}\end{matrix}\,b_2\,\begin{matrix}a_1\\ {}\end{matrix}\,b_1\,\begin{matrix} a_t\\ {}\end{matrix}\,a_{0\theta}\,\begin{matrix}a_{2,1}\\ a_{0}\end{matrix};\frac{1}{q},t,\frac{z}{t}\right). \ee
Likewise the Frobenius solutions around $t$ are
\be \psi_{t,\theta}(z)=
%\E^{\ri\pi(-\frac 1 2-\theta a_t)}
\left(\frac{1}{1-t}\right)^{-\frac 1 2-\theta a_t}\left(\frac{t}{t-1}\right)^{\theta a_{t} }\re^{\frac{\theta}{2}\partial_{a_t}F\left({1-t\over q-t},\frac{t}{t-1}\right)}\,\mathcal{F}\left( \begin{matrix} a_{q} \\ a_{\infty} \end{matrix} \, b_2 \, \begin{matrix} a_1  \\ \, \end{matrix} \,b_1 \, \begin{matrix} a_0 \\ \, \end{matrix} \,a_{t\theta} \, \begin{matrix} a_{2,1} \\ a_t \end{matrix} ; \frac{1-t}{q-t}, \frac{t}{t-1}, \frac{-z+t}{t} \right)~.\ee
%where we applied the following transformation on the block \be
%z~\to~{z-t\over 1-t} 
%\ee
By using \eqref{definitionF} and \eqref{jacobiank} we have
\begin{equation}
    \left(\frac{t}{t-1}\right)^{-\theta a_{t} }\re^{-\frac{\theta}{2}\partial_{a_t}F\left({1-t\over q-t},\frac{t}{t-1}\right)}
    %=t^{-a_{0\theta} }\E^{-\frac1 2\theta\partial_{a_0}F\left(\frac1 q , t\right)}
    =\E^{-\frac1 2\theta\partial_{a_t}\lim\limits_{b\rightarrow 0}b^2\log\mathfrak{F}\left(\begin{matrix}\alpha_q\\ \alpha_{\infty}\end{matrix}\,\beta_2\,\begin{matrix}\alpha_1\\ {}\end{matrix}\,\beta_1\,\begin{matrix}\alpha_0\\ \alpha_{t}\end{matrix};{1-t\over q-t},\frac{t}{t-1}\right)}.
\end{equation}
Since\footnote{To be more precise, the crossing symmetry applies to correlators as in \eqref{crosssympre}. Since we assume the symmetry passes to the integrand and all the structure constants $C$ are the same on the two sides of the identity, we can relate conformal blocks by the same Jacobian factor.}
\begin{align}
&\partial_{a_t}\lim\limits_{b\rightarrow 0}b^2\log\mathfrak{F}\left(\begin{matrix}\alpha_q\\ \alpha_{\infty}\end{matrix}\,\beta_2\,\begin{matrix}\alpha_1\\ {}\end{matrix}\,\beta_1\,\begin{matrix}\alpha_t\\ \alpha_{0}\end{matrix};{1\over q},t\right)\\
&=\partial_{a_t}\lim\limits_{b\rightarrow 0}b^2\left(\log\mathfrak{F}\left(\begin{matrix}\alpha_q\\ \alpha_{\infty}\end{matrix}\,\beta_2\,\begin{matrix}\alpha_1\\ {}\end{matrix}\,\beta_1\,\begin{matrix}\alpha_0\\ \alpha_{t}\end{matrix};{1-t\over q-t},\frac{t}{t-1}\right)\left(\frac{1}{1-t}\right)^{\Delta_0+\Delta_1+\Delta_t+\Delta_q+\Delta_\infty}\right)~,
\end{align}
we have
\begin{equation}  
\left(\frac{t}{(t-1)}\right)^{-\theta a_{t} }\re^{-\frac{\theta}{2}\partial_{a_t}F\left({1-t\over q-t},\frac{t}{t-1}\right)}=t^{-\theta a_t}\re^{-\frac{\theta}{2}\partial_{a_t}F\big({1\over q},t\big)}\left(\frac{1}{1-t}\right)^{-\theta a_t}.
\end{equation}
Hence, 
\be\label{connIt}  \ba  \psi_{t,\theta}(z)&=
%\E^{\ri\pi(-\frac 1 2-\theta a_t)} 
\left(\frac{1}{1-t}\right)^{-\frac 1 2-\theta a_t}t^{\theta a_t}\re^{\frac{\theta}{2}\partial_{a_t}F\big({1\over q},t\big)}\left(\frac{1}{1-t}\right)^{\theta a_t}\,\mathcal{F}\left( \begin{matrix} a_{q} \\ a_{\infty} \end{matrix} \, b_2 \, \begin{matrix} a_1  \\ \, \end{matrix} \,b_1 \, \begin{matrix} a_0 \\ \, \end{matrix} \,a_{t\theta} \, \begin{matrix} a_{2,1} \\ a_t \end{matrix} ; \frac{1-t}{q-t}, \frac{t}{t-1}, \frac{-z+t}{t} \right)\\
&=
%\E^{\ri\pi(-\frac 1 2-\theta a_t)} 
\left({1-t}\right)^{\frac 1 2}t^{\theta a_t}\re^{\frac{\theta}{2}\partial_{a_t}F\big({1\over q},t\big)}\,\mathcal{F}\left( \begin{matrix} a_{q} \\ a_{\infty} \end{matrix} \, b_2 \, \begin{matrix} a_1  \\ \, \end{matrix} \,b_1 \, \begin{matrix} a_0 \\ \, \end{matrix} \,a_{t\theta} \, \begin{matrix} a_{2,1} \\ a_t \end{matrix} ; \frac{1-t}{q-t}, \frac{t}{t-1}, \frac{-z+t}{t} \right),\ea\ee
Thanks to crossing symmetry we have
\begin{align}\label{connIblock}
&\E^{\ri\pi(-\frac 1 2-\theta a_t)} \left({1-t}\right)^{\frac 1 2}\mathcal{F}\left( \begin{matrix} a_{\infty} \\ a_{q} \end{matrix} \, b_2 \, \begin{matrix} a_1  \\ \, \end{matrix} \,b_1 \, \begin{matrix} a_0 \\ \, \end{matrix} \,a_{t\theta} \, \begin{matrix} a_{2,1} \\ a_t \end{matrix} ; \frac{1-t}{q-t}, \frac{t}{(t-1)}, \frac{-z+t}{t} \right)\\
&=\sum_{\theta'=\pm} \mathcal{M}_{\theta\theta'}(a_t,a_0;b_1) \mathcal{F}\left(\begin{matrix}a_q\\ a_{\infty}\end{matrix}\,b_2\,\begin{matrix}a_1\\ {}\end{matrix}\,b_1\,\begin{matrix} a_t\\ {}\end{matrix}\,a_{0\theta'}\,\begin{matrix}a_{2,1}\\ a_{0}\end{matrix};\frac{1}{q},t,\frac{z}{t}\right),
\end{align}
where  the extra phase factor $\E^{\ri\pi(-\frac 1 2-\theta a_t)}$ is chosen to match the phases on both sides.\footnote{We remark that this phase factor can change according to the choices of the (complex) values of the parameters. Here, we fixed it by matching the phases on both sides for real parameters, as shown in the Table below. We also show in \eqref{complexpar} an example involving complex parameters.}
%we explain the extra phase factor $\E^{\ri\pi(-\frac 1 2-\theta a_t)}$ in \autoref{appen:phase}.
The connection coefficients therefore read
\be\label{conn1tto0}
\psi_{t,\theta}(z)= \E^{\ri\pi\left(\frac{1}{2} + \theta a_t\right)} t^{\theta a_t} 
\E^{\frac{\theta}{2} \partial_{a_t} F\left(\frac{1}{q}, t\right)} 
\sum_{\theta'=\pm} \mathcal{M}_{\theta\theta'}(a_t, a_0; b_1)
t^{-\theta' a_0} \E^{-\frac{1}{2}\theta' \partial_{a_0} F\left(\frac{1}{q}, t\right)} \psi_{0,\theta'}.
\ee
Notice that the  matrix $\mathcal{M}_{\theta\theta'}$ can be simply reabsorbed in the one-loop part of the NS function as in \eqref{eq:fullns} and we have
\begin{empheq}[box=\fbox]{equation}\label{eq:conn1full}
\psi_{t,\theta}(z)= \E^{\ri\pi\left(\frac{1}{2} + \theta a_t\right)} t^{\theta a_t} 
\E^{\frac{\theta}{2} \partial_{a_t} F^{\rm NS}\left(\frac{1}{q}, t,-\theta a_t\right)} \Gamma\left(1+2 \theta a_t\right)
\sum_{\theta'=\pm} \Gamma\left(-2\theta' a_0\right)t^{-\theta' a_0}\re^{-{1\over 2}\theta' \partial_{a_0} F^{\rm NS} \left(\frac{1}{q}, t,-\theta a_t\right)} \psi_{0,\theta'}.
\end{empheq}
The function $F^{\rm NS}$  is defined in \eqref{eq:fullns}, and in \eqref{eq:conn1full} we have explicitly highlighted its dependence on $a_t$, as the dependences on the other $a_i$’s are fixed in the same manner as in \eqref{eq:conn1full}, see the comment around \eqref{eq:fnsconv}.
As an example we choose the following values for the parameters
\begin{equation}\label{eq:numtest}
\begin{aligned}
t&=\frac{1}{100},\quad
q=200,\quad
a_0=\frac{97}{70},\quad
a_1=\frac{7}{141},\quad
a_{\infty}=\frac{51}{40},\\
a_t&=\frac{10}{9},\quad
a_{q}=\frac{4}{3},\quad\quad
u_t=\frac{1}{50},\quad
u_{q}=\frac{3}{8},
\end{aligned}
\end{equation}
and we check the values of the connection coefficients obtained as \eqref{eq:conn1full}  against the ratios of Wronskians
\begin{equation}
\frac{W\left(\psi_{t,-}(z),\psi_{0,+}(z)\right)}{W\left(\psi_{0,-}(z),\psi_{0,+}(z)\right)}\quad\text{and}\quad\frac{W\left(\psi_{t,-}(z),\psi_{0,-}(z)\right)}{W\left(\psi_{0,+}(z),\psi_{0,-}(z)\right)},
\end{equation}
where we compute 100 coefficients $c_j$ for each Frobenius solution \eqref{frobenius0andt}.

We get the following results for the real and imaginary parts of the connection coefficients
\begin{table}[h]
    \centering
    \begin{tabular}{| c | c  c| }
    \hline
     Number of instantons & Real part  & Imaginary part  \\

    \hline   

    0  & -{\bf 14}2813.86214  & -{\bf 39}2377.86142  \\
        
    1  & -{\bf 1436}23.93208  & -{\bf 3946}03.51028  \\

    2  & -{\bf 143630}.89467  & -{\bf 394622}.63983 \\

    3  & -{\bf 143630.92}880  & -{\bf 394622.7}3363 \\
         \hline

    Wronskian & - 143630.92949 & - 394622.73552 \\
         \hline
    \end{tabular}
    \caption{Check of numerical values of $\mathcal{C}_{t,-}^{0,+}$ at the parameters \eqref{eq:numtest}}
\end{table}
\begin{table}[h]
    \centering
    \begin{tabular}{| c | c  c| }
    \hline
     Number of instantons & Real part  & Imaginary part \\

    \hline 

    0  & {\bf 0.0}3884950457  & {\bf 0.1}0673813655  \\

    1  & {\bf 0.0404}7079339 &  {\bf 0.111}19259100  \\

    2  & {\bf 0.040481}02795  & {\bf 0.111220}71021 \\

    3  & {\bf 0.0404810}7976  & {\bf 0.11122085}255 \\
         \hline
   Wronskian & 0.04048108053 & 0.11122085466 \\
         \hline
    \end{tabular}
    \caption{Check of numerical values of $\mathcal{C}_{t,-}^{0,-}$ at the parameters \eqref{eq:numtest}}
\end{table}

We finally show an example involving complex values of the parameters. Fixing
\begin{equation}\label{complexpar}
\begin{aligned}
t&=\frac{3}{100}+\frac{1}{1000}\,\mathrm{i},\quad
q=700+4\,\mathrm{i},\quad
a_0=\frac{13}{11}+\frac{1}{10}\,\mathrm{i},\quad
a_1=\frac{2}{7}-\frac{1}{12}\,\mathrm{i},\quad
a_{\infty}=\frac{9}{8}+\frac{1}{9}\,\mathrm{i},\\
a_t&=\frac{7}{6}-\frac{1}{11}\,\mathrm{i},\quad
a_{q}=\frac{11}{9},\quad\quad
u_t=\frac{3}{40},\quad
u_{q}=\frac{2}{5},
\end{aligned}
\end{equation}
at 3 instantons, replacing the phase factor of Eq.~\eqref{conn1tto0} with $\exp \left(i \pi  \left(a_t-\frac{1}{2}\right)\right)$, we find the results in Table 3.
\begin{table}[h]
    \centering
    \begin{tabular}{| c | c  c| }
    \hline
     Number of instantons & Real part  & Imaginary part  \\

    \hline   

    0  & -{\bf 38}80.678616  & {\bf 7}802.008331  \\
        
    1  & -{\bf 3872}.550280  & {\bf 7784}.393415  \\

    2  & -{\bf 3872.9}14312  & {\bf 7784.5}85712 \\

    3  & -{\bf 3872.925}641  & {\bf 7784.592}381 \\
         \hline

    Wronskian & -3872.925966 & 7784.592815 \\
         \hline
    \end{tabular}
    \caption{Check of numerical values of $\mathcal{C}_{t,-}^{0,+}$ at the parameters \eqref{complexpar}}
\end{table}

\subsection{Connection problem II} \label{sec:connII}
In this section, we study the connection formula for the solutions at $z=t$ and $z=1$, in the regime $|t|\ll 1 \ll |q|$, as in \autoref{fig:5point}.
Let us take the singularities $t$ and $q$ lying on the positive real axis\footnote{We use this hypothesis to perform the numerical checks, but we expect the formulae to hold for complex $t$ and $q$ too.}. For the local solutions around $1$, we use the conformal blocks of the form  
\begin{equation}\label{noncombdiagram}
\begin{tikzpicture}[baseline={(current bounding box.center)}, node distance=1cm and 1.5cm]
\coordinate[label=above:$\alpha_{\infty}$] (e1);
\coordinate[below=of e1] (aux1);
\coordinate[left=of aux1,label=left:$\alpha_{q}$] (e2);
\coordinate[right=1.5cm of aux1] (aux2);
\coordinate[above=of aux2,label=right:$\alpha_{1\theta_3}$] (aux4);
\coordinate[above=of aux4] (e3);
\coordinate[above=of e3,label=above:$\alpha_{1}$] (e6);
\coordinate[left=of e3,label=left:$\alpha_{2,1}$] (e7);
\coordinate[right=1.5cm of aux2] (aux3);
\coordinate[above=of aux3,label=above:$\alpha_{0}$] (e4);
\coordinate[right=of aux3,label=right:$\alpha_{t}$] (e5);

\draw (e1) -- (aux1);
\draw (aux1) -- (e2);
\draw (e3) -- (aux4);
\draw (aux2) -- (aux4);
\draw (e3) -- (e6);
\draw[dashed,red] (e3) -- (e7);
\draw (aux3) -- (e4);
\draw (aux3) -- (e5);
\draw (aux1) -- node[label=below:$\beta_2$] {} (aux2);
\draw (aux2) -- node[label=below:$\beta_{1}$] {} (aux3);
\end{tikzpicture}
\end{equation}
However, to our knowledge,  combinatorial expressions for such conformal blocks have not been written down. 
Nevertheless, the connection coefficient can still be calculated as described in \autoref{sec:crossing}. 
As a normalization for these blocks we propose
\begin{equation}\mathcal{F}\left( \begin{matrix} a_{\infty} \\ a_{q} \end{matrix} \, b_2 \, \begin{matrix} \begin{matrix} a_{1} & a_{2,1} \end{matrix} \\ a_{1\theta_3}  \\  \\ \, \end{matrix} \,b_1  \, \begin{matrix} a_{0} \\ a_t \end{matrix}\right)=\E^{-\frac{1}{2}\theta_3\partial_{a_1}F\big({1\over q},t\big)}\psi_{1,\theta_3}(z).
\end{equation}
Let us work out the connection problem for the solutions around $t$ and 1.
Schematically, we follow the steps
\begin{equation}\label{3steps1}
\begin{tikzpicture}[baseline={(current bounding box.center)}, node distance=1cm and 1.5cm]
\coordinate[label=above:$\alpha_{\infty}$] (e1);
\coordinate[below=of e1] (aux1);
\coordinate[left=of aux1,label=left:$\alpha_{q}$] (e2);
\coordinate[right=1.5cm of aux1] (aux2);
%%%%%%%%%%%%%%
\coordinate[below right=0.6cm and 0.75cm of aux2] (auxmd1);
%%%%%%%%%%%%%%
\coordinate[above=of aux2,label=above:$\alpha_{1}$] (e3);
\coordinate[right=1.5cm of aux2] (aux3);
\coordinate[above=of aux3,label=above:$\alpha_{0}$] (e4);
\coordinate[right=1.5cm of aux3] (aux4);
\coordinate[above=of aux4,label=above:$\alpha_{2,1}$] (e5);
\coordinate[right=of aux4,label=right:$\alpha_t$] (e6);
%%%%%%%%%%%%%%%%%%%%%%%%%%%%%%%%%%%%%%%%%%
\coordinate[below=3.0cm of e1,label=above:$\alpha_{\infty}$] (e12);
\coordinate[below=of e12] (aux12);
\coordinate[left=of aux12,label=left:$\alpha_{q}$] (e22);
\coordinate[right=1.5cm of aux12] (aux22);
\coordinate[above=of aux22,label=above:$\alpha_{1}$] (e32);
%%%%%%%%%%%%%%
\coordinate[below right=0.6cm and 0.75cm of aux22] (auxmd2);
\coordinate[above right=0.55cm and 0.75cm of e32] (auxmu2);
%%%%%%%%%%%%%%
\coordinate[right=1.5cm of aux22] (aux32);
\coordinate[above=of aux32,label=above:$\alpha_{2,1}$] (e42);
\coordinate[right=1.5cm of aux32] (aux42);
\coordinate[above=of aux42,label=above:$\alpha_{0}$] (e52);
\coordinate[right=of aux42,label=right:$\alpha_t$] (e62);
%%%%%%%%%%%%%%%%%%%%%%%%%%%%%%%%%%%%%
\coordinate[below right =5.0cm and 0.8cm of e12,label=above:$\alpha_{\infty}$] (e13);
\coordinate[below=of e13] (aux13);
\coordinate[left=of aux13,label=left:$\alpha_{q}$] (e23);
\coordinate[right=1.5cm of aux13] (aux23);
\coordinate[above=of aux23,label=right:$\alpha_{1\theta_3}$] (aux43);
\coordinate[above=of aux43] (e33);
\coordinate[above=of e33,label=above:$\alpha_{1}$] (e63);
\coordinate[left=of e33,label=left:$\alpha_{2,1}$] (e73);
\coordinate[right=1.5cm of aux23] (aux33);
\coordinate[above=of aux33,label=above:$\alpha_{0}$] (e43);
\coordinate[right=of aux33,label=right:$\alpha_{t}$] (e53);

\draw (e13) -- (aux13);
\draw (aux13) -- (e23);
\draw (e33) -- (aux43);
\draw (aux23) -- (aux43);
\draw (e33) -- (e63);
\draw[dashed,red] (e33) -- (e73);
\draw (aux33) -- (e43);
\draw (aux33) -- (e53);
\draw (aux13) -- node[label=below:$\beta_2$] {} (aux23);
\draw (aux23) -- node[label=below:$\beta_{1\theta_2}$] {} (aux33);

\draw (e1) -- (aux1);
\draw (aux1) -- (e2);
\draw (e3) -- (aux2);
\draw (e4) -- (aux3);
\draw[dashed,red] (aux4) -- (e5);
\draw (aux4) -- (e6);
\draw (aux1) -- node[label=below:$\beta_{2}$] {} (aux2);
\draw (aux2) -- node[label=below:$\beta_{1}$] {} (aux3);
\draw (aux3) -- node[label=below:$\alpha_{t\theta_1}$] {} (aux4);
%%%%%%%%%%%%%%%%%%%%%%%%%%
\draw (e12) -- (aux12);
\draw (aux12) -- (e22);
\draw (e32) -- (aux22);
\draw[dashed,red] (e42) -- (aux32);
\draw (aux42) -- (e52);
\draw (aux42) -- (e62);
\draw (aux12) -- node[label=below:$\beta_2$] {} (aux22);
\draw (aux22) -- node[label=below:$\beta_{1}$] {} (aux32);
\draw (aux32) -- node[label=below:$\beta_{1\theta_2}$] {} (aux42);
%%%%%%%%%%%%%%%%%%%%%%%%%%%%%

%%%%%%%%%%%%%%%%%%%%%%%%%%%%%

\draw[->,line width=0.05cm, orange] (auxmd1) -- (auxmu2);
\coordinate[above left = 0.55cm and 0.05cm of e63] (auxmun);
\draw[->,line width=0.05cm, orange] (auxmd2) -- (auxmun);

\end{tikzpicture}
\end{equation}

%%%%%%%%%%%%%%%%%%%%%%%%%%%%%
We find
 \begin{equation}
\label{connIIt1}
\begin{aligned}
\psi_{t,\theta_1}(z) &= t^{\theta_1 a_t} \E^{\frac{\theta_1}{2}\partial_{a_t}F\big({1\over q},t\big)}
\sum_{\theta_2,\theta_3=\pm} \mathcal{M}_{\theta_1\theta_2}(a_t, b_1, a_0) 
\mathcal{M}_{(-\theta_2)\theta_3}(b_1, a_1, b_2) \times \\
&\quad \times t^{\theta_2 b_1} \E^{-\frac{\theta_2}{2}\partial_{b_1}F\big({1\over q},t\big)}
\E^{\ri\pi\left(-\left(\frac{1}{2} + \theta_3 a_1\right)\right)}
\E^{-\frac{\theta_3}{2}\partial_{a_1}F\big({1\over q},t\big)}\psi_{1,\theta_3}(z)
\end{aligned}
\end{equation}
This problem was also recently discussed in \cite[App.B]{Liu:2024eut}.

By using the full NS functions we can write the following closed form expression
 \begin{empheq}[box=\fbox]{equation}
\label{connIIt1i}
\begin{aligned}
\psi_{t,\theta_1}(z) &= {2\re^{-\ri{\pi\over 2}} \pi t^{\theta_1 a_t} \E^{\frac{\theta_1}{2}\partial_{a_t}F^{\rm NS}\big({1\over q},t,-\theta_1 a_t\big)}\over \sin (2 \pi  b_1)} \Gamma (2 a_t \theta_1 +1)
\sum_{\theta_3=\pm} \E^{\ri\pi\left(\theta_3 a_1\right)}\Gamma (-2 a_1 \theta_3 )\\
&\quad
\sinh\left({1\over 2}\partial_{b_1}F^{\rm NS}\big({1\over q},t, -\theta_3 a_1, -\theta_1 a_t\big)\right)
\E^{-\frac{\theta_3}{2}\partial_{a_1}F^{\rm NS}\big({1\over q},t, -\theta_3 a_1 \big)}\psi_{1,\theta_3}(z)~,
\end{aligned}
\end{empheq}
where $F^{\rm NS}$ is given in \eqref{eq:fullns}. As before, we use the convention that in the arguments of the  $F^{\rm NS}$ function, we  explicitly write the dependence on the $a_i$ only when they have a different sign w.r.t.~\eqref{eq:fullns}, see  \eqref{eq:fnsconv}.

We check such expressions
at parameters 
\begin{equation}\label{paraconnIIt1}
\begin{aligned}
t&=\frac{1}{100},\quad
q=200,\quad
a_0=\frac{97}{70},\quad
a_1=\frac{7}{141},\quad
a_{\infty}=\frac{51}{40},\\
a_t&=\frac{10}{9},\quad
a_{q}=\frac{4}{3},\quad\quad
u_t=\frac{73}{50},\quad
u_{q}=\frac{3}{8}\ .
\end{aligned}
\end{equation}
The comparison of numerical and instanton results for the connection coefficient of $\psi_{t,-}$ and $\psi_{1,\pm}$ is in \autoref{tto11} and \autoref{tto12}.
\begin{table}
    \centering
 \begin{tabular}{| c | c  c| }
    \hline
     Number of instantons & Real part  & Imaginary part  \\

    \hline   

    0  & $-\textbf{16}9747.731$  & $\textbf{1.0}79527249\times 10^6$  \\
        
    1  & $-\textbf{1683}85.977$  & $\textbf{1.070}86704\times 10^6$  \\

    2  & $-\textbf{16837}0.680$  & $\textbf{1.07076}9760\times 10^6$ \\

    3  & $-\textbf{168370.2}87$  & $\textbf{1.0707672}60\times 10^6$ \\
         \hline

    Wronskian & $-168370.3$  &  $1.0707673\times 10^6$ \\
         \hline
    \end{tabular}
    \caption{Check of numerical values of $\mathcal{C}_{t,-}^{1,+}$ at the parameters \eqref{paraconnIIt1}. }
    \label{tto11}
\end{table}
\begin{table}
    \centering
 \begin{tabular}{| c | c  c| }
    \hline
     Number of instantons & Real part  & Imaginary part  \\

    \hline   

    0  & $-\textbf{14}9860.161$  & $-\textbf{9}53050.306$  \\
        
    1  & $-\textbf{1486}45.277$  & $-\textbf{945}324.135$ \\

    2  & $-\textbf{148630}.420$  & $-\textbf{94522}9.654 $ \\
    3  & $-\textbf{148630.1}02$  & $-\textbf{945227.6}27$ \\
         \hline
    Wronskian & $-148630.1$  &  $945227.7$ \\
        \hline
    \end{tabular}
    \caption{Check of numerical values of $\mathcal{C}_{t,-}^{1,-}$ at the parameters \eqref{paraconnIIt1}}
    \label{tto12}
\end{table}

\subsection{ Connection problem III} \label{sec:connIII}
In this section, we study the connection formula for the solutions at $z=1$ and $z=q$, in the regime $|t|\ll 1 \ll |q|$, as in \autoref{fig:5point}.
Let us take the singularities $t$ and $q$ lying on the positive real axis. For the local solutions around $1$, we use a conformal block of the form  \eqref{noncombdiagram}.

Schematically, we use the following steps
\begin{equation}\label{3steps2}
\begin{tikzpicture}[baseline={(current bounding box.center)}, node distance=1cm and 1.5cm]
\coordinate[label=above:$\alpha_{\infty}$] (e1);
\coordinate[below=of e1] (aux1);
\coordinate[left=of aux1,label=left:$\alpha_{q}$] (e2);
\coordinate[right=1.5cm of aux1] (aux2);
\coordinate[below=0.5cm of aux2] (aux2d);
\coordinate[above=of aux2,label=right:$\alpha_{1\theta_3}$] (aux4);
\coordinate[above=of aux4] (e3);
\coordinate[above=of e3,label=above:$\alpha_{1}$] (e6);
\coordinate[left=of e3,label=left:$\alpha_{2,1}$] (e7);
\coordinate[right=1.5cm of aux2] (aux3);
\coordinate[above=of aux3,label=above:$\alpha_{0}$] (e4);
\coordinate[right=of aux3,label=right:$\alpha_{t}$] (e5);
%%%%%%%%%%%%%%%%%%%%%%%%%%%%%%%%%%%%%
\coordinate[below left=3.0cm and 0.75cm of e1,label=above:$\alpha_{\infty}$] (e13);
\coordinate[below=of e13] (aux13);
\coordinate[left=of aux13,label=left:$\alpha_{q}$] (e23);
\coordinate[right=1.5cm of aux13] (aux23);
\coordinate[above=of aux23,label=above:$\alpha_{2,1}$] (e33);
%%%%%%%%%%%%%%
\coordinate[below right=0.6cm and 0.75cm of aux23] (auxmd3);
\coordinate[above right=0.55cm and 0.75cm of e33] (auxmu3);
%%%%%%%%%%%%%%
\coordinate[right=1.5cm of aux23] (aux33);
\coordinate[above=of aux33,label=above:$\alpha_{1}$] (e43);
\coordinate[right=1.5cm of aux33] (aux43);
\coordinate[above=of aux43,label=above:$\alpha_{0}$] (e53);
\coordinate[right=of aux43,label=right:$\alpha_t$] (e63);
%%%%%%%%%%%%%%%%%%%%%%%%%%%%%%%%%%%%%
\coordinate[below=3.0cm of e13,label=above:$\alpha_{2,1}$] (e15);
\coordinate[below=of e15] (aux15);
\coordinate[left=of aux15,label=left:$\alpha_{q}$] (e25);
\coordinate[right=1.5cm of aux15] (aux25);
\coordinate[above=of aux25,label=above:$\alpha_{\infty}$] (e35);
%%%%%%%%%%%%%%
\coordinate[below right=0.6cm and 0.75cm of aux25] (auxmd5);
\coordinate[above right=0.55cm and 0.75cm of e35] (auxmu5);
%%%%%%%%%%%%%%
\coordinate[right=1.5cm of aux25] (aux35);
\coordinate[above=of aux35,label=above:$\alpha_{1}$] (e45);
\coordinate[right=1.5cm of aux35] (aux45);
\coordinate[above=of aux45,label=above:$\alpha_{0}$] (e55);
\coordinate[right=of aux45,label=right:$\alpha_t$] (e65);
%%%%%%%%%%%%%%%%%%%%%%%%%%%%%%%%%%%%%%%%%%%
\coordinate[below=3.0cm of e15,label=above:$\alpha_{0}$] (e14);
\coordinate[below=of e14] (aux14);
\coordinate[left=of aux14,label=left:$\alpha_{t}$] (e24);
\coordinate[right=1.5cm of aux14] (aux24);
\coordinate[above=of aux24,label=above:$\alpha_{1}$] (e34);
%%%%%%%%%%%%%%
\coordinate[below right=0.6cm and 0.75cm of aux24] (auxmd4);
\coordinate[above right=0.55cm and 0.75cm of e34] (auxmu4);
%%%%%%%%%%%%%%
\coordinate[right=1.5cm of aux24] (aux34);
\coordinate[above=of aux34,label=above:$\alpha_{\infty}$] (e44);
\coordinate[right=1.5cm of aux34] (aux44);
\coordinate[above=of aux44,label=above:$\alpha_{2,1}$] (e54);
\coordinate[right=of aux44,label=right:$\alpha_{q}$] (e64);

\draw (e1) -- (aux1);
\draw (aux1) -- (e2);
\draw (e3) -- (aux4);
\draw (aux2) -- (aux4);
\draw (e3) -- (e6);
\draw[dashed,red] (e3) -- (e7);
\draw (aux3) -- (e4);
\draw (aux3) -- (e5);
\draw (aux1) -- node[label=below:$\beta_2$] {} (aux2);
\draw (aux2) -- node[label=below:$\beta_{1}$] {} (aux3);
%%%%%%%%%%%%%%%%%%%%%%%%%%%%%
\draw (e13) -- (aux13);
\draw (aux13) -- (e23);
\draw[dashed,red] (e33) -- (aux23);
\draw (e43) -- (aux33);
\draw (aux43) -- (e53);
\draw (aux43) -- (e63);
\draw (aux13) -- node[label=below:$\beta_{2}$] {} (aux23);
\draw (aux23) -- node[label=below:$\beta_{2\theta_4}$] {} (aux33);
\draw (aux33) -- node[label=below:$\beta_{1}$] {} (aux43);
%%%%%%%%%%%%%%%%%%%%%%%%%%%%%
\draw[dashed,red] (e15) -- (aux15);
\draw (aux15) -- (e25);
\draw (e35) -- (aux25);
\draw (e45) -- (aux35);
\draw (aux45) -- (e55);
\draw (aux45) -- (e65);
\draw (aux15) -- node[label=below:$\alpha_{q\theta_5}$] {} (aux25);
\draw (aux25) -- node[label=below:$\beta_{2\theta_4}$] {} (aux35);
\draw (aux35) -- node[label=below:$\beta_{1}$] {} (aux45);
%%%%%%%%%%%%%%%%%%%%%%%%%%%%%%%%%%%%
\draw (e14) -- (aux14);
\draw (aux14) -- (e24);
\draw (e34) -- (aux24);
\draw (e44) -- (aux34);
\draw[dashed,red] (aux44) -- (e54);
\draw (aux44) -- (e64);
\draw (aux14) -- node[label=below:$\beta_{1}$] {} (aux24);
\draw (aux24) -- node[label=below:$\beta_{2\theta_4}$] {} (aux34);
\draw (aux34) -- node[label=below:$\alpha_{q\theta_5}$] {} (aux44);
%%%%%%%%%%
\draw[->,line width=0.05cm, orange] (aux2d) -- (auxmu3);
\draw[->,line width=0.05cm, orange] (auxmd3) -- (auxmu5);
\draw[->,line width=0.05cm, orange] (auxmd5) -- (auxmu4);
\end{tikzpicture}
\end{equation}
We find
\be
\label{paraconnII1qi}
\ba
\psi_{1,\theta_3}(z) &= \E^{\frac{\theta_3}{2}\partial_{a_1}F\big({1\over q},t\big)}
\sum_{\theta_4,\theta_5=\pm} \mathcal{M}_{\theta_3\theta_4}(a_1, b_2, b_1)
\mathcal{M}_{(-\theta_4)\theta_5}(b_2, a_q, a_\infty) \times \\
&\quad \times q^{-\theta_4 b_2} \E^{-\frac{\theta_4}{2}\partial_{b_2}F\big({1\over q},t\big)}
q^{-\theta_5 a_q} \E^{-\frac{\theta_5}{2}\partial_{a_q}F\big({1\over q},t\big)}
\E^{\ri\pi\left(-\left(\frac{1}{2} + \theta_5 a_q\right)\right)}\psi_{q,\theta_5}(z).
\ea
\ee
As before, we can write this by using the full NS function and we get
\begin{empheq}[box=\fbox]{equation}
\label{paraconnII1q}
\begin{aligned}
\psi_{1,\theta_3}(z) &= {2 \pi \re^{-{\ri \pi\over 2} }\E^{\frac{\theta_3}{2}\partial_{a_1}F^{\rm NS}\big({1\over q},t,\theta_3 a_1\big)} \over \sin (2 \pi  b_2)}\Gamma (1+2 \theta_3 a_1  )\sum_{\theta_5=\pm}  \Gamma (-2\theta_5 a_q)  \re^{- \ri \pi  \theta_5 a_q}\\
& \sinh\left({1\over 2}\partial_{b_2}F^{\rm NS}\big({1\over q},t,\theta_3 a_1,\theta_5 a_q \big)\right)\re^{-\frac{\theta_5}{2}\partial_{a_q}F^{\rm NS}\big({1\over q},t, \theta_5 a_q\big)}
q^{-\theta_5 a_q}\psi_{q,\theta_5}(z)~,
\end{aligned}
\end{empheq}
where $F^{\rm NS}$ is given in \eqref{eq:fullns}.
In \eqref{paraconnII1q}, in the argument of the $F^{\rm NS}$ function, we explicitly wrote the dependence only for the mass parameters  $a_i$  that are subject to sign changes, see  conventions in \eqref{eq:fnsconv}. 
We check this identity
at the same parameter \eqref{paraconnIIt1}.
The comparison of numerical and instanton results for the connection coefficient of $\psi_{1,-}$ and $\psi_{q,\mp}$ is in \autoref{1toq1} and \autoref{1toq2}.
\begin{table}[h]
\centering
 \begin{tabular}{| c | c  c| }
    \hline
     Number of instantons & Real part  & Imaginary part  \\

    \hline   

    0  & $\textbf{8}03.328333$  & $-\textbf{}463.801830$  \\
        
    1  & $\textbf{77}2.013681$  & $-\textbf{44}5.722307$  \\

    2  & $\textbf{773.9}98712$  & $-\textbf{446}.868365 $ \\
    3  & $\textbf{773.95}7904$  & $-\textbf{446.8}44804$ \\
         \hline
    Wronskian & $773.96$  &  $-446.85$ \\
        \hline
    \end{tabular}
    \caption{Check of numerical values of $\mathcal{C}_{1,-}^{q,-}$ at the parameters \eqref{paraconnIIt1}}
    \label{1toq1}
\end{table}
\begin{table}[h]
\centering
 \begin{tabular}{| c | c  c| }
    \hline
     Number of instantons & Real part  & Imaginary part  \\

    \hline   

    0  & $\textbf{0.0006}44596393$  & $\textbf{}0.0003721579010$  \\
        
    1  & $\textbf{0.00059}7936507$  & $\textbf{0.00034}5218803$  \\

    2  & $\textbf{0.000599}876502$  & $\textbf{0.0003463}38860 $ \\
    
    3  & $\textbf{0.0005998}22354$  & $\textbf{0.0003463}07598 $ \\
         \hline
    Wronskian & $0.0005998$  &  $0.0003463$ \\
        \hline
    \end{tabular}
    \caption{Check of numerical values of $\mathcal{C}_{1,-}^{q,+}$ at the parameters \eqref{paraconnIIt1}}
        \label{1toq2}
\end{table}

\subsection{Connection problem IV}\label{sec:connIV}

In this section, we study another connection formula for the solutions at $z=t$ and $z=q$, in the regime $|t|\ll 1 \ll |q|$, as in \autoref{fig:5point}.

The Frobenius solutions around $z=t$ are related to the semiclassical limit of the conformal blocks by
\be\ba
\label{conn_t}\psi_{t,\theta}(z)=&t^{\theta a_t}\exp\left(\frac{\theta}{2}\partial_{a_t}F\big({1\over q},t\big)\right)
(z-q)\left(\frac{1-t}{(t-q)(1-q)}\right)^{1/2}\\
&\,\mathcal{F}\left( \begin{matrix} a_{\infty} \\ a_{q} \end{matrix} \, b_2 \, \begin{matrix} a_1  \\ \, \end{matrix} \,b_1 \, \begin{matrix} a_0 \\ \, \end{matrix} \,a_{t\theta} \, \begin{matrix} a_{2,1} \\ a_t \end{matrix} ; \frac{1-t}{1-q}, \frac{t(1-q)}{q(1-t)}, \frac{q(z-t)}{t(z-q)} \right),
\ea\ee
where we applied the following coordinate transform to the conformal block \eqref{semist}.\footnote{We can also use \eqref{connIt} for the solution around $t$. The choice \eqref{conn_t} is to explain the calculation of connection coefficients more intuitively.}
\begin{equation}
z\rightarrow\frac{(z-t)(1-q)}{(z-q)(1-t)}.
\end{equation}
The Frobenius solutions around $z=q$ are related to the semiclassical limit of the conformal blocks by
\be\ba
\label{conn_q}\psi_{q,\theta}(z)=&q^{\theta a_q}\exp\left(\frac{\theta}{2}\partial_{a_q}F\big({1\over q},t\big)\right)(z-t)\left(\frac{(1-q)}{(1-t)(q-t)}\right)^{\frac{1}{2}} \\
&\mathcal{F}\left( \begin{matrix} a_{0} \\ a_{t} \end{matrix} \, b_{1} \, \begin{matrix} a_{1}  \\ \, \end{matrix} \,b_{2} \, \begin{matrix} a_{\infty} \\ \, \end{matrix} \,a_{q\theta} \, \begin{matrix} a_{2,1} \\ a_{q} \end{matrix} ; \frac{t(1-q)}{q(1-t)},\frac{1-t}{1-q}, \frac{z-q}{z-t}\right),
\ea\ee
where we applied the following coordinate transform to the conformal blocks \eqref{semist} 
\be z\rightarrow\frac{(z-q)(1-t)}{(z-t)(1-q)}. \ee
As discussed in \eqref{conn_state}, different phase factors must be considered depending on the arguments of  $t$  and  $q$.  Let us now work out a few examples in detail.

\subsubsection{$q<t<0$}\label{connII:qt0m}
To find the connection coefficients, we go a step back to the conformal blocks before the semiclassical limit is taken. The conformal block around $z=t$ is presented by
\begin{equation}\label{blockt2}
\begin{tikzpicture}[baseline={(current bounding box.center)}, node distance=1cm and 1.5cm]
\coordinate[label=above:$\alpha_{\infty}$] (e1);
\coordinate[below=of e1] (aux1);
\coordinate[left=of aux1,label=left:$\alpha_{q}$] (e2);
\coordinate[right=1.5cm of aux1] (aux2);
\coordinate[above=of aux2,label=above:$\alpha_{1}$] (e3);
\coordinate[right=1.5cm of aux2] (aux3);
\coordinate[above=of aux3,label=above:$\alpha_{0}$] (e4);
\coordinate[right=1.5cm of aux3] (aux4);
\coordinate[above=of aux4,label=above:$\alpha_{2,1}$] (e5);
\coordinate[right=of aux4,label=right:$\alpha_t$] (e6);

\draw (e1) -- (aux1);
\draw (aux1) -- (e2);
\draw (e3) -- (aux2);
\draw (e4) -- (aux3);
\draw[dashed,red] (aux4) -- (e5);
\draw (aux4) -- (e6);
\draw (aux1) -- node[label=below:$\beta_2$] {} (aux2);
\draw (aux2) -- node[label=below:$\beta_1$] {} (aux3);
\draw (aux3) -- node[label=below:$\alpha_{t\theta}$] {} (aux4);
\end{tikzpicture}
\end{equation}
Explicitly, the conformal block together with the Jacobian factors for \eqref{blockt2} is
\begin{equation}\label{5pointarounz=t}
\begin{aligned}
&q^{-2\Delta_{0}}(z-q)^{-2\Delta_{2,1}}(1-t)^{\Delta_{q}-\Delta_{\infty}-\Delta_{2,1}-\Delta_0-\Delta_t-\Delta_1}(1-q)^{\Delta_{\infty}+\Delta_{2,1}+\Delta_0+\Delta_t-\Delta_1-\Delta_{q}}\times\\
&\times (t-q)^{\Delta_{\infty}+\Delta_{2,1}+\Delta_0+\Delta_1-\Delta_t-\Delta_{q}}\mathfrak{F}\left( \begin{matrix} \alpha_{\infty} \\ \alpha_{q} \end{matrix} \, \beta_2 \, \begin{matrix} \alpha_1  \\ \, \end{matrix} \,\beta_1 \, \begin{matrix} \alpha_0 \\ \, \end{matrix} \,\alpha_{t\theta} \, \begin{matrix} \alpha_{2,1} \\ \alpha_t \end{matrix} ; \frac{1-t}{1-q},\frac{t}{q}\frac{1-q}{1-t},\frac{q(z-t)}{t(z-q)} \right).
\end{aligned}
\end{equation}
Similarly, we study the conformal block around $z=q$
\begin{equation}\label{blockq}
\begin{tikzpicture}[baseline={(current bounding box.center)}, node distance=1cm and 1.5cm]
\coordinate[label=above:$\alpha_{0}$] (e1);
\coordinate[below=of e1] (aux1);
\coordinate[left=of aux1,label=left:$\alpha_{t}$] (e2);
\coordinate[right=1.5cm of aux1] (aux2);
\coordinate[above=of aux2,label=above:$\alpha_{1}$] (e3);
\coordinate[right=1.5cm of aux2] (aux3);
\coordinate[above=of aux3,label=above:$\alpha_{\infty}$] (e4);
\coordinate[right=1.5cm of aux3] (aux4);
\coordinate[above=of aux4,label=above:$\alpha_{2,1}$] (e5);
\coordinate[right=of aux4,label=right:$\alpha_{q}$] (e6);

\draw (e1) -- (aux1);
\draw (aux1) -- (e2);
\draw (e3) -- (aux2);
\draw (e4) -- (aux3);
\draw[dashed,red] (aux4) -- (e5);
\draw (aux4) -- (e6);
\draw (aux1) -- node[label=below:$\beta_{1}$] {} (aux2);
\draw (aux2) -- node[label=below:$\beta_{2}$] {} (aux3);
\draw (aux3) -- node[label=below:$\alpha_{q\,\theta}$] {} (aux4);
\end{tikzpicture}
\end{equation}
Explicitly, the conformal block together with the Jacobian factors for \eqref{blockq} is 
\begin{equation}\label{5pointaroundz=q}
\begin{aligned}
&t^{-2\Delta_0}(z-t)^{-2\Delta_{2,1}}(1-t)^{\Delta_{\infty}+\Delta_{2,1}+\Delta_0+\Delta_{q}-\Delta_t-\Delta_1}(1-q)^{\Delta_{t}-\Delta_{\infty}-\Delta_{2,1}-\Delta_0-\Delta_{q}-\Delta_1}\times\\
&\times(q-t)^{\Delta_{\infty}+\Delta_{2,1}+\Delta_0+\Delta_1-\Delta_t-\Delta_{q}}\mathfrak{F}\left( \begin{matrix} \alpha_{0} \\ \alpha_{t} \end{matrix} \, \beta_1 \, \begin{matrix} \alpha_1  \\ \, \end{matrix} \,\beta_2 \, \begin{matrix} \alpha_{\infty} \\ \, \end{matrix} \,\alpha_{q\,\theta} \, \begin{matrix} \alpha_{2,1} \\ \alpha_{q} \end{matrix} ; \frac{t}{q}\frac{1-q}{1-t},\frac{1-t}{1-q},\frac{z-q}{z-t} \right).
\end{aligned}
\end{equation}
In order to obtain the connection coefficients connecting \eqref{5pointarounz=t} and \eqref{5pointaroundz=q}, we follow the steps
\begin{equation}\label{4steps}
\begin{tikzpicture}[baseline={(current bounding box.center)}, node distance=1cm and 1.5cm]
\coordinate[label=above:$\alpha_{\infty}$] (e1);
\coordinate[below=of e1] (aux1);
\coordinate[left=of aux1,label=left:$\alpha_{q}$] (e2);
\coordinate[right=1.5cm of aux1] (aux2);
%%%%%%%%%%%%%%
\coordinate[below right=0.6cm and 0.75cm of aux2] (auxmd1);
%%%%%%%%%%%%%%
\coordinate[above=of aux2,label=above:$\alpha_{1}$] (e3);
\coordinate[right=1.5cm of aux2] (aux3);
\coordinate[above=of aux3,label=above:$\alpha_{0}$] (e4);
\coordinate[right=1.5cm of aux3] (aux4);
\coordinate[above=of aux4,label=above:$\alpha_{2,1}$] (e5);
\coordinate[right=of aux4,label=right:$\alpha_t$] (e6);
%%%%%%%%%%%%%%%%%%%%%%%%%%%%%%%%%%%%%%%%%%
\coordinate[below=3.0cm of e1,label=above:$\alpha_{\infty}$] (e12);
\coordinate[below=of e12] (aux12);
\coordinate[left=of aux12,label=left:$\alpha_{q}$] (e22);
\coordinate[right=1.5cm of aux12] (aux22);
\coordinate[above=of aux22,label=above:$\alpha_{1}$] (e32);
%%%%%%%%%%%%%%
\coordinate[below right=0.6cm and 0.75cm of aux22] (auxmd2);
\coordinate[above right=0.55cm and 0.75cm of e32] (auxmu2);
%%%%%%%%%%%%%%
\coordinate[right=1.5cm of aux22] (aux32);
\coordinate[above=of aux32,label=above:$\alpha_{2,1}$] (e42);
\coordinate[right=1.5cm of aux32] (aux42);
\coordinate[above=of aux42,label=above:$\alpha_{0}$] (e52);
\coordinate[right=of aux42,label=right:$\alpha_t$] (e62);
%%%%%%%%%%%%%%%%%%%%%%%%%%%%%%%%%%%%%
\coordinate[below=3.0cm of e12,label=above:$\alpha_{\infty}$] (e13);
\coordinate[below=of e13] (aux13);
\coordinate[left=of aux13,label=left:$\alpha_{q}$] (e23);
\coordinate[right=1.5cm of aux13] (aux23);
\coordinate[above=of aux23,label=above:$\alpha_{2,1}$] (e33);
%%%%%%%%%%%%%%
\coordinate[below right=0.6cm and 0.75cm of aux23] (auxmd3);
\coordinate[above right=0.55cm and 0.75cm of e33] (auxmu3);
%%%%%%%%%%%%%%
\coordinate[right=1.5cm of aux23] (aux33);
\coordinate[above=of aux33,label=above:$\alpha_{1}$] (e43);
\coordinate[right=1.5cm of aux33] (aux43);
\coordinate[above=of aux43,label=above:$\alpha_{0}$] (e53);
\coordinate[right=of aux43,label=right:$\alpha_t$] (e63);
%%%%%%%%%%%%%%%%%%%%%%%%%%%%%%%%%%%%%
\coordinate[below=3.0cm of e13,label=above:$\alpha_{2,1}$] (e15);
\coordinate[below=of e15] (aux15);
\coordinate[left=of aux15,label=left:$\alpha_{q}$] (e25);
\coordinate[right=1.5cm of aux15] (aux25);
\coordinate[above=of aux25,label=above:$\alpha_{\infty}$] (e35);
%%%%%%%%%%%%%%
\coordinate[below right=0.6cm and 0.75cm of aux25] (auxmd5);
\coordinate[above right=0.55cm and 0.75cm of e35] (auxmu5);
%%%%%%%%%%%%%%
\coordinate[right=1.5cm of aux25] (aux35);
\coordinate[above=of aux35,label=above:$\alpha_{1}$] (e45);
\coordinate[right=1.5cm of aux35] (aux45);
\coordinate[above=of aux45,label=above:$\alpha_{0}$] (e55);
\coordinate[right=of aux45,label=right:$\alpha_t$] (e65);
%%%%%%%%%%%%%%%%%%%%%%%%%%%%%%%%%%%%%%%%%%%
\coordinate[below=3.0cm of e15,label=above:$\alpha_{0}$] (e14);
\coordinate[below=of e14] (aux14);
\coordinate[left=of aux14,label=left:$\alpha_{t}$] (e24);
\coordinate[right=1.5cm of aux14] (aux24);
\coordinate[above=of aux24,label=above:$\alpha_{1}$] (e34);
%%%%%%%%%%%%%%
\coordinate[below right=0.6cm and 0.75cm of aux24] (auxmd4);
\coordinate[above right=0.55cm and 0.75cm of e34] (auxmu4);
%%%%%%%%%%%%%%
\coordinate[right=1.5cm of aux24] (aux34);
\coordinate[above=of aux34,label=above:$\alpha_{\infty}$] (e44);
\coordinate[right=1.5cm of aux34] (aux44);
\coordinate[above=of aux44,label=above:$\alpha_{2,1}$] (e54);
\coordinate[right=of aux44,label=right:$\alpha_{q}$] (e64);

\draw (e1) -- (aux1);
\draw (aux1) -- (e2);
\draw (e3) -- (aux2);
\draw (e4) -- (aux3);
\draw[dashed,red] (aux4) -- (e5);
\draw (aux4) -- (e6);
\draw (aux1) -- node[label=below:$\beta_{2}$] {} (aux2);
\draw (aux2) -- node[label=below:$\beta_{1}$] {} (aux3);
\draw (aux3) -- node[label=below:$\alpha_{t\theta_1}$] {} (aux4);
%%%%%%%%%%%%%%%%%%%%%%%%%%
\draw (e12) -- (aux12);
\draw (aux12) -- (e22);
\draw (e32) -- (aux22);
\draw[dashed,red] (e42) -- (aux32);
\draw (aux42) -- (e52);
\draw (aux42) -- (e62);
\draw (aux12) -- node[label=below:$\beta_2$] {} (aux22);
\draw (aux22) -- node[label=below:$\beta_{1}$] {} (aux32);
\draw (aux32) -- node[label=below:$\beta_{1\theta_2}$] {} (aux42);
%%%%%%%%%%%%%%%%%%%%%%%%%%%%%
\draw (e13) -- (aux13);
\draw (aux13) -- (e23);
\draw[dashed,red] (e33) -- (aux23);
\draw (e43) -- (aux33);
\draw (aux43) -- (e53);
\draw (aux43) -- (e63);
\draw (aux13) -- node[label=below:$\beta_{2}$] {} (aux23);
\draw (aux23) -- node[label=below:$\beta_{2\theta_3}$] {} (aux33);
\draw (aux33) -- node[label=below:$\beta_{1\theta_2}$] {} (aux43);
%%%%%%%%%%%%%%%%%%%%%%%%%%%%%
\draw[dashed,red] (e15) -- (aux15);
\draw (aux15) -- (e25);
\draw (e35) -- (aux25);
\draw (e45) -- (aux35);
\draw (aux45) -- (e55);
\draw (aux45) -- (e65);
\draw (aux15) -- node[label=below:$\alpha_{q\theta_4}$] {} (aux25);
\draw (aux25) -- node[label=below:$\beta_{2\theta_3}$] {} (aux35);
\draw (aux35) -- node[label=below:$\beta_{1\theta_2}$] {} (aux45);
%%%%%%%%%%%%%%%%%%%%%%%%%%%%%%%%%%%%
\draw (e14) -- (aux14);
\draw (aux14) -- (e24);
\draw (e34) -- (aux24);
\draw (e44) -- (aux34);
\draw[dashed,red] (aux44) -- (e54);
\draw (aux44) -- (e64);
\draw (aux14) -- node[label=below:$\beta_{1\theta_2}$] {} (aux24);
\draw (aux24) -- node[label=below:$\beta_{2\theta_3}$] {} (aux34);
\draw (aux34) -- node[label=below:$\alpha_{q\theta_4}$] {} (aux44);
%%%%%%%%%%
\draw[->,line width=0.05cm, orange] (auxmd1) -- (auxmu2);
\draw[->,line width=0.05cm, orange] (auxmd2) -- (auxmu3);
\draw[->,line width=0.05cm, orange] (auxmd3) -- (auxmu5);
\draw[->,line width=0.05cm, orange] (auxmd5) -- (auxmu4);
\end{tikzpicture}
\end{equation}
In each step, we used \eqref{Mfactors} for moving $\alpha_{2,1}$ around. For example, the first step can be written as
\begin{equation}
\begin{aligned}
&\begin{tikzpicture}[baseline={(current bounding box.center)}, node distance=1cm and 1.5cm]
\coordinate[label=above:$\alpha_{\infty}$] (e1);
\coordinate[below=of e1] (aux1);
\coordinate[left=of aux1,label=left:$\alpha_{q}$] (e2);
\coordinate[right=1.5cm of aux1] (aux2);
\coordinate[above=of aux2,label=above:$\alpha_{1}$] (e3);
\coordinate[right=1.5cm of aux2] (aux3);
\coordinate[above=of aux3,label=above:$\alpha_{0}$] (e4);
\coordinate[right=1.5cm of aux3] (aux4);
\coordinate[above=of aux4,label=above:$\alpha_{2,1}$] (e5);
\coordinate[right=of aux4,label=right:$\alpha_t$] (e6);

\draw (e1) -- (aux1);
\draw (aux1) -- (e2);
\draw (e3) -- (aux2);
\draw (e4) -- (aux3);
\draw[dashed,red] (aux4) -- (e5);
\draw (aux4) -- (e6);
\draw (aux1) -- node[label=below:$\beta_2$] {} (aux2);
\draw (aux2) -- node[label=below:$\beta_1$] {} (aux3);
\draw (aux3) -- node[label=below:$\alpha_{t\theta_1}$] {} (aux4);
\end{tikzpicture}=\\
&=\sum_{\theta_2=\pm}\mathcal{M}_{\theta_1\theta_2}(b\alpha_t,b\beta_1;b\alpha_0)\,\begin{tikzpicture}[baseline={(current bounding box.center)}, node distance=1cm and 1.5cm]
\coordinate[label=above:$\alpha_{\infty}$] (e1);
\coordinate[below=of e1] (aux1);
\coordinate[left=of aux1,label=left:$\alpha_{q}$] (e2);
\coordinate[right=1.5cm of aux1] (aux2);
\coordinate[above=of aux2,label=above:$\alpha_{1}$] (e3);
\coordinate[right=1.5cm of aux2] (aux3);
\coordinate[above=of aux3,label=above:$\alpha_{2,1}$] (e4);
\coordinate[right=1.5cm of aux3] (aux4);
\coordinate[above=of aux4,label=above:$\alpha_{0}$] (e5);
\coordinate[right=of aux4,label=right:$\alpha_t$] (e6);

\draw (e1) -- (aux1);
\draw (aux1) -- (e2);
\draw (e3) -- (aux2);
\draw[dashed,red] (e4) -- (aux3);
\draw (aux4) -- (e5);
\draw (aux4) -- (e6);
\draw (aux1) -- node[label=below:$\beta_2$] {} (aux2);
\draw (aux2) -- node[label=below:$\beta_1$] {} (aux3);
\draw (aux3) -- node[label=below:$\beta_{1\theta_2}$] {} (aux4);
\end{tikzpicture}
\end{aligned}
\end{equation}
Putting the steps all together, the connection formula between the conformal blocks reads\footnote{For now we ignore the phase factor. We will fix it when we write down the explicit form that we use.}
\begin{equation}
\begin{aligned}
&q^{-2\Delta_{0}}(z-q)^{-2\Delta_{2,1}}(1-t)^{\Delta_{q}-\Delta_{\infty}-\Delta_{2,1}-\Delta_0-\Delta_t-\Delta_1}(1-q)^{\Delta_{\infty}+\Delta_{2,1}+\Delta_0+\Delta_t-\Delta_1-\Delta_{q}}\times\\
&\times (t-q)^{\Delta_{\infty}+\Delta_{2,1}+\Delta_0+\Delta_1-\Delta_t-\Delta_{q}}\mathfrak{F}\left( \begin{matrix} \alpha_{\infty} \\ \alpha_{q} \end{matrix} \, \beta_2 \, \begin{matrix} \alpha_1  \\ \, \end{matrix} \,\beta_1 \, \begin{matrix} \alpha_0 \\ \, \end{matrix} \,\alpha_{t\theta_1} \, \begin{matrix} \alpha_{2,1} \\ \alpha_t \end{matrix} ; \frac{1-t}{1-q},\frac{t}{q}\frac{1-q}{1-t},\frac{q(z-t)}{t(z-q)} \right)=\\
&=\sum_{\theta_2,\theta_3,\theta_4=\pm}\mathcal{M}_{\theta_1\theta_2}(b\alpha_t,b\beta_1;b\alpha_0)\mathcal{M}_{(-\theta_2)\theta_3}(b\beta_1,b\beta_2;b\alpha_1)\mathcal{M}_{(-\theta_3)\theta_4}(b\beta_2,b\alpha_{q};b\alpha_{\infty})\times\\
&\times t^{-2\Delta_0}(z-t)^{-2\Delta_{2,1}}(1-t)^{\Delta_{\infty}+\Delta_{2,1}+\Delta_0+\Delta_{q}-\Delta_t-\Delta_1}(1-q)^{\Delta_{t}-\Delta_{\infty}-\Delta_{2,1}-\Delta_0-\Delta_{q}-\Delta_1}\times\\
&\times(q-t)^{\Delta_{\infty}+\Delta_{2,1}+\Delta_0+\Delta_1-\Delta_t-\Delta_{q}}\mathfrak{F}\left( \begin{matrix} \alpha_{0} \\ \alpha_{t} \end{matrix} \, \beta_{1\,\theta_2} \, \begin{matrix} \alpha_1  \\ \, \end{matrix} \,\beta_{2\,\theta_3} \, \begin{matrix} \alpha_{\infty} \\ \, \end{matrix} \,\alpha_{q\,\theta_4} \, \begin{matrix} \alpha_{2,1} \\ \alpha_{q} \end{matrix} ; \frac{t}{q}\frac{1-q}{1-t},\frac{1-t}{1-q},\frac{z-q}{z-t} \right),
\end{aligned}
\end{equation}
which in the semiclassical limit becomes
\begin{equation}\label{connIIblock}
\begin{aligned}
&(z-q)(t-q)^{-\frac{1}{2}}(1-q)^{-\frac{1}{2}}(1-t)^{\frac{1}{2}}\mathcal{F}\left( \begin{matrix} a_{\infty} \\ a_{q} \end{matrix} \, b_2 \, \begin{matrix} a_1  \\ \, \end{matrix} \,b_1 \, \begin{matrix} a_0 \\ \, \end{matrix} \,a_{t\theta_1} \, \begin{matrix} a_{2,1} \\ a_t \end{matrix} ; \frac{1-t}{1-q}, \frac{t(1-q)}{q(1-t)}, \frac{(z-t)q}{(z-q)t} \right)=\\
&\sum_{\theta_2,\theta_3,\theta_4=\pm}\E^{\ri\pi(-\frac1 2-\theta_4 a_q)}\mathcal{M}_{\theta_1,\theta_2}(a_t,b_1;a_0)\mathcal{M}_{(-\theta_2),\theta_3}(b_1,b_2;a_1)\mathcal{M}_{(-\theta_3),\theta_4}(b_2,a_{q};a_{\infty})\times\\
&\left(t\right)^{\theta_2b_1}\left(q\right)^{-\theta_3b_2}\exp\left(-\frac{\theta_2}{2}\partial_{b_1}F\big({1\over q},t\big)-\frac{\theta_3}{2}\partial_{b_2}F\big({1\over q},t\big)\right)\times\\
&(z-t)(q-t)^{-\frac{1}{2}}(1-q)^{\frac{1}{2}}(1-t)^{-\frac{1}{2}} \mathcal{F}\left( \begin{matrix} a_{0} \\ a_{t} \end{matrix} \, b_{1} \, \begin{matrix} a_{1}  \\ \, \end{matrix} \,b_{2} \, \begin{matrix} a_{\infty} \\ \, \end{matrix} \,a_{q\theta_4} \, \begin{matrix} a_{2,1} \\ a_{q} \end{matrix} ; \frac{t(1-q)}{q(1-t)},\frac{1-t}{1-q}, \frac{z-q}{z-t}\right),
\end{aligned}
\end{equation}
where we used \eqref{momentumshift}. 
Written in terms of Frobenius solutions (up to phases), 
\begin{equation}\label{connII}
\begin{aligned}
&\psi_{t,\theta_1}(z)=t^{\theta_1 a_t}\exp\left(\frac{1}{2}\theta_1\partial_{a_t}F\big({1\over q},t\big)\right)\\
&\times\sum_{\theta_2,\theta_3,\theta_4=\pm}\mathcal{M}_{\theta_1\theta_2}(a_t,b_1;a_0)\mathcal{M}_{(-\theta_2)\theta_3}(b_1,b_2;a_1)\mathcal{M}_{(-\theta_3)\theta_4}(b_2,a_{q};a_{\infty})\times\\
&t^{\theta_2b_1}q^{-\theta_3b_2}\exp\left(-\frac{\theta_2}{2}\partial_{b_1}F\big({1\over q},t\big)-\frac{\theta_3}{2}\partial_{b_2}F\big({1\over q},t\big)\right)\times\\
&\times q^{-\theta_4 a_q}\exp\left(-\frac{1}{2}\theta_4\partial_{a_q}F\big({1\over q},t\big)\right)\psi_{q,\theta_4}(z).
\end{aligned}
\end{equation}
In the case when $q<t<0$, by appropriately choosing the phase factor,  \eqref{connII} becomes
\begin{equation}
\label{conn2eqn}
\begin{aligned}
&\psi_{t,\theta_1}(z) = t^{\theta_1 a_t}\exp\left(\frac{1}{2}\theta_1\partial_{a_t}F\big({1\over q},t\big)\right)\\
&\times \sum_{\theta_2,\theta_3,\theta_4=\pm} \E^{\ri\pi\left(\frac{1}{2} - \theta_2 b_1 + \theta_3 b_2 + \theta_4 a_q\right)}
\mathcal{M}_{\theta_1\theta_2}(a_t, b_1; a_0) 
\mathcal{M}_{(-\theta_2)\theta_3}(b_1, b_2; a_1) 
\mathcal{M}_{(-\theta_3)\theta_4}(b_2, a_q; a_{\infty}) \\
&\times t^{\theta_2 b_1} q^{-\theta_3 b_2} 
\exp\left(-\frac{\theta_2}{2}\partial_{b_1}F\big({1\over q},t\big) 
-\frac{\theta_3}{2}\partial_{b_2}F\big({1\over q},t\big)\right)  q^{-\theta_4 a_q} 
\exp\left(-\frac{1}{2}\theta_4\partial_{a_q}F\big({1\over q},t\big)\right)
\psi_{q,\theta_4}(z).
\end{aligned}
\end{equation}

As before, we can write it in closed form using the full NS function: \begin{empheq}[box=\fbox]{equation}
\label{conn2eqni}
\begin{aligned}
\psi_{t,\theta_1}(z) &={ \pi \re^{\ri {\pi\over 2}}t^{\theta_1 a_t}\re^{\frac{1}{2}\theta_1\partial_{a_t}F^{\rm NS}\big({1\over q},t,-\theta_1 a_t\big)}\over \sin (2 \pi  b_1) \sin (2 \pi  b_2)}\Gamma (2 a_t\theta_1+1)\\
&\times \sum_{\theta_2,\theta_3,\theta_4=\pm} \E^{\ri\pi\left(- \theta_2 b_1 + \theta_3 b_2 + \theta_4 a_q+{\theta_2-\theta_3\over 2}\right)} q^{-\theta_4 a_q}  \Gamma (-2 \theta_4 a_q)  \cos\left(\pi (a_1 + b_1 \theta_2 + b_2 \theta_3)\right)
 \\
&\times 
\re^{-\frac{\theta_2}{2}\partial_{b_1}F^{\rm NS}\big({1\over q},t,-\theta_1 a_t, \theta_4 a_q\big) 
-\frac{\theta_3}{2}\partial_{b_2}F^{\rm NS}\big({1\over q},t,-\theta_1 a_t, \theta_4 a_q\big)-\frac{1}{2}\theta_4\partial_{a_q}F^{\rm NS}\big({1\over q},t,-\theta_1 a_t, \theta_4 a_q\big)}  
\psi_{q,\theta_4}(z),
\end{aligned}
\end{empheq}
where $F^{\rm NS}$ is defined in \eqref{eq:fullns}. As before, in the argument of the $F^{\rm NS}$ function, we explicitly wrote the dependence only for the mass parameters  $a_i$  that are subject to sign changes.

We check \eqref{conn2eqn} at the parameters
\begin{equation}\label{eq:numtest2}
\begin{aligned}
&t=-\frac{1}{100},\quad
q=-200,\quad
a_0=\frac{97}{70},\quad
a_1=\frac{7}{141},\quad
a_{\infty}=\frac{51}{40},\\
&a_t=\frac{10}{9},\quad\quad
a_{q}=\frac{4}{3},\quad\quad
u_t=\frac{73}{50},\quad
u_{q}=\frac{3}{8}.
\end{aligned}
\end{equation}
The comparison of numerical and instanton results for the connection coefficients of $\psi_{t,-}$ and $\psi_{q,\mp}$ for the values in \eqref{eq:numtest2} is in \autoref{ttoqnegative1} and \autoref{ttoqnegative2}.
\begin{table}[h]
\centering
 \begin{tabular}{| c | c  c| }
    \hline
     Number of instantons & Real part  & Imaginary part  \\

    \hline   

    0  & $\textbf{4}.82094126\times 10^8$  & $\textbf{1.3}24542726\times 10^9$  \\
        
    1  & $\textbf{4.7}1162611\times 10^8$  & $\textbf{1.29}4508635\times 10^9$  \\

    2  & $\textbf{4.706}14908\times 10^8$  & $\textbf{1.2930}03833\times 10^9$ \\

    3  & $\textbf{4.70608}452\times 10^8$  & $\textbf{1.29298}6095\times 10^9$ \\
         \hline

    Wronskian & $4.70608\times 10^8$  &  $1.29299\times 10^9$ \\
         \hline
    \end{tabular}
    \caption{Check of numerical values of $\mathcal{C}_{t,-}^{q,-}$ at the parameters \eqref{eq:numtest2}}
    \label{ttoqnegative1}
\end{table}
\begin{table}[h!]
\centering
    \begin{tabular}{| c | c  c| }
    \hline
     Number of instantons & Real part  & Imaginary part \\

    \hline 

    0  & $-\textbf{36}7.655967$  & $-\textbf{10}10.126467$  \\

    1  & $-\textbf{372}.690209$ &  $-\textbf{1023}.957933$ \\

    2  & $-\textbf{372.5}22917$  & $-\textbf{1023.4}98302$ \\

    3  & $-\textbf{372.532}782$ & $-\textbf{1023.52}541$ \\
         \hline
   Wronskian & $-372.533$ & $-1023.53$ \\
         \hline
    \end{tabular}
    \caption{Check of numerical values of $\mathcal{C}_{t,-}^{q,+}$ at the parameters \eqref{eq:numtest2}}
    \label{ttoqnegative2}
\end{table}

\subsubsection{$q>t>0$}\label{connII:qt0g}

In this example, the singularities $t$ and $q$ lie on the positive real axis. We separate the problem into two parts by studying the connection problem for the solutions around $t$ and 1 or the solutions around 1 and $q$, respectively, and we use \autoref{sec:connII} and \autoref{sec:connIII}.  All together, we write down the connection coefficients for the connection problem relating the Frobenius solutions around $t$ and $q$ along an upper semicircle contour avoiding the singularity at 1. We start with \eqref{connIIt1} and plug in \eqref{paraconnII1q}:
\begin{equation}
\begin{aligned}
\psi_{t,\theta_1}(z)&=t^{\theta_1 a_t}\E^{\frac{\theta_1}{2}\partial_{a_t}F\big({1\over q},t\big)}\sum_{\theta_2,\theta_3=\pm}\mathcal{M}_{\theta_1\theta_2}(a_t,b_1,a_0)\mathcal{M}_{(-\theta_2)\theta_3}(b_1,a_1,b_2)\times\\
&\quad\ \times\quad\ t^{\theta_2 b_1}\E^{-\frac{\theta_2}{2}\partial_{b_1}F\big({1\over q},t\big)}\E^{\ri\pi\left(-(\frac1 2+\theta_3 a_1)\right)}\E^{-\frac{\theta_3}{2}\partial_{a_1}F\big({1\over q},t\big)}\E^{\frac{\theta_3}{2}\partial_{a_1}F\big({1\over q},t\big)}\\
&\quad\ \times\sum_{\theta_4,\theta_5=\pm}\mathcal{M}_{\theta_3\theta_4}(a_1,b_2,b_1)\mathcal{M}_{(-\theta_4)\theta_5}(b_2,a_q,a_\infty)\times\\
&\quad\ \times q^{-\theta_4 b_2}\E^{-\frac{\theta_4}{2}\partial_{b_2}F\big({1\over q},t\big)}q^{-\theta_5 a_q}\E^{-\frac{\theta_5}{2}\partial_{a_q}F\big({1\over q},t\big)}\E^{\ri\pi\left(-(\frac1 2 +\theta_5 a_q)\right)}\psi_{q,\theta_5}(z)\\
&=t^{\theta_1 a_t}\E^{\frac{\theta_1}{2}\partial_{a_t}F\big({1\over q},t\big)}\sum_{\theta_2,\theta_3,\theta_4,\theta_5=\pm}\mathcal{M}_{\theta_1\theta_2}(a_t,b_1,a_0)\mathcal{M}_{(-\theta_2)\theta_3}(b_1,a_1,b_2)\times\\
&\quad\ \times\E^{\ri\pi\left(-(\frac1 2+\theta_3 a_1)\right)}\mathcal{M}_{\theta_3\theta_4}(a_1,b_2,b_1)\mathcal{M}_{(-\theta_4)\theta_5}(b_2,a_q,a_\infty) t^{\theta_2 b_1}\E^{-\frac{\theta_2}{2}\partial_{b_1}F\big({1\over q},t\big)}\times\\\
&\quad\ \times q^{-\theta_4 b_2}\E^{-\frac{\theta_4}{2}\partial_{b_2}F\big({1\over q},t\big)}q^{-\theta_5 a_q}\E^{-\frac{\theta_5}{2}\partial_{a_q}F\big({1\over q},t\big)}\E^{\ri\pi\left(-(\frac1 2 +\theta_5 a_q)\right)}\psi_{q,\theta_5}(z).
\end{aligned}
\end{equation}
Applying the identity
\begin{equation}
    \sum_{\theta_3=\pm}\mathcal{M}_{(-\theta_2)\theta_3}(b_1,a_1,b_2)\E^{\ri\pi\left(-(\frac1 2+\theta_3 a_1)\right)}\mathcal{M}_{\theta_3\theta_4}(a_1,b_2,b_1)=\mathcal{M}_{(-\theta_2)\theta_4}(b_1,b_2,a_1)\E^{\ri\pi(\theta_4 b_2-\theta_2 b_1)}
\end{equation}
we get
\begin{equation}
\begin{aligned}
\psi_{t,\theta_1}(z)
&=t^{\theta_1 a_t}\E^{\frac{\theta_1}{2}\partial_{a_t}F\big({1\over q},t\big)}\times\\
&\quad\ \times\sum_{\theta_2,\theta_4,\theta_5=\pm}\mathcal{M}_{\theta_1\theta_2}(a_t,b_1,a_0)\mathcal{M}_{(-\theta_2)\theta_4}(b_1,b_2,a_1)\E^{\ri\pi(\theta_4 b_2-\theta_2 b_1)}\mathcal{M}_{(-\theta_4)\theta_5}(b_2,a_q,a_\infty) t^{\theta_2 b_1}\times\\
&\quad\ \times \E^{-\frac{\theta_2}{2}\partial_{b_1}F\big({1\over q},t\big)} q^{-\theta_4 b_2}\E^{-\frac{\theta_4}{2}\partial_{b_2}F\big({1\over q},t\big)}q^{-\theta_5 a_q}\E^{-\frac{\theta_5}{2}\partial_{a_q}F\big({1\over q},t\big)}\E^{\ri\pi\left(-(\frac1 2 +\theta_5 a_q)\right)}\psi_{q,\theta_5}(z).
\end{aligned}
\end{equation}
Hence,
\begin{equation}\label{connIItq}
\begin{aligned}
\psi_{t,\theta_1}(z)
&= t^{\theta_1 a_t} \E^{\frac{\theta_1}{2}\partial_{a_t}F\big({1\over q},t\big)} \times \\
&\quad \times \sum_{\theta_2,\theta_3,\theta_4=\pm} 
\mathcal{M}_{\theta_1\theta_2}(a_t,b_1,a_0) 
\mathcal{M}_{(-\theta_2)\theta_3}(b_1,b_2,a_1) 
\mathcal{M}_{(-\theta_3)\theta_4}(b_2,a_q,a_\infty) 
t^{\theta_2 b_1} 
\E^{-\frac{\theta_2}{2}\partial_{b_1}F\big({1\over q},t\big)} \\
&\quad \times q^{-\theta_3 b_2} 
\E^{-\frac{\theta_3}{2}\partial_{b_2}F\big({1\over q},t\big)} 
q^{-\theta_4 a_q} 
\E^{-\frac{\theta_4}{2}\partial_{a_q}F\big({1\over q},t\big)} 
\E^{\ri\pi\left(\theta_3 b_2 - \theta_2 b_1\right)} 
\E^{\ri\pi\left(-\left(\frac{1}{2} + \theta_4 a_q\right)\right)} 
\psi_{q,\theta_4}(z).
\end{aligned}
\end{equation}
We check this at the parameters in \eqref{paraconnIIt1}.
The comparison of numerical and instanton results for the connection coefficients of $\psi_{t,-}$ and $\psi_{q,\mp}$ is in \autoref{ttoqpositive1} and \autoref{ttoqpositive2}.
\begin{table}[h]
\centering
 \begin{tabular}{| c | c  c| }
    \hline
     Number of instantons & Real part  & Imaginary part  \\

    \hline   

    0  & $\textbf{6}.5011735\times 10^8$  & $\textbf{1}.24167927\times 10^9$\,i  \\
        
    1  & $\textbf{6.8}270759\times 10^8$  & $\textbf{1.26}157640\times 10^9$\,i  \\

    2  & $\textbf{6.810}4579\times 10^8$  & $\textbf{1.260}54896\times 10^9 $\,i \\
    3  & $\textbf{6.8107}137\times 10^8$  & $\textbf{1.260557}70\times 10^9$\,i \\
         \hline
    Wronskian & $6.8107\times 10^8$  &  $1.260558\times 10^9$\,i \\
        \hline
    \end{tabular}
    \caption{Check of numerical values of $\mathcal{C}_{t,-}^{q,-}$ at the parameters \eqref{paraconnIIt1}}
    \label{ttoqpositive1}
\end{table}

\begin{table}[h]
\centering
 \begin{tabular}{| c | c  c| }
    \hline
     Number of instantons & Real part  & Imaginary part  \\

    \hline   

    0  & $-\textbf{5}86.254383$  & $\textbf{9}09.002256$\,i  \\
        
    1  & $-\textbf{565}.545454$  & $\textbf{90}4.228989$\,i  \\

    2  & $-\textbf{565.9}33719$  & $\textbf{903}.463207 $\,i \\
    3  & $-\textbf{565.90}7003$  & $\textbf{903.44}6140$\,i \\
         \hline
    Wronskian & $-565.908$  &  $903.447$\,i \\
        \hline
    \end{tabular}
    \caption{Check of numerical values of $\mathcal{C}_{t,-}^{q,+}$ at the parameters \eqref{paraconnIIt1}}
    \label{ttoqpositive2}
\end{table}

\section{A few applications}\label{sec:applications}
   We now discuss some applications of our results in the context of black hole perturbation theory.

There are many problems in black hole perturbation theory in which the perturbation equation has five regular singularities. 
These include (generic) massive scalar perturbations of Schwarzschild-(A)dS in four and seven dimensions, as well as of Kerr-(A)dS in four dimensions and of Reissner-Nordstr\"om-(A)dS in four dimensions. In the rotating and higher-dimensional cases, the black hole solutions to Einstein equations are not unique as there can be multiple rotation axes and different regimes in which the singularity structure of the perturbation equation changes \cite{Emparan:2003sy, Cardoso:2004cj, Emparan:2008eg, Caldarelli:2008pz, Hennigar:2015gan, Ponglertsakul:2020ufm}, see \autoref{app:example} for some examples. Among these, we find that the massive scalar perturbation has five regular singularities in the seven-dimensional case with a single rotation parameter, or in the hyperbolic membrane limit, and in the nine-dimensional case in the black brane and ultraspinning regime. The same singularity structure arises also when considering different types of perturbations. For example, this happens for the scalar-sector of gravitational perturbations in Schwarzschild-(A)dS in four dimensions as studied in \cite{Aminov:2023jve}, and for the angular equation satisfied by the quasinormal modes of the charged C-metric \cite{Lei:2023mqx}.
For charged black holes, perturbations of electromagnetic fields are non-trivially coupled to the vector modes of metric perturbations. In \cite{Kodama:2003kk}, it was shown that it is possible to reduce the system of perturbation equations to two decoupled ODEs by taking appropriate combinations of the gauge-invariant quantities.
In five dimensions, such perturbation equation has the singularity structure of our interest, as we will analyze in \autoref{sec:AdS5}. We refer to \cite{Loganayagam:2022teq} for a recent review of the singularity structure of equations governing black hole perturbation theory.

Once the Seiberg-Witten to gravity dictionary is established, several applications can be explored. Here we  consider  applications in the framework of the AdS/CFT correspondence, where a black hole in AdS$_{d+1}$ is dual to a $d$-dimensional CFT at finite temperature. In this setting, the ratio of connection coefficients in the wave equation directly computes the thermal two-point functions in the dual CFT \cite{Son:2002sd, Nunez:2003eq, Policastro:2002se,Kovtun:2005ev}. 
By further assuming the Eigenstate Thermalization Hypothesis and focusing on the large-spin regime, we can use the thermal two-point function  to extract both the anomalous dimensions and the three-point functions for a class of double-twist operators \cite{Lashkari:2016vgj,Kulaxizi:2018dxo,Karlsson:2019qfi,Karlsson:2019dbd,Karlsson:2021duj,Dodelson:2022eiz}. This was used in \cite{Dodelson:2022yvn} where, in the example of a scalar perturbation, an explicit solution to the heavy-light light-cone bootstrap in four dimensions was provided via the NS functions. See also \cite{Bhatta:2022wga, Bhatta:2023qcl,He:2023wcs,Ren:2024ifh,Jia:2024zes,BarraganAmado:2024tfu,Arnaudo:2024sen} for an analogous computation of the thermal-two point functions in different backgrounds or under other types of perturbations. 
In this section,  we use the results of \autoref{sec:five} to extend the analysis to include scalar perturbations in Schwarzschild-AdS\(_7\) black holes and  perturbations decribing 
electromagnetic fields  coupled to  vector  modes of the metric
in Reissner-Nordström-AdS\(_5\) black holes.

\subsection{Scalar perturbations Schwarzschild-AdS\texorpdfstring{$_7$}{} black hole}\label{sec:AdS7} 

The Schwarzschild-AdS\(_7\) black hole metric is given by
\begin{equation}\label{eq:ads7}
\mathrm{d}s^2=-f(r)\mathrm{d}t^2+f(r)^{-1}\mathrm{d}r^2+r^2\mathrm{d}\Omega^2_5,
\end{equation}
with
\begin{equation}
f(r)=1-\frac{\mu}{r^4}-\frac{\Lambda}{15}r^2,
\end{equation}
where $\mu$ is related to the mass $M$ of the black hole via \footnote{{We are setting $G_N=1$.}}
\begin{equation}
M=\frac{5\,\pi^2\,\mu}{16}.
\end{equation}
In what follows, we will normalize the AdS radius $L$ to equal 1, or, equivalently, $\Lambda=-15$.

We further perturb the background \eqref{eq:ads7} by a massive scalar field of mass $\nu$. 
The corresponding Klein-Gordon equation can be encoded, after mode decomposition, in the following radial equation 
\begin{equation}\label{eq:qnmeq}
\left(\partial^2_r + \frac{f'(r)}{f(r)} \partial_r +\frac{\omega^2-V(r)}{f(r)^2}\right)\Phi(r)=0,
\end{equation}
where the potential is  
\begin{equation}
V(r)= f(r)\left(\frac{\ell(\ell+4)+\frac{15}{4}}{r^2}+\frac{25\,\mu}{4 r^6}+\frac{35}{4}+\nu^2\right)~,
\end{equation}
see \cite{review} and references therein.
 
Let us start by working out the SW theory to gravity dictionary. In the variable $y=r^2$, the function $y^2f(y)$ is a third-degree polynomial in $y$ with one positive real root representing the BH horizon, $y_1$, and two real negative or complex conjugated roots (depending on the value of the black hole mass), $y_2,y_3$. The two roots $y_2,y_3$ can be written in terms of $y_1$ as
\begin{equation}
y_2=\frac{-1-y_1-\sqrt{1-2y_1-3y_1^2}}{2},\quad y_3=\frac{-1-y_1+\sqrt{1-2y_1-3y_1^2}}{2},
\end{equation}
and the full expression for $y_1$ is
\be y_1= {1\over 3}\left({\rm D} +\frac{1}{\rm D}-1\right), \quad {\rm D}^3= \frac{27 \mu }{2}+\frac{3}{2} \sqrt{3} \sqrt{\mu  (27 \mu -4)}-1~>0~.\ee
For a generic value of the BH mass $\mu$, the wave equation \eqref{eq:qnmeq} has five regular singularities at $y=0, y_1, y_2, y_3,\infty$. However, for  $\mu = \frac{4}{27}$, the singularities at  $y_2$  and $y_3$ merge, resulting in a collision of two singular points. Consequently, the wave equation is transformed: it now has three regular singularities and one irregular singularity. It is not clear to us what is the meaning of this point\footnote{It does not seem to be related to BH thermodynamics, e.g.~to the Hawking-Page point.}. In the following, we will assume that $\mu$ does not lie in a neighborhood of this point. Specifically, we assume $\mu\notin[36/343, 3/8]$ so that $1/|q|<1$ and we have the configuration of \autoref{fig:5point}. It would be interesting to explore what implications this change in configurations might have on the black hole side (if any), but we will not pursue that direction further here.

Under this assumption, we can introduce the new variable 
\begin{equation}
z=\frac{y(y_3-y_2)}{y_3(y-y_2)},
\end{equation}
and redefining the wave function as
\begin{equation}
\Phi(z)=\frac{z^{3/4}}{\sqrt{1-z}\, \sqrt[4]{y_2+y_3 (z-1)}\, \sqrt{y_2 y_3 z-y_1 (y_2+y_3 (z-1))}}\,\psi(z),
\end{equation}
the differential equation satisfied by $\psi(z)$ can be written in the form \eqref{diffeq5normal} with the following dictionary\footnote{We remark that in principle other dictionaries can be found, since the differential equation only depends on the square of the $a$-parameters and there are symmetries w.r.t.~some permutations of the $a_i$'s parameters.}:
\begin{equation}\label{theoryparameter}
\begin{aligned}
t=&\,\frac{y_1 (y_3-y_2)}{y_3(y_1-y_2)},\quad
q=\,1-\frac{y_2}{y_3},\quad
a_0=\,0,\quad
a_t=\,\frac{i\,y_1^{3/2}\omega}{2(y_1-y_2)(y_1-y_3)},\\
a_1=&\,\frac{i\,y_3^{3/2}\omega}{2(y_1-y_3)(y_2-y_3)},\quad
a_{q}=\,\frac{\sqrt{9+\nu^2}}{2},\quad
a_{\infty}=\,\frac{i\,y_2^{3/2}\omega}{2(y_1-y_2)(y_2-y_3)},\\
u_t=&\,t\,\Bigg(\frac{\omega ^2 y_1^2 y_3 (y_1-3 y_3)}{4 y_2 (y_1-y_3)^3 (y_2-y_3)}+\frac{\nu ^2 y_1 y_3 (y_2-y_1)}{4 y_2 (y_1-y_3) (y_2-y_3)}+\\
&\,-\frac{\ell (\ell+4) y_3 (y_1-y_2)}{4 y_2 (y_1-y_3) (y_2-y_3)}-\frac{y_3 (y_1-y_2) (3 y_3 (y_1-y_2)+y_1 (14 y_1+y_2+10))}{4 y_1 y_2 (y_1-y_3) (y_2-y_3)}\Bigg),\\
u_{q}=&\,q\,\frac{y_3 \left(4 \ell (\ell+4)-4 \omega ^2+4 \nu ^2 (y_1-y_2+y_3)+31 y_1-33 y_2+31 y_3+15\right)}{16 y_2 (y_2-y_3)}.
\end{aligned}
\end{equation}
In the $z$ variable, the black hole horizon is located at $z=t$ and the AdS boundary is at $z=q$. 
As boundary conditions for the QNMs, we impose the vanishing Dirichlet boundary condition at $z=q$, and the presence of only ingoing modes at the horizon $z=t$. In terms of the wave function $\psi$, this means 
\begin{equation}\label{bcAdS7}
\begin{aligned}
&\psi(z)=\psi_{t,-}(z)\sim (z-t)^{\frac{1}{2}-\frac{i\omega y_1^{3/2}}{2(y_1-y_2)(y_1-y_3)}} &\text{for}\ z\sim t,\\
&\psi(z)=\psi_{q,+}(z)\sim (z-q)^{\frac{1}{2}+\frac{\sqrt{9+\nu^2}}{2}} &\text{for}\ z\sim q.
\end{aligned}
\end{equation}
which are exactly the Frobenius solutions.

The connection formula between the Frobenius solutions expanded around $z=t$ and $z=q$ is given in \eqref{conn2eqn}. 
According to the boundary conditions \eqref{bcAdS7}, the quantization condition is given by
\begin{equation}\label{eq:qcads7full}
\begin{aligned}
&\sum_{\theta_2,\theta_3=\pm}\mathcal{M}_{-\theta_2}(a_t,b_1;a_0)\mathcal{M}_{(-\theta_2)\theta_3}(b_1,b_2;a_1)\mathcal{M}_{(-\theta_3)-}(b_2,a_{q};a_{\infty})\times\\
&t^{\theta_2b_1}\,q^{-\theta_3b_2}\exp\left(-\frac{\theta_2}{2}\partial_{b_1}F\big({1\over q},t\big)-\frac{\theta_3}{2}\partial_{b_2}F\big({1\over q},t\big)\right)=0.
\end{aligned}
\end{equation}

In the framework of the AdS/CFT correspondence \cite{adscft,Witten:1998qj,Gubser:1998bc},
 a massive scalar field of mass $\nu$ in an $\mathrm{AdS}_7$ background is dual  a  scalar operator of dimension $\Delta$ with
\be
\nu^2 = \Delta(\Delta - 6).
\ee
A notable example of AdS$_7$/CFT$_6$ duality is the correspondence between M-theory on AdS$_7 \times S^4$ and the six-dimensional $\mathcal{N}=(2,0)$ superconformal field theory. This theory lacks of a Lagrangian description and very little is known about it. From this perspective, an analytic treatment of the perturbation equation in AdS$_7$ is particularly compelling, since, by following \cite{Dodelson:2022yvn}, it can be used to extract  OPE data for a class of operators in the 6d $\mathcal{N} = (2,0)$ theory. 

In this approach, a key role is played by the scalar thermal two-point function which, in the language of the AdS$_7$ wave equation, is simply given by the ratio of the connection coefficient \cite{Son:2002sd, Nunez:2003eq, Policastro:2002se}. From \eqref{conn2eqni} we get 
 \be\label{eq:thermalscal}\ba 
 G_R(\omega, \ell)=&-\left(\frac{y_2 (y_3-y_2)}{ y_3}\right)^{2 a_q}{W\left[\psi_{t,-}, \psi_{q,-}\right]\over W\left[\psi_{t,-},\psi_{q,+}\right]}\\
 &=\left(-y_2\right)^{2 a_q} {\Gamma(-2 a_q)\over \Gamma(2 a_q)}{\mathcal{G} (a_0,a_1,a_t,a_q,a_{\infty}, t,q)\over \mathcal{G}(a_0,a_1,a_t,-a_q,a_{\infty}, t,q)}~\Big|_{\text{\eqref{theoryparameter}}}~,
\ea \ee
where the overall factor is due to the change of variables from $r$ to $z$ and
\begin{equation}
\begin{aligned}\label{eq:gg7d}
\mathcal{G} (a_0, a_1, a_t, a_q, a_{\infty}, t, q)& = 
\re^{-\frac{1}{2} \partial_{a_q}F^{\rm NS}(1/q, t)} \Bigg(
\sinh\left(\frac{1}{2} \partial_{b_1}F^{\rm NS}\left(\frac{1}{q}, t\right)\right) 
\sinh\left(\frac{1}{2} \partial_{b_2}F^{\rm NS}\left(\frac{1}{q}, t, -a_1\right)\right) \\
&- \re^{-2 \ri a_1 \pi} 
\sinh\left(\frac{1}{2} \partial_{b_1}F^{\rm NS}\left(\frac{1}{q}, t, -a_1\right)\right) 
\sinh\left(\frac{1}{2} \partial_{b_2}F^{\rm NS}\left(\frac{1}{q}, t\right)\right)
\Bigg)~.
\end{aligned}  
\end{equation}
The QNM's frequencies are then identified as the poles of the thermal two point function.
Inspired by recent developements in the light-cone heavy-light bootstrap program \cite{Lashkari:2016vgj,Kulaxizi:2018dxo,Karlsson:2019qfi,Karlsson:2019dbd,Karlsson:2021duj,Dodelson:2022eiz}, we consider the small $\mu\over \ell^2$ expansion.
In this limit we have 
\begin{equation}
b_1^{(0,0)}=\ell\left(\frac{1}{2\sqrt{2}}+\mathcal{O}(\mu\ell^{-1})\right),\quad b_2^{(0,0)}=\ell\left(\frac{1}{2}+\mathcal{O}(\mu\ell^{-1})\right),
\end{equation}
and since within our range of parameters 
\begin{equation}\label{instantonbehavior}
\begin{aligned}
0<|t|<1,\quad\quad 0<\frac{1}{|q|}<1,
\end{aligned}
\end{equation}
 it follows that in the large $\ell$ regime
\begin{equation}
\begin{aligned}
t^{b_1}\to 0\,,\quad\quad
q^{-b_2}\to 0\,
\end{aligned}
\end{equation}
are both exponentially suppressed. 
Therefore, if we neglect non-perturbative effects at large \(\ell\), equation \eqref{eq:qcads7full} simplifies, as the terms with \(\theta_2 = 1\) and \(\theta_3 = -1\) are non-perturbative in \(1/\ell\). The quantization condition then reduces to:
\begin{equation}\label{quantcondAdS7}
\begin{aligned}
\frac{\Gamma\left(2b_1\right)\Gamma\left(1-2a_t\right)\Gamma\left(1+2b_1\right)\Gamma\left(2b_2\right)\Gamma\left(1+2b_2\right)\Gamma\left(2a_{q}\right)}{\prod_{\sigma=\pm}\Gamma\left(\frac{1}{2}-a_t+b_1+\sigma\,a_0\right)\Gamma\left(\frac{1}{2}+b_1+b_2+\sigma\,a_1\right)\Gamma\left(\frac{1}{2}+b_2+a_{q}+\sigma\,a_{\infty}\right)}=0.
\end{aligned}
\end{equation}
This condition translates into requiring the presence of poles in  $\Gamma\left(\frac{1}{2}+b_2+a_{q}+\sigma\,a_{\infty}\right)$, the only ones with arguments whose leading order in $\mu$ depends on $\omega$. Note that the different sign of $a_{\infty}$ results in a different sign in the real part of $\omega$. Without loss of generality, we choose the quantization condition
\begin{equation}\label{QCb2}
\frac{1}{2}+b_2+a_{q}+a_{\infty}=-n,\quad n\in\mathbb{Z}_{\ge 0}.
\end{equation}
As a quick reminder, $b_2$ is actually a function of the parameters in \eqref{theoryparameter}, which can be obtained by inverting the Matone relation
\begin{equation}
u_{q}=-\frac{1}{4}+b_2^2-a_{\infty}^2+a_{q}^2+q\frac{\partial F\big({1\over q},t\big)}{\partial q}.
\end{equation}
In the inversion process, we also need the instanton expansion of $b_1$, which is obtained by inverting the other Matone relation
\begin{equation}
u_{t}=-\frac{1}{4}- b_1^2 + a_0^2 + a_t^2+t\frac{\partial F\big({1\over q},t\big)}{\partial t}.
\end{equation}
Hence, upon using \eqref{theoryparameter}, equation \eqref{QCb2} becomes a quantization condition for the frequency $\omega$. 

Note that, from \eqref{theoryparameter}, we have the following behavior
\begin{equation}
t\sim -1+3\sqrt{\mu},\quad\quad q\sim -\frac{1}{\sqrt{\mu}}+\frac{3}{2}+\mathcal{O}\left(\mu\right)\quad\quad\text{as}\quad \mu\sim 0~.
\end{equation}
Since  $|t|$  is not parametrically small, it is not immediately clear whether the approach based on the NS functions is effective. Nevertheless, thanks to the triangular-like expansion in  $b_2$  given in \eqref{matone7d}, it suffices for  $1/|q|$   to be parametrically small, which is indeed the case.\footnote{For the thermal two point functions $G_R(\omega, \ell)$ instead we need not only $b_2$ but the full NS function. 
Consequently, an efficient expansion in  $\mu$  is not possible for  $G_R(\omega, \ell)$ , as the instanton counting parameter  $t$ in this case  cannot be made parametrically small.} Therefore, we can use the NS function to invert the quantization condition \eqref{QCb2} and extract the QNM frequencies.

Since the coefficients in the $b_2$ expansion \eqref{matone7d} behave as
\begin{equation}
b_2^{(K,K)}\left(\frac{t}{q}\right)^{K}\propto \mu^{K/2},
\end{equation}
we expect the  solutions to \eqref{QCb2} to be of the form
\begin{equation}\label{eq:omegar}
\omega = \sum_{k \geq 0} \omega_k \left( \frac{\mu}{\ell^2} \right)^{k/4}.
\end{equation}
However, upon applying the SW to gravity dictionary, we find that  all non-integer powers vanish because of some non-trivial cancellation which only happens when we use the SW to gravity dictionary in \eqref{theoryparameter}. For the first few coefficients, we find
\begin{equation}\label{QNMresults}
\begin{aligned}
\omega_0=&\,\ell+2n+\Delta,\\
\omega_{1}=&\omega_{2}=\omega_3=0\\
\omega_4=&\,-\frac{\ell^2}{2 (\ell+1) (\ell+2) (\ell+3)}\Bigl\{3 \Delta ^2 \left(\ell (4 n-1)+4 n^2+6 n-3\right)+\Delta ^3 (\ell+2 n+3)+\\
&\quad+2 \Delta  \{3 \ell n (5 n-7)+\ell+n [n (10 n+9)-35]+3\}+10 (n-2) (n-1) n (2 \ell+n+3)\Bigr\}\\
=&-\frac{1}{2} (\Delta +2 n-2) \left((\Delta -1) \Delta +10 n^2+10 (\Delta -2) n\right)+O\left(\left(\frac{1}{\ell}\right)^1\right)~.
\end{aligned}
\end{equation}
In the context  the AdS/CFT correspondence, the QNM expansion \eqref{eq:omegar} should match the the anomalous dimensions of heavy-light double-twist operators in the dual six-dimensional CFTs \cite{Kulaxizi:2018dxo,Karlsson:2019qfi,Karlsson:2019dbd,Karlsson:2021duj,Dodelson:2022eiz}. In fact, we can explicitly verify that the result \eqref{QNMresults} is in agreement with the anomalous dimensions computed in \cite{Li:2020dqm} hence providing an additional test of this correspondence.
Higher order terms in \eqref{eq:omegar} can be computed by going to higher order in the NS function \eqref{FNS}. This should be made more systematic either by using  generalizations of Zamolodchikov  recursion \cite{zamorecursion,Cho:2017oxl} or by extending the method of \cite[App.A]{Lisovyy:2022flm}. However we do not explore this further here.

\subsection{Electromagnetic and Gravitational perturbations Reissner-Nordstr\"om-AdS\texorpdfstring{$_5$}{} }\label{sec:AdS5}
 
We now turn to an application of the connection formula \eqref{paraconnII1q}. To this end, we consider a Reissner-Nordstr\"om-AdS$_5$ black hole (RN), whose metric is given by:
\begin{equation}
\mathrm{d}s^2=-f(r)\mathrm{d}t^2+f(r)^{-1}\mathrm{d}r^2+r^2\mathrm{d}\Omega^2_3,
\end{equation}
with
\begin{equation}
f(r)=1+r^2-\frac{2M}{r^2}+\frac{Q^2}{r^4},
\end{equation}
where $M$ and $Q$ are the mass and the charge of the black hole, and where the AdS radius was normalized to 1 (that is, the cosmological constant is set to $\Lambda=-6$). In the variable $y=r^2$, the function $y^2f(y)$ is a third-degree polynomial in $y$ with  three roots 
\be y^2f(y)= (y-y_h)(y-y_i)(y+1+y_h+y_i)~.\ee
Above extremity, we have two positive real roots, the bigger one, $y_h$, representing the event horizon and the smaller one, $y_i$, representing the inner horizon, and one real negative root, $y_n$.  
The two roots $y_i,y_n$ can be written in terms of $y_h$ as
\begin{equation}
y_i=\frac{-1-y_h+\sqrt{1-2y_h+8 M-3 y_h^2}}{2},\quad y_n=\frac{-1-y_h-\sqrt{1-2y_h+8 M-3 y_h^2}}{2}.
\end{equation}
The BH mass and charge are then given as
 \be M=-\frac{y_h\,y_i+y_h\,y_n+y_i\,y_n}{2} , \qquad Q^2=-y_i\,y_h\,y_n.\ee
The extremal limit of the geometry is reached in the confluence $y_h\to y_i$, that happens for the value of the charge
\begin{equation}
Q_{\text{ext}}=\frac{1}{3} \sqrt{\frac{2}{3}} \sqrt{(6 M+1)^{3/2}-9 M-1}.
\end{equation}
Massive scalar perturbations in this background are described by a second order differential equation with four regular singularities, hence the analysis is done exactly as in \cite{Dodelson:2022yvn}. 
Electromagnetic and gravitational perturbations, on the other hand, are more challenging due to their non-trivial coupling. In \cite{Kodama:2003kk, Kodama:2007ph}, the authors showed that these perturbation equations can be transformed into a set of two decoupled equations by introducing master variables written as linear combinations of gauge-invariant variables. Without sources, the differential equations read \cite[eqs. (4.25)-(4.29)]{Kodama:2007ph}, \cite[eq.~(4.37)]{Kodama:2003kk}
\begin{equation}\label{eq:masterRN}
f(r)\frac{\mathrm{d} }{\mathrm{d} r}\left(f(r) \frac{\mathrm{d} \Phi_{\pm} (r)}{\mathrm{d} r}\right)+\left(\omega ^2-V_{\pm}(r)\right)\Phi_{\pm} (r)=0,
\end{equation}
where the potential is 
\begin{equation}
V_{\pm}(r)= \frac{f(r)}{r^2}\left(k_V^2+\frac{39 Q^2}{4 r^4}+\frac{\mu_{\pm}}{r^2}+\frac{3 r^2}{4}+\frac{7}{4}\right),
\end{equation}
with $k_V^2=\ell(\ell+2)-1$ and 
\begin{equation}
\mu_{\pm}=-\frac{11}{2}M\pm\left[64M^2+12(k_V^2-2)Q^2\right]^{1/2}.
\end{equation}
When $Q=0$, $\Phi_+$ reduces to the electromagnetic perturbation while $\Phi_-$ to the standard vector-type gravitational perturbation in Schwarzshild AdS$_5$ background, \cite[eq.~(4.40)]{Kodama:2003kk}.

For a generic value of the charge $Q$, the master equation \eqref{eq:masterRN} has five regular singularities at $0,y_i,y_h, y_n$ and $\infty$. For $Q=Q_{\rm ext}$, $y_h\to y_i$ and  the equation develops an irregular singularity at the horizon, in addition to the regular singularities at $y=0$, $y=y_n$, and $y=\infty$. On the gauge theory side this correspond to the decoupling limit of the quiver theory.

Let us take the differential equation satisfied by $\Phi_+$ (for the $\Phi_-$ case  the procedure is the same). Introducing the new variable 
\begin{equation}
z=\frac{y}{y_h}\frac{y_h-y_n}{y-y_n}
\end{equation}
and redefining the wave function as
\begin{equation}
\Phi_{+}(z)=\frac{z^{3/4}}{\sqrt{1-z} \sqrt[4]{y_h (z-1)+y_n} \sqrt{y_h (-y_i z+y_i+y_n z)-y_i y_n}}\,\psi(z),
\end{equation}
the differential equation satisfied by $\psi(z)$ can be written in the form \eqref{diffeq5normal}
with the following dictionary:
\begin{equation}\label{eq:dicRN}
\begin{aligned}\small
t=&\,\frac{y_h y_i-y_i y_n}{y_h (y_i-y_n)},\quad
q=\,\frac{y_h-y_n}{y_h},\quad
a_0=\,2,\quad a_{q}=\,\frac{1}{2},\\
a_t=&\,\frac{i\,\omega\,y_i^{3/2}}{2 (y_h-y_i) (y_n-y_i)},\quad
a_1=\,\frac{i\,\omega\, y_h^{3/2}}{2 (y_h-y_i) (y_h-y_n)},\quad
a_{\infty}=\,\frac{i\,\omega\,y_n^{3/2}}{2 (y_h-y_n) (y_i-y_n)},\\
u_t=&\,\frac{3 y_h-y_i}{4 (y_h-y_i)}-\frac{y_i \left[y_i \left(4 k_V^2+12 y_h-y_i+7\right)-22 M\right]+39 Q^2}{16 y_i y_n (y_h-y_i)}\\
&-\frac{\sqrt{3 \left(k_V^2-2\right) Q^2+16 M^2}}{2 y_n (y_h-y_i)}+\frac{ \omega ^2 y_i^3 (y_i-3 y_h)}{4 y_n (y_h-y_i)^3 (y_i-y_n)},\\
u_{q}=&\,\frac{-4 k_V^2+4 \omega ^2-8}{16 y_n}.
\end{aligned}
\end{equation}
The differential equation satisfied by $\Phi_-$ has almost the same dictionary, but with a different sign in front of the square root term in $u_t$:
\begin{equation}
\begin{aligned}
u_t=&\,\frac{3 y_h-y_i}{4 (y_h-y_i)}-\frac{y_i \left[y_i \left(4 k_V^2+12 y_h-y_i+7\right)-22 M\right]+39 Q^2}{16 y_i y_n (y_h-y_i)}\\
&+\frac{\sqrt{3 \left(k_V^2-2\right) Q^2+16 M^2}}{2 y_n (y_h-y_i)}+\frac{ \omega ^2 y_i^3 (y_i-3 y_h)}{4 y_n (y_h-y_i)^3 (y_i-y_n)}.
\end{aligned}
\end{equation}
In the $z$ variable, the black hole horizon is located at $z=1$ and the AdS boundary is at $z=q$.  Therefore, the relevant problem is the one in \autoref{sec:connIII}. As boundary conditions for the QNMs, we impose the vanishing Dirichlet boundary condition at $z=q$, and the presence of only ingoing modes at the horizon $z=1$. In terms of the original wave function $\Phi_+(r)$ and of the tortoise coordinate $r_*=\int \mathrm{d}r/f(r)$, we require 
\begin{equation}
\begin{aligned}
&\Phi_+(r)\sim \exp(-i\,\omega\,r_*)  &\text{for}\ r\sim r_h,\\
&\Phi_+(r)\sim r^{-3/2}&\text{for}\ r\to \infty.
\end{aligned}
\end{equation}
In terms of the wave function $\psi$, this translates to
\begin{equation}\label{bcAdS5}
\begin{aligned}
&\psi(z)=\psi_{1,-}(z)\sim (z-1)^{\frac{1}{2}-a_1}  &\text{for}\ z\sim 1,\\
&\psi(z)=\psi_{q,+}(z)\sim (z-q)^1 &\text{for}\ z\sim q~.
\end{aligned}
\end{equation} Note that $a_q = \frac{1}{2}$ and $a_0 = 2$  correspond to logarithmic points of the differential equation, and the connection coefficients computed in \autoref{sec:connIII} are singular at this points. However, as discussed in \cite{Jia:2024zes,unpGGDC}, the quantization condition and the connection coefficients can be extracted from the non-logarithmic case $a_i \notin \mathbb{Z}/2$ by analytic continuation,  which is equivalent to retaining the regular part as  $a_i \to \mathbb{Z}/2$. This is the approach we  adopt here.
Therefore, the quantization condition arises from the connection formula between the semiclassical conformal blocks expanded around $z=1$ and the ones around $z=q$, and is given by the requirement that the coefficient in front of the non-decaying solution at infinity is zero:
\begin{equation}\label{eq:qcch}
\begin{aligned}
\sum_{\theta=\pm}\mathcal{M}_{-\theta}(a_1,b_2;b_1)\mathcal{M}_{(-\theta)-}(b_2,a_{q};a_{\infty})q^{-\theta\,b_2}\exp\left(-\frac{\theta}{2}\partial_{b_2}F\big({1\over q},t\big)\right)=0.
\end{aligned}
\end{equation}
From the perspective of AdS/CFT, the wave equation associated with gravitational and electromagnetic perturbations encodes information about the two-point functions of (mixed)  stress-energy tensor and conserved currents, respectively \cite{Son:2002sd,Kovtun:2005ev}. As in the scalar case the relevant quantity from which we extract these two-point functions is the ratio of connection coefficients 
\be\ba G_R=&-\left(\frac{y_n (y_h-y_n)}{y_h}\right)^{2a_q}{W\left[\psi_{1,-},\psi_{q,-}\right]\over W\left[ \psi_{1,-},\psi_{q,+}\right] }.\ea\ee
By using the \eqref{paraconnII1q} this reads
\be  \label{eq:grgrav} \ba G_R=& y_n^{2a_q} {\Gamma\left(-2 a_q\right)\over \Gamma\left(2 a_q\right)}{\mathcal{G}(a_0,a_1,a_t,a_q,a_{\infty},t,q)\over \mathcal{G}(a_0,a_1,a_t,-a_q, a_{\infty},t,q)}~\Big|_{\text{\eqref{eq:dicRN}}}\ea \ee
where 
\be \mathcal{G}(a_0,a_1,a_t,a_q,a_{\infty},t,q)=\cosh\left({1\over 2}\partial_{b_2}F^{\rm NS}\big({1\over q},t,- a_1, a_q \big)\right)\re^{-\frac{1}{2}\partial_{a_q}F^{\rm NS}\big({1\over q},t, a_1, a_q\big)}.
\ee
As previously mentioned, for the black hole geometry under consideration, we set  $a_q = \frac{1}{2}$, which corresponds to a logarithmic singularity of the differential equation. Consequently, the equation \eqref{eq:grgrav} requires regularization. Following the approach in \cite{Jia:2024zes, unpGGDC}, this is achieved by extracting the finite part. %

Similarly to the case of scalar perturbations, the expressions here also simplify significantly in the small $\frac{M}{\ell} $ expansion.
It is convenient to introduce
 \be \tilde{Q}= {Q\over M}~.\ee 
Let us first consider the quantization condition \eqref{eq:qcch}.
Note that
\be
b_2^{(0,0)} = \ell \left(\frac{1}{2} + \mathcal{O}\left(\frac{M}{\ell}, \frac{\tilde{Q}}{\ell}\right)\right).
\ee
Using the correspondence provided in \eqref{eq:dicRN}, we note that
\be
\frac{1}{|q|} < 1.
\ee
This implies that the term proportional to $q^{-b_2}$ in \eqref{eq:qcch},  is exponentially suppressed for large $\ell$.
This result holds both for $\Phi_-$ and $\Phi_+$.
 Therefore, by neglecting non-perturbative effects in $\ell$, the quantization condition simplifies to
\begin{equation}\label{eq:qcsimple}
\frac{1}{2}+b_2+a_{q}\pm a_{\infty}=-n,\quad n\in\mathbb{Z}_{\ge 0}.
\end{equation}
where the different $\pm$ signs in \eqref{eq:qcsimple} correspond to a sign change in the real part of the QNM frequencies.
The two instanton parameter behave as  
\begin{equation}  \ba 
q^{-1}=&2M +\mathcal{O}(M^2, M \tilde{Q}^2 )~,\\
t = & {1\over 4}\tilde{Q}^2 + \mathcal{O}(M\tilde{Q}^2,\tilde{Q}^4)~,  
\ea
\end{equation}  
suggesting that the natural approach would be a double expansion in \( M \) and \(\tilde{Q}\). However, due to the triangular structure of the Matone relation \eqref{matone7d}, an expansion at small \(\tilde{Q}\) is unnecessary for the QNMs.  As a consequence, we consider the following expansion for the frequencies:
\begin{equation}\label{omegaexpansionRNAdS5}
\omega^{\pm}(Q,M,\ell,n)=\sum_{j\ge 0}\omega_{j}^{\pm}(\ell,n,\tilde{Q})\,\left({M\over \ell}\right)^j,
\end{equation}
where $\pm$ refer to the $\Phi_{\pm}$ perturbation.

Taking the quantization condition in \eqref{eq:qcsimple} with the minus sign in front of $a_{\infty}$, we find for the first few orders
$\omega_{0}^{\pm}(\ell,n,\tilde{Q})=3 + \ell + 2 n$ and 
\be \ba \omega_1^{\pm}(\ell,n, \tilde{Q})=&\,\frac{-2 (n+1) }{(\ell+1) (\ell+2)}\biggl[3 \ell^2 (n+1)\mp(\ell+n+2) \sqrt{3 (\ell-1) (\ell+3) \tilde{Q}^2+16}+\ell (6 n+11)+5 (n+2)\biggr].%,\\
\ea\ee
 More orders are given in  \autoref{app:frequencies}.
 When $\tilde{Q}=0$ the structure simplifies significantly, and, in particular, no square roots are involved.
We find  that  $\omega_{j}^{\pm}(\ell,n,0)$ are polynomials in $n$ of degree $2j$ and meromorphic functions in $\ell$ with poles at integers  $\ell$ for $-j-1\leq\ell\leq j-1$\footnote{Note that these poles in $\ell$ are artifacts of our expansion and are resolved once the non-perturbative effects in $\ell$ are taken into account. }. This behavior is very similar to the one of scalar perturbations in Schwarzschild AdS$_5$, see \cite{Dodelson:2022eiz}.
 
 In  the large $\ell$ limit we find
\be \label{omlagq}\omega_j^{\pm}(\ell,n,\tilde Q)=p_j^{\pm}(n,\tilde{Q})+\mathcal{O}\left({1\over \ell}\right), \ee where \(p_j^\pm(n,\tilde{Q})\)  is a polynomial of degree \(j+1\) in $n$  and of degree $j$ in $\tilde{Q}$. 
 In addition,  for the leading term in \eqref{omlagq}, we have 
\be p_j^+(n,0)=p_j^-(n,0)=p_j(n) \ee
where 
$p_j(n)$  is the same polynomial as for scalar QNMs frequencies if we set the dimension of the scalar operator to be $\Delta=3$   in \cite{Karlsson:2020ghx, Li:2019zba, Li:2020dqm,Dodelson:2022yvn}.

Let us now consider the ratio of connection coefficients \eqref{eq:grgrav}.
Neglecting the non-perturbative contributions in $\ell$  we find 
\begin{equation}
\begin{aligned}\label{GRcharge}
G_R&\simeq 
y_n^{2a_q}\exp\left(-\partial_{a_{q}}F\big({1\over q},t\big)\right)\frac{\Gamma\left(-2a_{q}\right)\Gamma\left(\frac{1}{2}+b_2+a_{q}+a_{\infty}\right)\Gamma\left(\frac{1}{2}+b_2+a_q-a_{\infty}\right)}{\Gamma\left(2a_{q}\right)\Gamma\left(\frac{1}{2}+b_2-a_{q}+a_{\infty}\right)\Gamma\left(\frac{1}{2}+b_2-a_q-a_{\infty}\right)}.
\end{aligned}
\end{equation}
We look for the residue of \eqref{GRcharge} at the poles of the $\Gamma$-function in the numerator given by
\begin{equation}
\frac{1}{2}+b_2+a_q-a_{\infty}=-n,
\end{equation}
which correspond to the QNMs in \eqref{omegaexpansionRNAdS5}.
We find 
\begin{equation}
\begin{aligned}
\mathrm{Res}\left(G_R,\omega_{\text{QNM}}^{\pm}\right)&
=y_n\,\exp\left(-\partial_{a_{q}}F\big({1\over q},t\big)\right)\,(n+1)\,(2a_{\infty}-n-1)\left[\frac{\mathrm{d}\left(b_2-a_{\infty}\right)}{\mathrm{d}\,\omega}\right]^{-1}\bigg|_{\omega=\text{\eqref{omegaexpansionRNAdS5}}}.
\end{aligned}
\end{equation}
For simplicity let us consider the case $Q=0$. We have
\begin{equation}
\begin{aligned}
\mathrm{Res}\left(G_R,\omega=\text{\eqref{omegaexpansionRNAdS5}}\right)&=\sum_{j\ge 0}c_{j}^\pm(\ell,n)M^j,
\end{aligned}
\end{equation}
where for the first few orders  
\be\ba 
  c_0^+(\ell,n)&=c_0^-(\ell,n)=2(n+1)\,(\ell+n+2),  \\
c_1^+(\ell,n)&=-\frac{2(n+1)}{\ell (\ell+1) (\ell+2)}\times\\
&\left[9 \ell^3 (n+1)+\ell^2 \left(12 n^2+51 n+41\right)+\ell \left(24 n^2+72 n+52\right)+2 \left(2 n^2+7 n+6\right)\right],
\\
c_1^-(\ell,n)&=c_1^+(\ell,n)-\frac{16 (n+1) (\ell+n+2) (\ell+4 n+5)}{\ell (\ell+1) (\ell+2)}.
\ea\ee
More results are given in \autoref{app:frequencies}.
As expected, the leading terms in the large $\ell$ expansion have the form
\be c_i^{\pm}(\ell,n)={p_i(n)\over \ell^i}+\mathcal{O}\left({1\over\ell^{i+1}}\right),\ee
where $p_i(n)$ are polynomials in $n$ of degree $i+1$ and they coincide with the results in the $\mu$-expansion in \cite{Dodelson:2022yvn} with $\Delta=3$ taking into account the additional overall factor $\frac{\ell+1}{2(\Delta-1)(\Delta-2)}=\frac{\ell+1}{4}$ from formula (38) in \cite{Dodelson:2022yvn} and the fact that $\mu=2M$.

The quantities above should be related to the OPE data of double-twist operators involving the stress tensor and a heavy operator. However, we are not aware of any available results on the bootstrap side for the case of the black hole geometry considered here. It would be interesting to take the black brane limit and make connections with, e.g.,~\cite{Karlsson:2022osn}. We leave this for future work.

\section{Spectral Network}\label{sec:sn}

Starting with \cite{Aminov:2020yma}, there has been a growing number of applications of the NS functions in the context of black hole perturbation theory. In these applications, the NS functions are typically computed using instanton expansions, relying on expressions such as those provided in \autoref{appendixA} and \autoref{appendix5+1}. However, certain questions in gravity require expansions around specific values of the instanton counting parameters, as discussed in \autoref{sec:AdS5} and \autoref{sec:AdS7}. In these cases, the instanton counting representation of the NS function proves useful, provided we operate in a regime where the instanton counting parameters (e.g., $t$ and $1/q$) are parametrically small. However, this condition is not always satisfied.
For instance, when studying massive scalar perturbations of an $\mathrm{AdS}_4$ Schwarzschild black hole, the wave equation takes the form \eqref{diffeq5normal},  with instanton counting parameters
\be t=\frac{R_h (R_--R_+)}{R_- (R_h-R_+)}~,\quad
{1\over q}={1\over 1-\frac{R_+}{R_-}}~,\ee
where
\be R_{\pm}=\frac{-R_h\pm i\sqrt{4R_h^2+3}}{2}~,\ee
and $R_h$ is the BH horizon.  Hence $|t|\in \left[0 ,{8\over \sqrt{65}}\right]$ and ${1\over |q|}\in \left[{1 \over 2} , \frac{\sqrt{5}}{4}\right]$. 
Since $\frac{1}{q}$ is not parametrically small, we cannot even exploit the triangular structure of the Matone relation \eqref{matone7d}, making the approach based on instanton counting inefficient  to explore the regime of interest for the light-cone bootstrap.   Likewise, when studying the hydrodynamic limit \cite{Policastro:2002se}, one needs to take a scaling limit where the black hole mass goes to infinity. In this case, the instanton counting parameters are usually fixed to specific values (e.g., $q = 1/2$ or $t = 1/2$). Although these values lie within the radius of convergence of the instanton counting expressions, they are not parametrically small, making it difficult to extract useful information.

Therefore, in such situations, it would be valuable to have an alternative method for computing the NS functions 
that does not rely on instanton counting. This could be done via the NLIE approach discussed in \cite{ns,kt2,Meneghelli:2013tia}, the Zamolodchikov-TBA formalism of  \cite{post-zamo,Fioravanti:2021dce,Fioravanti:2019vxi} or via
 the conformal limit of the GMN TBA \cite{Gaiotto:2009hg,Gaiotto:2014bza,Hollands:2017ahy,Hollands:2013qza,Hollands:2019wbr,Grassi:2019coc,Grassi:2021wpw,Imaizumi:2021cxf}.
 The advantage of these methods is that they do not require the instanton counting parameters to be small.  However, at least the  GMN   approach, is more practical when applied to the so-called strong-coupling region of the moduli space \cite{seiberg1994, fb, selfdual}, as this region is characterized by a finite set of BPS particles and therefore a finite set of TBA equations. 
To transition from the strong to the weak coupling region, one must cross the so-called wall of marginal stability, where BPS particles can decay or appear. Once the weak coupling region is reached, the spectrum changes and it includes an infinite number of BPS particles, hence to compute the NS functions via the GMN approach one would need an infinite towers of coupled TBA equations.

In the black hole setting, the gauge theory parameters are often constrained by non-trivial relations. It is therefore natural to ask whether the spectral problems associated with black hole perturbation theory lie within the weak-coupling or strong-coupling region. This  depends on the value of the frequency  $\omega$ as well as the other black hole parameters. By setting $\omega$ to  one of the quasinormal modes (QNMs), we provide examples below where we show whether the relevant region corresponds to the strong-coupling region or weak-coupling region.

To determine the relevant region in the moduli space we use the spectral network approach \cite{Gaiotto:2012rg}, see  \cite{Hollands:2021itj} for a recent review on their relevance in the context of  $\mathcal{N}=2$ theories.
A spectral network $\cW^\vartheta$ depends on a phase factor $\vartheta$ and a quadratic differential $\lambda$  which is defined by the ODE data in \eqref{eq:ode}, that is
\begin{equation}\label{SWdiff}
    \lambda^2=-P(z)\rd z^2,
\end{equation}
where $P(z)$ is defined in \eqref{eq:ode}. There are 3 walls emanated from each branch point $b$, which is denoted by an orange cross. And each wall comes with a label $ij$, where $i,j=1,2$, $i\neq j$ denoting the two sheets of \eqref{SWdiff}
\begin{equation}
    \lambda^{(i)}=(-1)^i\sqrt{-P(z)}\rd z
\end{equation}
and is given by a trajectory of $z$ satisfying
\begin{equation}
    \arg\left(-\int_{b}^z \left(\lambda^{(i)}-\lambda^{(j)}\right) \right)=\vartheta.
\end{equation}
In this section, we will neglect this label for convenience.

Spectral networks can be used to detect crucial information about the system. For example, an important application of it is to count BPS states. Graphically, BPS states show up as 
degenerate walls. In the $SU(2)$ theories, a hypermultiplet consists of 2 head to head single walls -- usually referred to as a saddle. On the other hand, a vectormultiplets corresponds to a ring domain whose boundary are degenerate walls. 
If we let the phase $\vartheta$ vary from $0$ to $2\pi$ and record all the degenerate walls, this gives us the BPS spectrum of the supersymmetric gauge theory.
Actually, the counting of BPS states 
also gives an important piecewise invariant of the theory, known as the BPS index \cite{Gaiotto:2009hg,Hollands:2016kgm,Hao:2019ryd}, and
the BPS index can tell us whether the parameters of supersymmetric gauge theory is in the strong coupling region or the weak coupling region.

We plot the spectral networks for scalar perturbations of  
Schwarzschild black hole, corresponding to $SU(2)$ $N_f=3$ theory,  the extremal Kerr black hole, corresponding to $SU(2)$ $N_f=2$ theory, as well as   AdS$_5$- black brane, corresponding to $SU(2)$ $N_f=4$ theory. 

%In the $SU(2)$ $N_f=2,3$ examples, we see a vectormultiplet which suggest that the parameter is in the weak coupling region. For the black brane case, varying the phase $\vartheta$, we see no vector multiplets. This indicates that such a spectral network belongs to a strong coupling region. 

\begin{figure}
    \centering
    \includegraphics[width=0.35\linewidth]{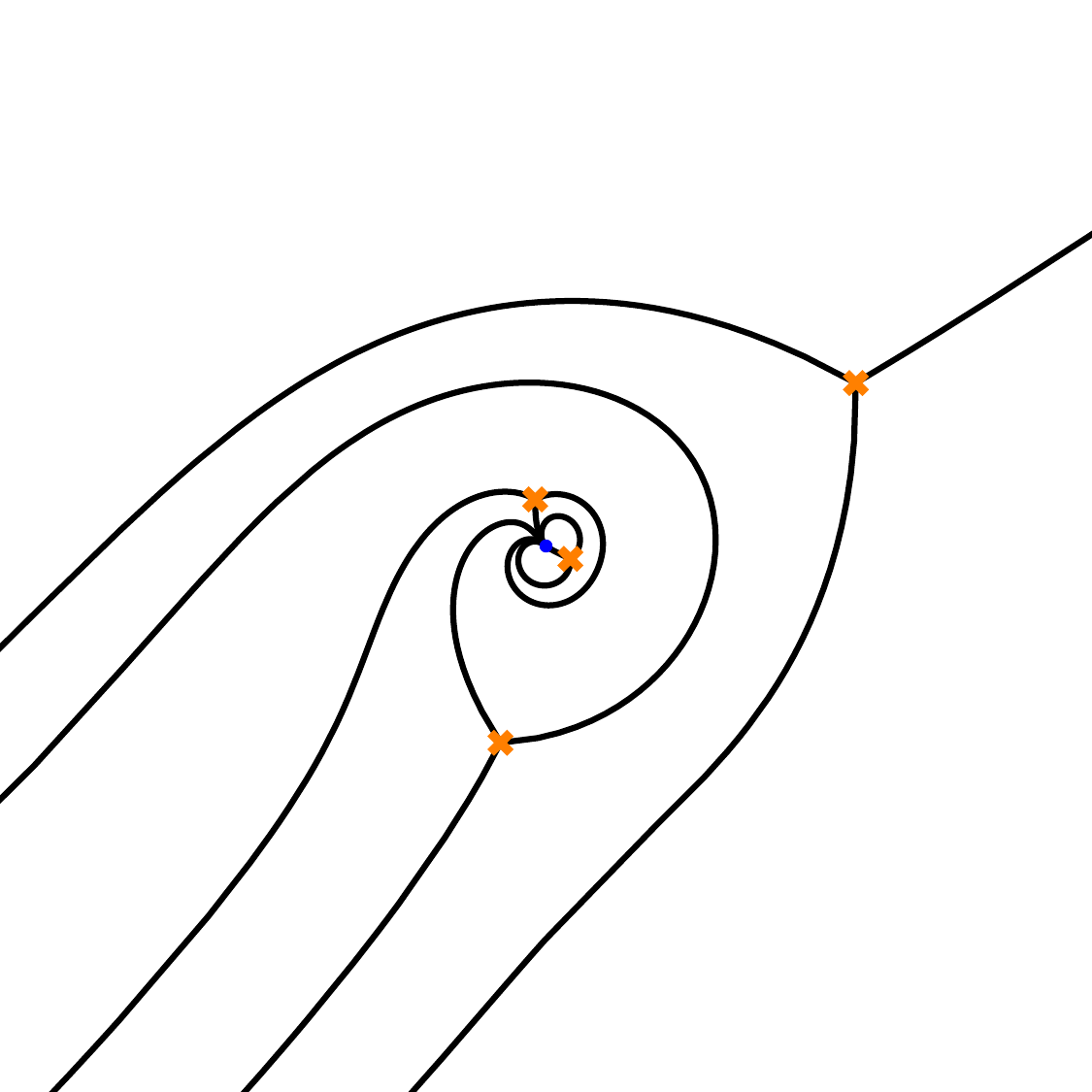} \qquad \includegraphics[width=0.35\linewidth]{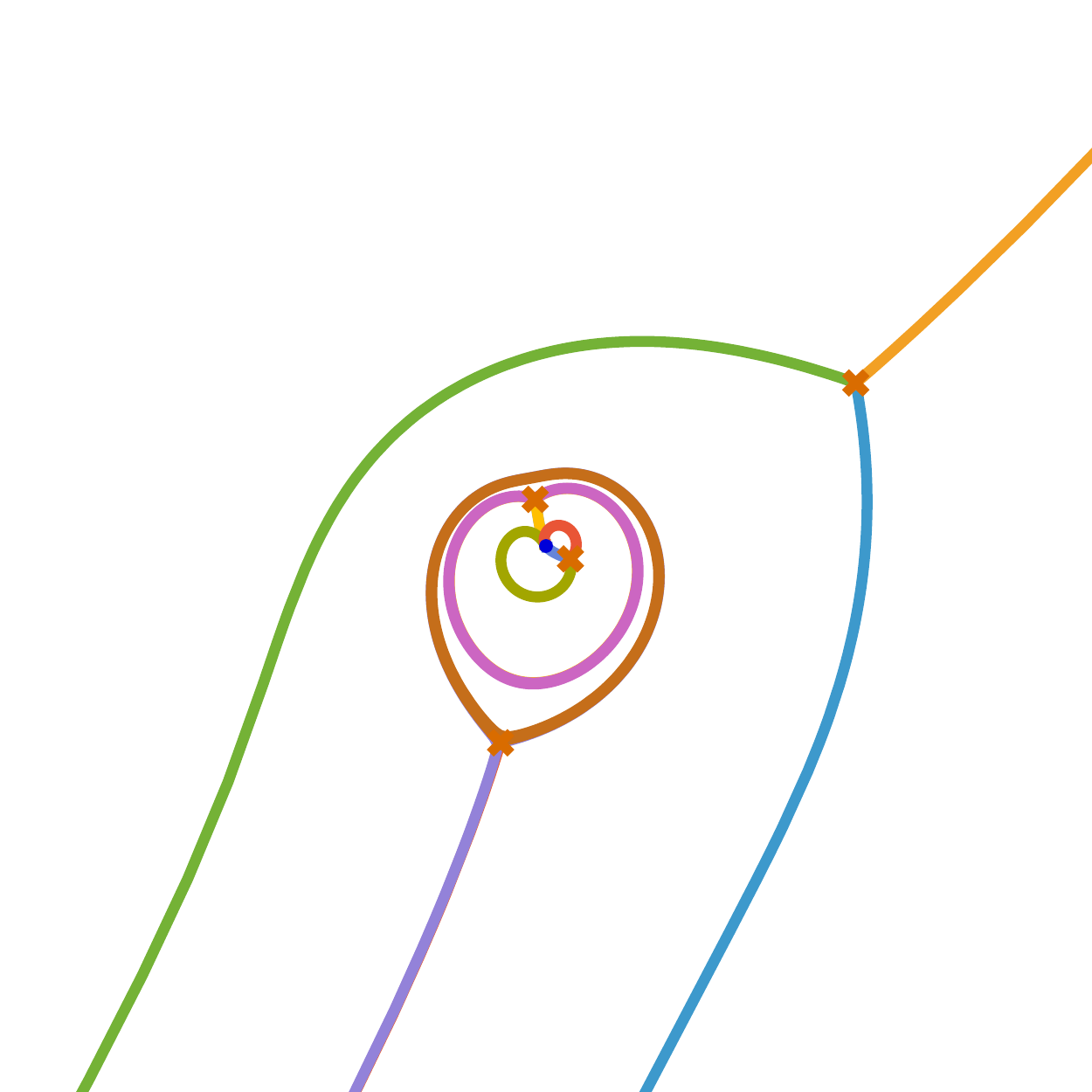}
    \caption{Spectral networks for extremal Kerr black hole with parameters in \cite{Aminov:2020yma}. The BH quantum numbers are $l=s=m=0$. On the left we set  $\vartheta=0$ while on the right we set $\vartheta\approx 0.0872\pi$. In the second case we see a ring domain in between the brown and pink degenerate walls. }
    \label{fig:Nf2}
\end{figure}

\begin{figure}
    \centering
    \includegraphics[width=0.23\textwidth]{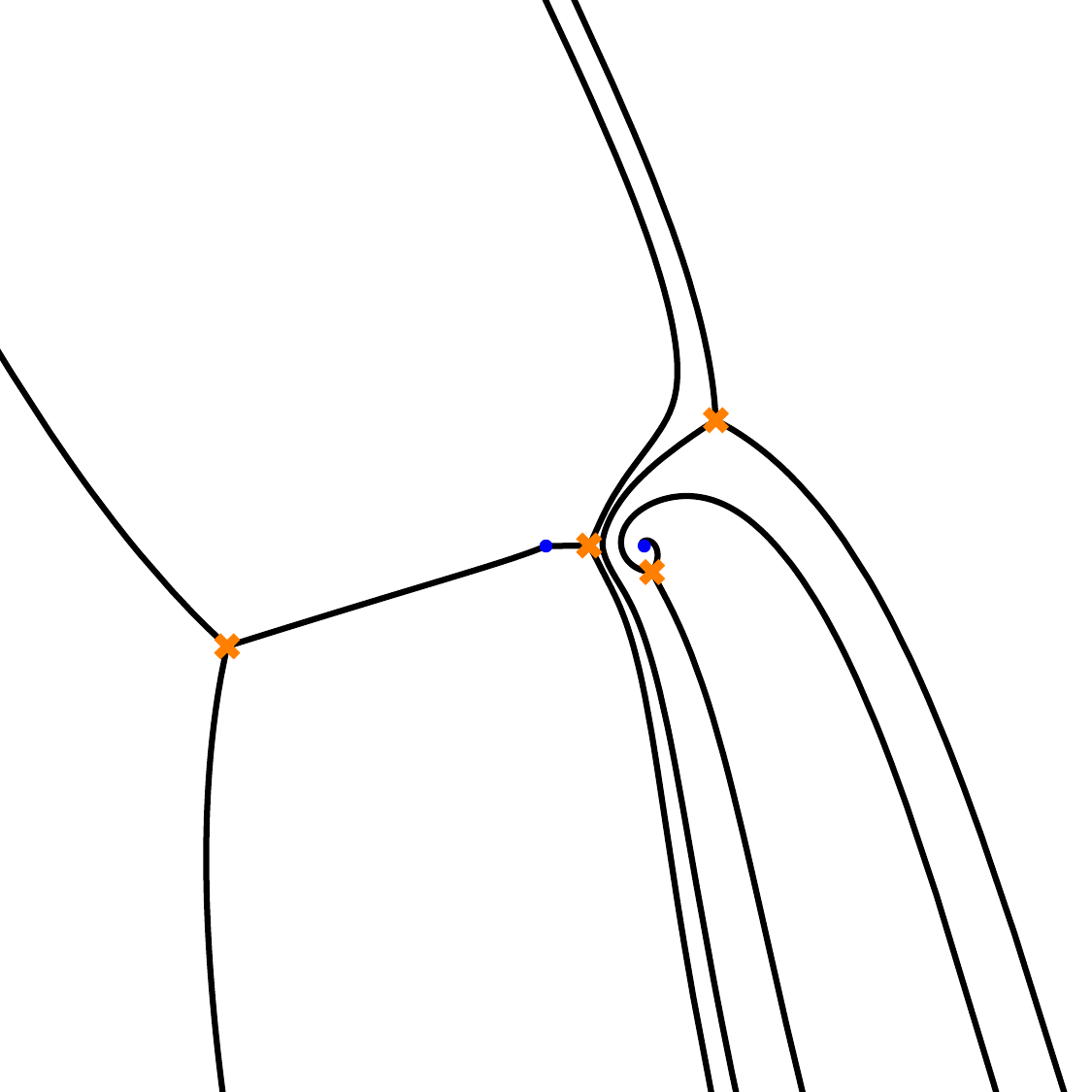}
    \hspace{1cm}
    \includegraphics[width=0.23\textwidth]{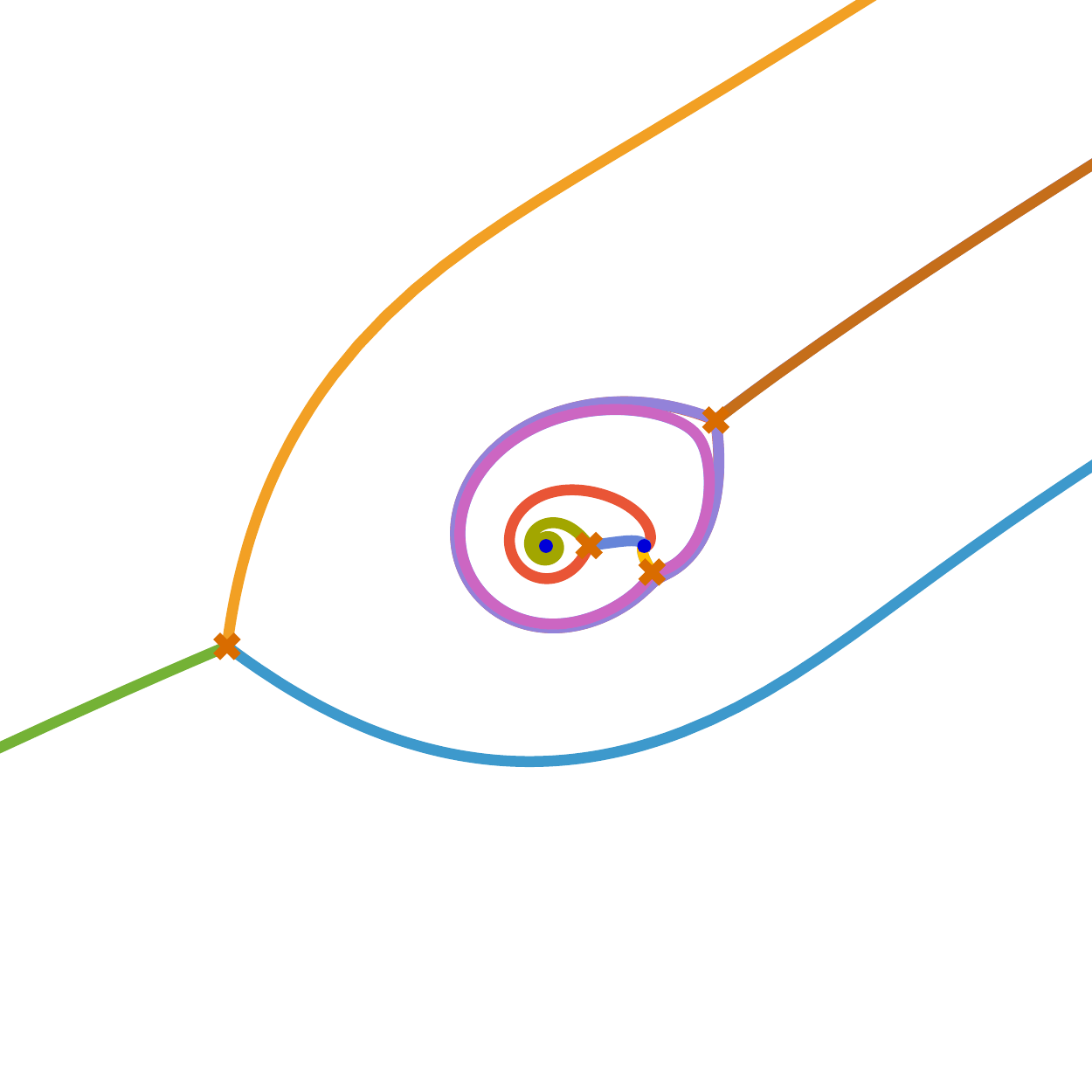}
    \hspace{1cm}
    \includegraphics[width=0.23\textwidth]{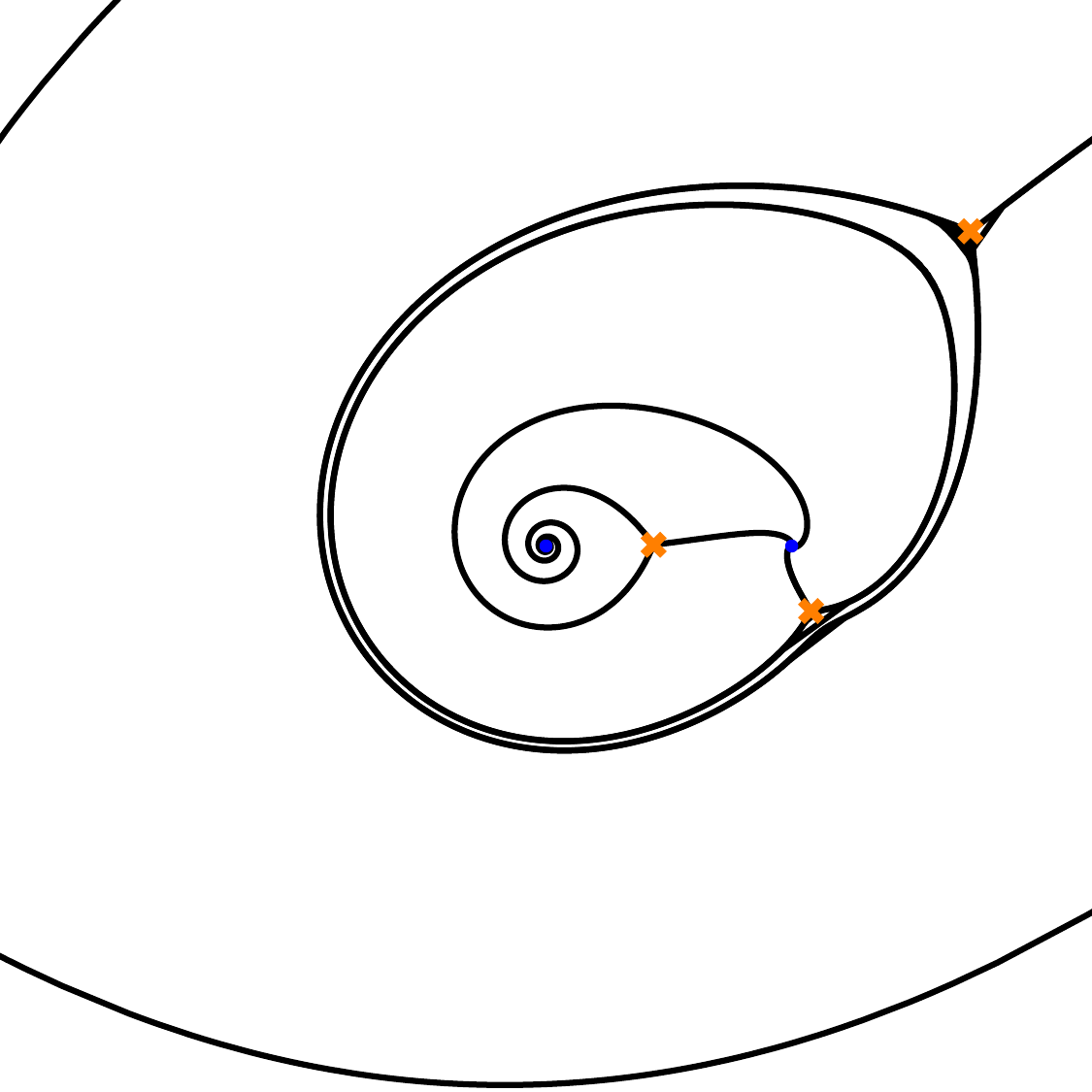}
    \caption{Spectral networks for Schwarzschild black hole corresponding to $N_f=3$ theory with parameters in \cite{Aminov:2020yma}. The BH quantum number are $n=0$, $l=s=1$. In the left picture we set $\vartheta=0$ while in the middle and right pictures we set $\vartheta\approx 0.5449\pi$. We see a ring domain in the middle picture in between the pink and purple degenerate walls. We zoom in further on the right picture.}
    \label{fig:Nf3list}
\end{figure}

The spectral network for the extremal Kerr black  is shown in \autoref{fig:Nf2}. Varying the phase $\vartheta$, we see a ring domain approximately at $\vartheta\approx 0.0872\pi$, which is shown in \autoref{fig:Nf2}. This suggests that the black hole parameters lie in the weak coupling region. 
Likewise, the spectral network for the scalar perturbations of Schwarzschild black hole is shown in \autoref{fig:Nf3list}.   Varying the phase $\vartheta$, we see a ring domain approximately at $\vartheta\approx 0.5449\pi$. This indicates that the black hole parameters lie in the weak coupling region. 

\begin{figure}[h!]
    \centering
    \includegraphics[width=0.23\textwidth]{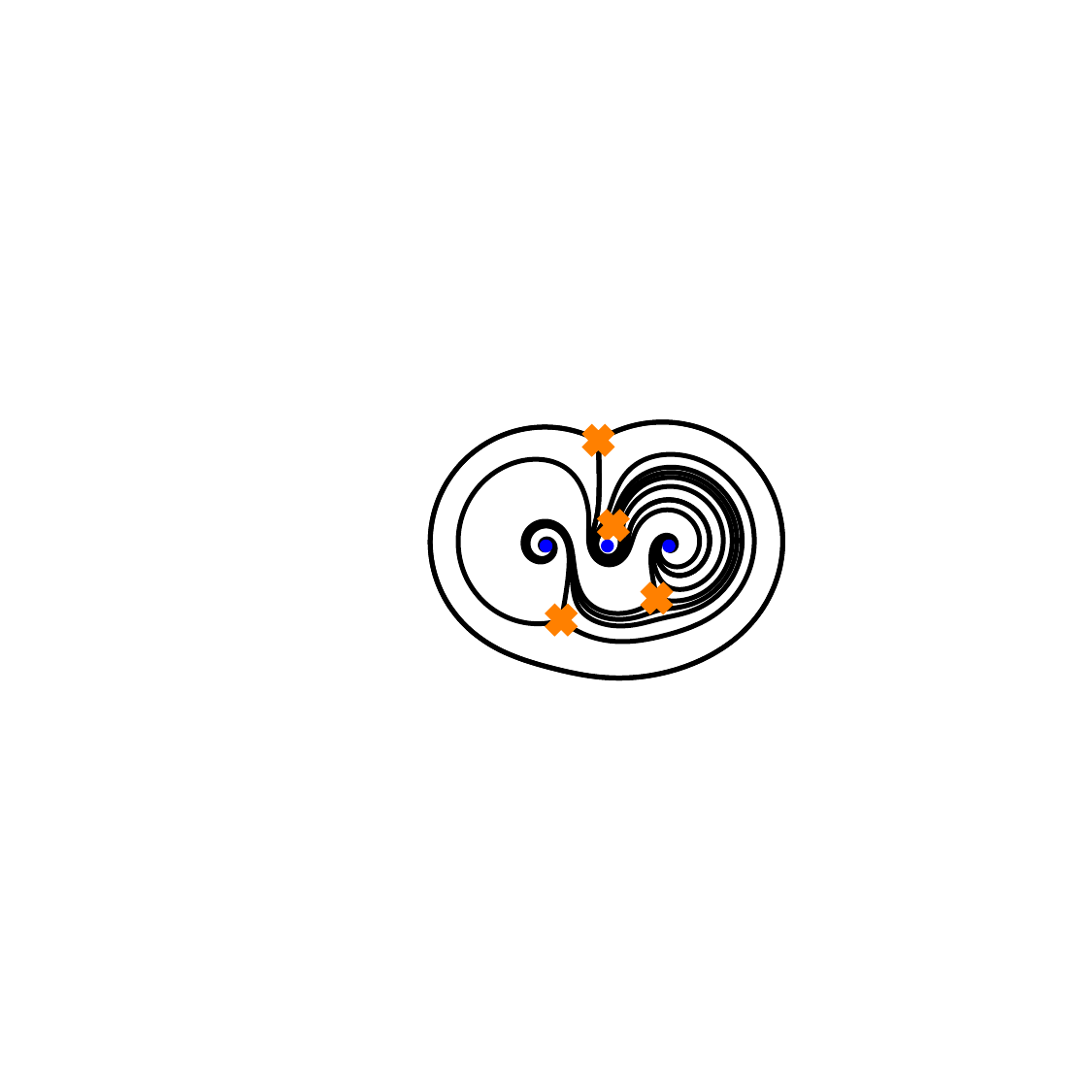} 
    \includegraphics[width=0.23\textwidth]{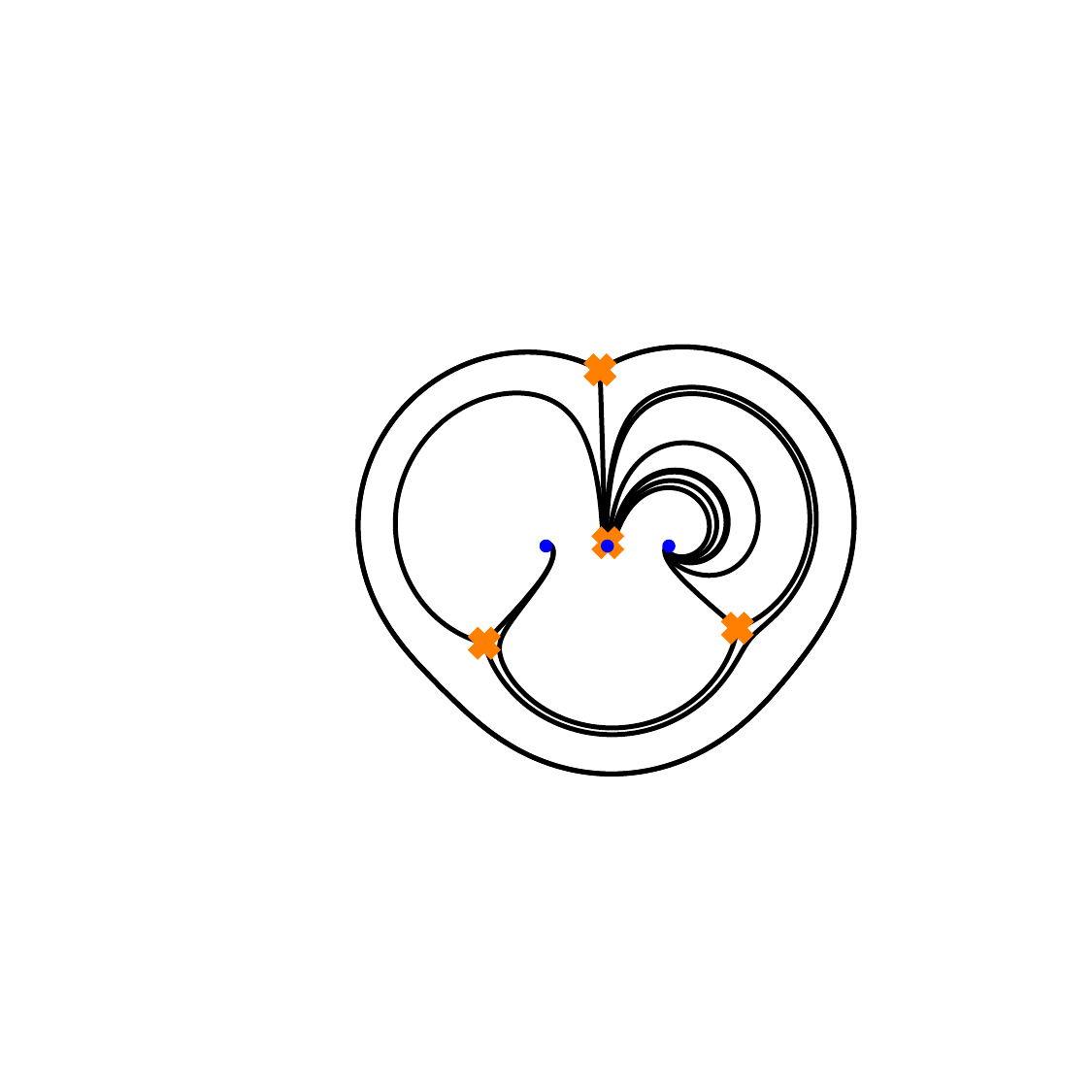}
    \includegraphics[width=0.23\textwidth]{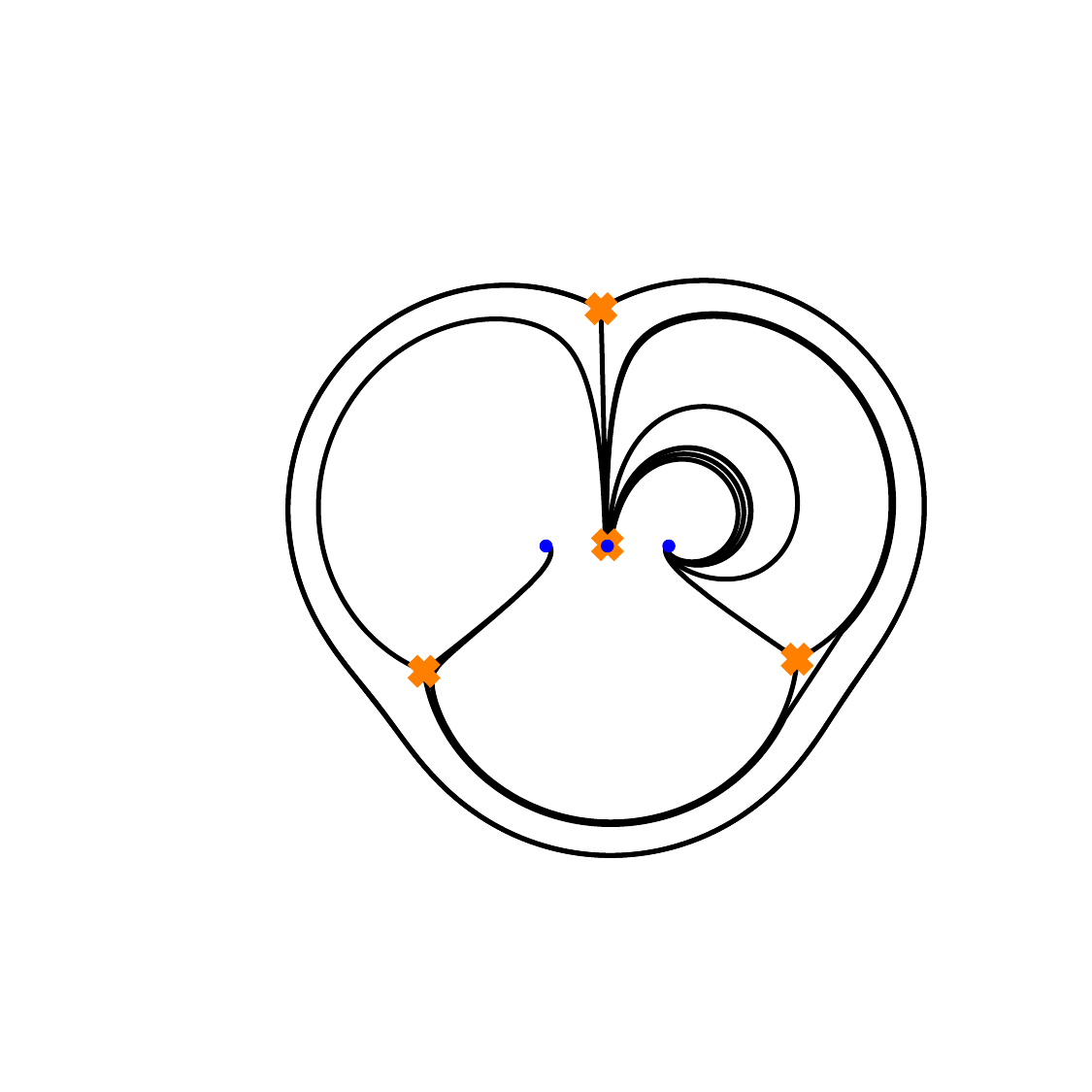}
    \includegraphics[width=0.23\textwidth]{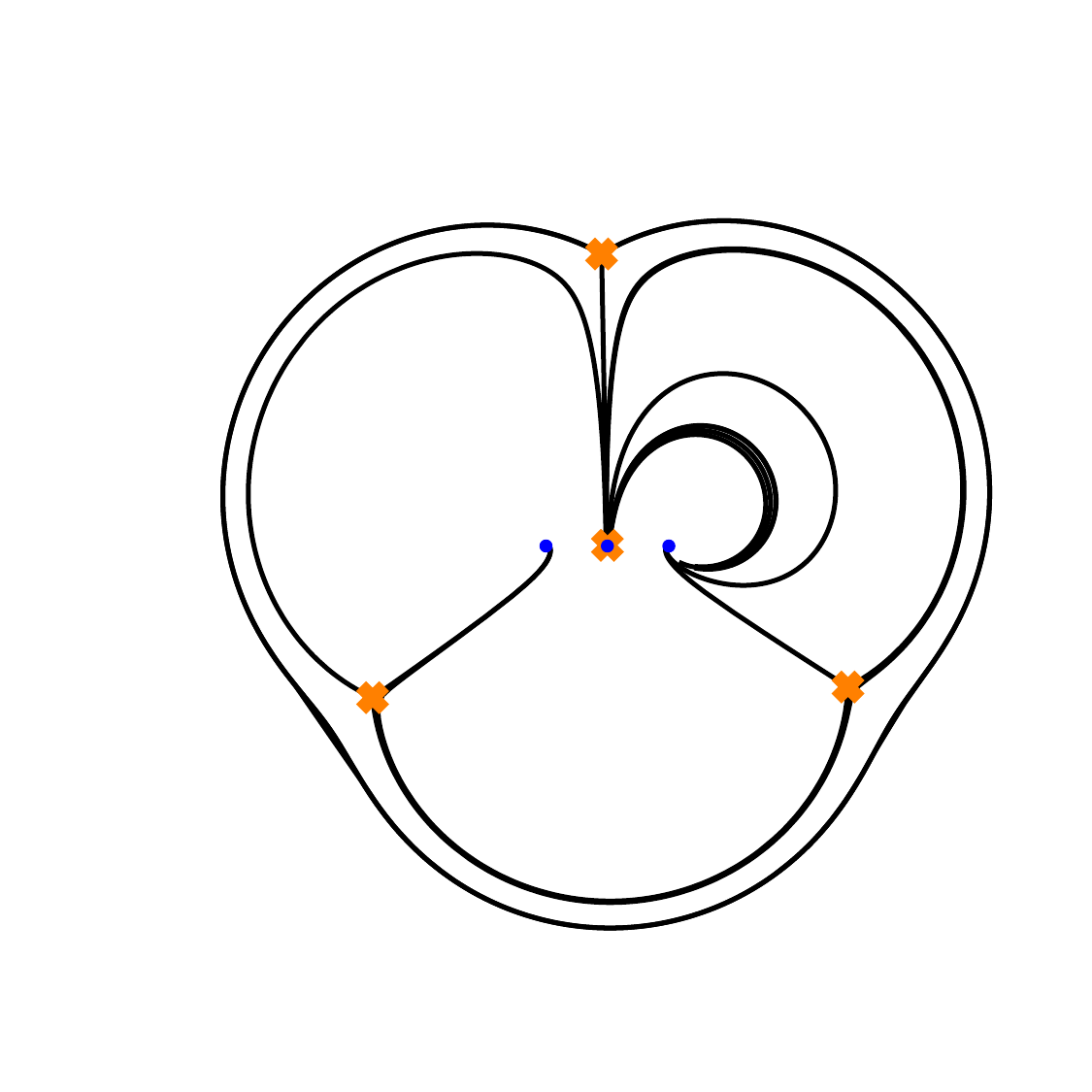}
    \caption{The first 4 quasinormal modes corresponding to $\Delta=2, \mathfrak{q}=0$ in the table 1 of \cite{Nunez:2003eq}.}
\label{fig:hydro}
\end{figure}

We also consider the AdS$_5$ black brane which is relevant for the hydrodynamic regime \cite{Nunez:2003eq}, whose wave equation is the one corresponding to the $SU(2)$ $N_f=4$ theory.  
The spectral networks corresponding to the first 4 quasinormal modes at $\Delta=2$ in the table 1 of \cite{Nunez:2003eq} are shown in \autoref{fig:hydro}.   Varying the phase $\vartheta$, we see no ring domain. This indicates that such a spectral network belongs to a strong coupling region. Moreover we can see from \autoref{fig:hydro} that the spectral networks also asymptotic to certain topology  which is identified with an hypermultiplet. Similar results can be found using other $\Delta$.
Let us note that this system was also studied in a beautiful subsequent work \cite{Jia:2026ryl}\footnote{This work appeared approximately one year after the first version of the present manuscript.} in the large-$\omega$, large-${\mathfrak q}$ regime with fixed $\zeta={\mathfrak q}/\omega$. 
The results presented there suggest the existence of two phase transitions: one occurring as ${\mathfrak q}$ is turned on from zero, and another when $\zeta$ exceeds unity. Both transitions bear some resemblance to wall-crossing phenomena, and it would be interesting to understand this connection more precisely.

More generally, the appearance of string-coupling phases, open an intriguing avenue for exploring the interplay between GMN TBA and black hole perturbation theory, a direction we leave for future investigations. 
Moreover, we expect that by varying the frequency and black hole parameters, it may be possible to transition from the strong coupling regime to the weak coupling regime. It would be  interesting to understand the implications of crossing the wall of marginal stability from the perspective of black holes.  However, we leave this investigation for future work.

\section*{Declarations}
\section*{Conflict of interest statement}
The authors have no competing interests to declare that are relevant to the content of this article.
\section*{Data availability statement}
The study didn’t involve any data.

%%%%%%%%%%%
\appendix
\section{NS function without defect: the 5-point block}\label{appendixA}

In this appendix, we define the NS partition function associated with the $SU(2) \times SU(2)$  linear quiver gauge theory discussed in \autoref{sec:gauge}, without any defects. The first step consists of defining the Nekrasov function.

For every Young diagram $Y$, we denote with $(Y_1\ge Y_2\ge\dots)$ the heights of its columns and with $(Y'_1\ge Y'_2,\dots)$ its rows' lengths. For every box $s=(i,j)$, we denote the arm length and the leg length of $s$ with respect to the diagram $Y$ as
\begin{equation}
A_Y(i, j) = Y_j -  i, \quad L_Y(i, j) =Y'_i - j.
\end{equation}
We introduce the main contributions crucial for the definition of the instanton partition function 
with fundamental matter. We specify the notations that we use for the black hole problems, where the five singular points are $0,1,\infty,t,q$, with $|t|,1/|q|\ll 1$. 

We denote with $\vec{Y}=\left( Y_1, Y_2 \right)$ a pair of Young diagrams and with $\vec{b}_i=(b_{i,1},b_{i,2})$ the v.e.v. of the adjoint scalar in the vector multiplet for each gauge factor $SU(2)_i$, $i=1,2$, and with $\epsilon_1,\epsilon_2$ the parameters characterizing the $\Omega$-background.  

Let us introduce the main blocks involved in the definition of the instanton part of the NS function, that is the hypermultiplet, vector, and bifundamental contributions:
\begin{equation}
\begin{aligned}
&E \left( b, Y_1, Y_2, (i,j) \right) = b - \epsilon_1 L_{Y_2} (i,j) + \epsilon_2 \left( A_{Y_1} (i,j) + 1 \right) \,,\\
&z_{\text{hyp}} \left( \vec{b}, \vec{Y}, m \right) = \prod_{k= 1,2} \prod_{(i,j) \in Y_k} \left( b_k + m + \epsilon_1 \left( i - \frac{1}{2} \right) + \epsilon_2 \left( j - \frac{1}{2} \right) \right) \,, \\
&z_{\text{vec}} \left( \vec{b}, \vec{Y} \right) = \prod_{k,l = 1,2} \prod_{(i,j) \in Y_k}  E^{-1} \left( b_k - b_l, Y_k, Y_l, (i, j) \right) \prod_{(i',j') \in Y_l} \left( \epsilon_1+\epsilon_2 - E \left( b_l - b_k, Y_l, Y_k, (i', j') \right) \right)^{-1}\,, \\
&z_{\text{bif}}\left( \vec{b_1},\vec{Y},\vec{b_2},\vec{W};m_{\mathrm{bif}} \right)=\\
&\prod_{k,l = 1,2} \prod_{(i,j) \in Y_k}  \left(E \left( (b_1)_k - (b_2)_l, Y_k, W_l, (i, j) \right)-\left(\frac{\epsilon_1+\epsilon_2}{2}+m_{\mathrm{bif}}\right)  \right) \\
&\prod_{(i',j') \in W_l} \left( - E \left( (b_2)_l - (b_1)_k, W_l, Y_k, (i', j') \right)+\left(\frac{\epsilon_1+\epsilon_2}{2}-m_{\mathrm{bif}}\right) \right)\,.
\end{aligned}
\end{equation}
The Nekrasov function is then defined as
\begin{equation}
\begin{aligned}\label{eq:nekfunction}
&\mathfrak{F}\left(\begin{matrix}\alpha_q\\ \alpha_{\infty}\end{matrix}\,\beta_2\,\begin{matrix}\alpha_1\\ {}\end{matrix}\,\beta_1\,\begin{matrix}\alpha_t\\ \alpha_{0}\end{matrix};\frac{1}{q},t\right)\\
=&\,q^{\Delta_{\infty}-\Delta_q-\Delta_{\beta_2}} t^{\Delta_{\beta_1} - \Delta_t - \Delta_{0}}\left(1-\frac{1}{q}\right)^{-2\left(\frac{Q}{2}+\alpha_q\right)\left(\frac{Q}{2}-\alpha_1\right)}(1-t)^{-2\left(\frac{Q}{2}+ \alpha_1\right)\left(\frac{Q}{2}-\alpha_t\right)}\left(1-\frac{t}{q}\right)^{-2\left(\frac{Q}{2}+\alpha_q\right)\left(\frac{Q}{2}+\alpha_t\right)} \times\\
&\sum_{\vec{Y},\vec{W}} \frac{t^{| \vec{Y} |}}{q^{|\vec{W}|}} z_{\text{vec}} \left( \vec{\beta_1}, \vec{Y} \right)z_{\text{vec}} \left( \vec{\beta_2}, \vec{W} \right) z_{\text{bif}} \left( \vec{\beta_1},\vec{Y},\vec{\beta_2},\vec{W};-\alpha_1 \right)\times\\
&\qquad\prod_{\sigma = \pm}  z_{\text{hyp}} \left( \vec{\beta_1}, \vec{Y}, \alpha_t + \sigma a_0 \right) z_{\text{hyp}} \left( \vec{\beta_2}, \vec{W}, \alpha_{q} + \sigma \alpha_{\infty} \right).
\end{aligned}
\end{equation}
Taking $\epsilon_1=1$ and $\vec{b_i}=(b_i,-b_i)$, $i=1,2$, the instanton part of the NS function is defined by  taking $\epsilon_2\to 0$, that is
\begin{equation}\label{FNS}
\begin{aligned}
&\FIVc{a_{\infty}}{a_{q}}{b_2}{a_1}{b_1}{a_t}{a_0}{\frac{1}{q}}{t}=\lim_{\epsilon_2\to 0}\epsilon_2\log\Biggl[(1-t)^{-\frac{2 \left(a_1+\frac{1}{2}\right) \left(a_t+\frac{1}{2}\right)}{\epsilon_2}} \left(1-\frac{1}{q}\right)^{-\frac{2 \left(\frac{1}{2}-a_1\right) \left(a_{q}+\frac{1}{2}\right)}{\epsilon_2}} \left(1-\frac{t}{q}\right)^{-\frac{2 \left(a_t+\frac{1}{2}\right) \left(a_{q}+\frac{1}{2}\right)}{\epsilon_2}}\times\\
&\sum_{\vec{Y},\vec{W}} \frac{t^{| \vec{Y} |}}{q^{|\vec{W}|}} z_{\text{vec}} \left( \vec{b_1}, \vec{Y} \right)z_{\text{vec}} \left( \vec{b_2}, \vec{W} \right) \prod_{\sigma = \pm}  z_{\text{hyp}} \left( \vec{b_1}, \vec{Y}, a_t + \sigma a_0 \right) z_{\text{hyp}} \left( \vec{b_2}, \vec{W}, a_{q} + \sigma a_{\infty} \right) z_{\text{bif}} \left( \vec{b_1},\vec{Y},\vec{b_2},\vec{W};-a_1 \right)\Biggr].
\end{aligned}
\end{equation}
The first instanton contribution reads
\begin{equation}
\begin{aligned}
F(1/q,t)=&\left[t\,\frac{\left(4a_0^2-4a_t^2-4b_1^2+1\right) \left(4a_1^2+4b_1^2-4b_2^2-1\right)}{32b_1^2-8}+\frac{1}{q}\,\frac{\left(4a_1^2-4b_1^2+4b_2^2-1\right) \left(4a_{\infty}^2-4a_{q}^2-4b_2^2+1\right)}{32b_2^2-8}\right]\\
&+\mathcal{O}(t^2, q^{-2},t/q).
\end{aligned}
\end{equation}
The full prepotential inculdes a classical and a 1-loop contribution in addition to the instanton part 
\begin{equation}
\begin{aligned}\label{eq:fullns}
F^{\text{NS}}(1/q,t)=&\,F(1/q,t)-b_1^2\,\log(t)-b_2^2\,\log(1/q)+\sum_{\theta=\pm}\psi^{(-2)}(1+2\theta\,b_1)+\sum_{\theta=\pm}\psi^{(-2)}(1+2\theta\,b_2)\\
&-\sum_{\theta_1,\theta_2=\pm}\psi^{(-2)}\left(\frac{1}{2}+a_1+\theta_1\,b_1+\theta_2\,b_2\right)\\
&-\sum_{\theta,\sigma=\pm}\psi^{(-2)}\left(\frac{1}{2}+\theta\,b_1-a_t+\sigma\,a_0\right)-\sum_{\theta,\sigma=\pm}\psi^{(-2)}\left(\frac{1}{2}+\theta\,b_2-a_q+\sigma\,a_{\infty}\right).
\end{aligned}
\end{equation}
This can be obtained from \cite[App.~B]{Alday:2009aq} and by using the NS limit of the $\Gamma_2$ functions which can be found for instance in \cite[(A.30)-(A.31)]{Jeong:2018qpc}~.
In \eqref{FNS} and \eqref{eq:fullns}, we have omitted the dependence on the $a_i$ parameters for the sake of simplicity in notation. However, in the main text, we occasionally use different signs for these parameters. This does not affect \eqref{FNS} but it does affect \eqref{eq:fullns}.
Hence, in such cases, we explicitly indicate the dependence. For example, we note
\be \label{eq:fnsconv}F^{\rm NS}\left({1\over q},t, \theta a_k\right)=\text{\eqref{eq:fullns}}\Big|_{a_k\to \theta a_k}, \quad \theta=\pm 1~.\ee

\section{NS function with defect: the degenerate 6-point block}\label{appendix5+1}

The Nekrasov function in presence of a canonical surface defect is defined by
\begin{equation}
\begin{aligned}\label{blockaround0}
    &\mathfrak{F}\left(\begin{matrix}\alpha_q\\ \alpha_{\infty}\end{matrix}\,\beta_2\,\begin{matrix}\alpha_1\\ {}\end{matrix}\,\beta_1\,\begin{matrix}\alpha_t\\ {}\end{matrix}\,\alpha_{0\theta}\,\begin{matrix}\alpha_{2,1}\\ \alpha_{0}\end{matrix};\frac{1}{q},t,\frac{z}{t}\right)=q^{\Delta_{\infty}-\Delta_q-\Delta_{\beta_2}} t^{\Delta_{\beta_1} - \Delta_t - \Delta_{0\theta}} z^{\frac{bQ}{2}+\theta b \alpha_0}\left(1-\frac{1}{q}\right)^{-2\left(\frac{Q}{2}+\alpha_q\right)\left(\frac{Q}{2}-\alpha_1\right)}\times\\
    & (1-t)^{-2\left(\frac{Q}{2}+ \alpha_1\right)\left(\frac{Q}{2}-\alpha_t\right)}\left(1-\frac{z}{t}\right)^{-2(\frac{Q}{2}+ \alpha_t)\left(\frac{Q}{2}+\alpha_{2,1}\right)}
    \left(1-\frac{t}{q}\right)^{-2\left(\frac{Q}{2}+\alpha_q\right)\left(\frac{Q}{2}-\alpha_t\right)}(1-z)^{-2\left(\frac{Q}{2}+ \alpha_1\right)\left(\frac{Q}{2}+\alpha_{2,1}\right)}\times\\
    &\left(1-\frac{z}{q}\right)^{-2\left(\frac{Q}{2}+\alpha_q\right)\left(\frac{Q}{2}+\alpha_{2,1}\right)}\sum_{\vec{Y}_1,\vec{Y}_2,\vec{Y}_3} \left(\frac{1}{q}\right)^{| \vec{Y}_1 |}t^{| \vec{Y}_2 |} \left(\frac{z}{t}\right)^{|\vec{Y}_3|} z_{\text{vec}} \left( \vec{\beta}_2, \vec{Y}_1 \right)z_{\text{vec}} \left( \vec{\beta}_1, \vec{Y}_2 \right)z_{\text{vec}} \left( \vec{\alpha_{0\theta}}, \vec{Y}_3 \right)\times\\
    &  z_{\text{bifund}} \left(\vec{\beta}_1,\vec{Y}_2,\vec{\beta}_2,\vec{Y}_1;-\alpha_1 \right)z_{\text{bifund}} \left( \vec{\beta}_1,\vec{Y}_2,\vec{\alpha_{0\theta}},\vec{Y}_3;\alpha_t \right)\prod_{\sigma = \pm}  z_{\text{hyp}} \left( \vec{\beta}_2, \vec{Y}_1, \alpha_q + \sigma \alpha_\infty \right) z_{\text{hyp}} \left( \vec{\alpha_{0\theta}}, \vec{Y}_3, \alpha_{2,1} + \sigma \alpha_0 \right).
\end{aligned}
\end{equation}
The NS function in presence of a canonical surface defect is then defined as
\begin{equation}\label{semist}
\begin{aligned}
&\mathcal{F}\left(\begin{matrix}a_q\\ a_{\infty}\end{matrix}\,b_2\,\begin{matrix}a_1\\ {}\end{matrix}\,b_1\,\begin{matrix} a_t\\ {}\end{matrix}\,a_{0\theta}\,\begin{matrix}a_{2,1}\\ a_{0}\end{matrix};\frac{1}{q},t,\frac{z}{t}\right)=\frac{\mathfrak{F}\left(\begin{matrix}\alpha_q\\ \alpha_{\infty}\end{matrix}\,\beta_2\,\begin{matrix}\alpha_1\\ {}\end{matrix}\,\beta_1\,\begin{matrix}\alpha_t\\ {}\end{matrix}\,\alpha_{0\theta}\,\begin{matrix}\alpha_{2,1}\\ \alpha_{0}\end{matrix};\frac{1}{q},t,\frac{z}{t}\right)}{\mathfrak{F}\left(\begin{matrix}\alpha_q\\ \alpha_{\infty}\end{matrix}\,\beta_2\,\begin{matrix}\alpha_1\\ {}\end{matrix}\,\beta_1\,\begin{matrix}\alpha_t\\ \alpha_{0}\end{matrix};\frac{1}{q},t\right)}\\
    &=t^{-a_{0\theta} }\E^{-\frac1 2\theta\partial_{a_0}F\left(\frac1 q , t\right)}(1-z)^{\frac{1}{2}+a_1}\left(1-\frac z q\right)^{\frac1 2 +a_q}\left(1-\frac z t\right)^{\frac 1 2 -a_t}z^{\frac1 2+a_0 \theta}\left(1+\mathcal O\left(\frac{z}{t}\right)\right)\\
    &= t^{-a_{0\theta} }\E^{-\frac1 2\theta\partial_{a_0}F\left(\frac1 q , t\right)} \psi_{0,\theta}(z),
\end{aligned}
\end{equation}
where $F\left(\frac1 q , t\right)$ is the shortcut for the NS free energy $\FIVc{a_{\infty}}{a_{q}}{b_2}{a_1}{b_1}{a_t}{a_0}{\frac{1}{q}}{t}$ in \eqref{FNS}.

Note that $t^{-a_{0\theta} }\E^{-\frac1 2\theta\partial_{a_0}F\left(\frac1 q , t\right)}$ in \eqref{semist} can be thought of as
\begin{equation}
    t^{-a_{0\theta} }\E^{-\frac1 2\theta\partial_{a_0}F\left(\frac1 q , t\right)}=\E^{-\frac1 2\theta\partial_{a_0}\lim\limits_{b\rightarrow 0}b^2\log\mathfrak{F}\left(\begin{matrix}\alpha_q\\ \alpha_{\infty}\end{matrix}\,\beta_2\,\begin{matrix}\alpha_1\\ {}\end{matrix}\,\beta_1\,\begin{matrix}\alpha_t\\ \alpha_{0}\end{matrix};\frac{1}{q},t\right)}.
\end{equation}

\section{Frequencies and residue expansions}\label{app:frequencies}

Some results for generic charge $\tilde{Q}$.
\begin{equation}\footnotesize
\begin{aligned}
\omega_2^+(\ell,n,\tilde{Q})&=\,\frac{(n+1)}{(\ell-1) \ell (\ell+1)^3 (\ell+2)^3 (\ell+3)} \biggl\{\ell^9 \left[2 n (n+2) \left(5 \tilde{Q}^2-17\right)+9 \tilde{Q}^2-39\right]\\
&+\ell^8 \biggl[12 (n+1) \sqrt{3 (\ell-1) (\ell+3) \tilde{Q}^2+16}+5 n^3 \left(\tilde{Q}^2-7\right)+3 n^2 \left(35 \tilde{Q}^2-137\right)+n \left(214 \tilde{Q}^2-802\right)\\
&+105 \tilde{Q}^2-471\biggr]+2 \ell^7 \biggl\{\left(18 n^2+84 n+74\right) \sqrt{3 (\ell-1) (\ell+3) \tilde{Q}^2+16}+20 n^3 \left(\tilde{Q}^2-7\right)\\
&+3 n^2 \left(73 \tilde{Q}^2-341\right)+2 n \left(227 \tilde{Q}^2-983\right)+5 \left(51 \tilde{Q}^2-239\right)\biggr\}+\ell^6 \biggl\{n^3 \left(113 \tilde{Q}^2-905\right)\\
&+\left(30 n^3+342 n^2+984 n+784\right) \sqrt{3 (\ell-1) (\ell+3) \tilde{Q}^2+16}+15 n^2 \left(59 \tilde{Q}^2-363\right)\\
&+2 n \left(1025 \tilde{Q}^2-5291\right)+1446 \tilde{Q}^2-6770\biggr\}+\ell^5 \biggl\{2 n^3 \left(59 \tilde{Q}^2-755\right)+4 n^2 \left(273 \tilde{Q}^2-2179\right)\\
&+4 \left(45 n^3+335 n^2+784 n+576\right) \sqrt{3 (\ell-1) (\ell+3) \tilde{Q}^2+16}+8 n \left(427 \tilde{Q}^2-2228\right)+3057 \tilde{Q}^2-12135\biggr\}\\
&+\ell^4 \biggl\{4 \left(105 n^3+685 n^2+1437 n+987\right) \sqrt{3 (\ell-1) (\ell+3) \tilde{Q}^2+16}+n^3 \left(179 \tilde{Q}^2-1457\right)\\
&+n^2 \left(2127 \tilde{Q}^2-9425\right)+n \left(6274 \tilde{Q}^2-20518\right)+15 \left(359 \tilde{Q}^2-969\right)\biggr\}+2 \ell^3 \biggl\{26 n^3 \left(13 \tilde{Q}^2-19\right)\\
&+2 \left(120 n^3+705 n^2+1366 n+899\right) \sqrt{3 (\ell-1) (\ell+3) \tilde{Q}^2+16}+n^2 \left(2227 \tilde{Q}^2-3355\right)\\
&+n \left(4686 \tilde{Q}^2-7246\right)+3300 \tilde{Q}^2-5174\biggr\}+\ell^2 (n+2) \biggl[n^2 \left(723 \tilde{Q}^2+97\right)+3 n \left(723 \tilde{Q}^2+97\right)\\
&+\left(118 n^2+354 n+468\right) \sqrt{3 (\ell-1) (\ell+3) \tilde{Q}^2+16}+1956 \tilde{Q}^2-504\biggr]-2\ell (n+2) \biggl[-4 \left(9 \tilde{Q}^2+266\right)\\
&+\left(122 n^2+366 n+220\right) \sqrt{3 (\ell-1) (\ell+3) \tilde{Q}^2+16}+\left(n^2+3 n\right) \left(9 \tilde{Q}^2-565\right)\biggr]\\
&-12 (n+1) (n+2)^2 \left(10 \sqrt{3 (\ell-1) (\ell+3) \tilde{Q}^2+16}+9 \tilde{Q}^2-41\right)\biggr\},
\end{aligned}
\end{equation}

{\footnotesize\begin{align}
\nonumber\omega_2^-(\ell,n,\tilde{Q})&=\,-\frac{n+1}{(\ell-1) \ell (\ell+1)^3 (\ell+2)^3 (\ell+3)}\biggl\{\ell^9 \biggl[n^2 \left(34-10 \tilde{Q}^2\right)+n \left(68-20 \tilde{Q}^2\right)-9 \tilde{Q}^2+39\biggr]\\
\nonumber&+\ell^8 \biggl[12 (n+1) \sqrt{3 (\ell-1) (\ell+3) \tilde{Q}^2+16}-5 n^3 \left(\tilde{Q}^2-7\right)+n^2 \left(411-105 \tilde{Q}^2\right)\\
\nonumber&+n \left(802-214 \tilde{Q}^2\right)-105 \tilde{Q}^2+471\biggr]+\ell^7 \biggl[4 \left(9 n^2+42 n+37\right) \sqrt{3 (\ell-1) (\ell+3) \tilde{Q}^2+16}\\
\nonumber&-40 n^3 \left(\tilde{Q}^2-7\right)-6 n^2 \left(73 \tilde{Q}^2-341\right)+n \left(3932-908 \tilde{Q}^2\right)-510 \tilde{Q}^2+2390\biggr]\\
\nonumber&+\ell^6 \biggl[\left(30 n^3+342 n^2+984 n+784\right) \sqrt{3 (\ell-1) (\ell+3) \tilde{Q}^2+16}+n^3 \left(905-113 \tilde{Q}^2\right) \\
\nonumber&+n^2 \left(5445-885 \tilde{Q}^2\right)+n \left(10582-2050 \tilde{Q}^2\right)-1446 \tilde{Q}^2+6770\biggr]\\
\nonumber&+\ell^5 \biggl[4 \left(45 n^3+335 n^2+784 n+576\right) \sqrt{3 (\ell-1) (\ell+3) \tilde{Q}^2+16}-2 n^3 \left(59 \tilde{Q}^2-755\right)\\
\nonumber&+n^2 \left(8716-1092 \tilde{Q}^2\right)-8 n \left(427 \tilde{Q}^2-2228\right)-3057 \tilde{Q}^2+12135\biggr]\\
\nonumber&+\ell^4 \biggl[4 \left(105 n^3+685 n^2+1437 n+987\right) \sqrt{3 (\ell-1) (\ell+3) \tilde{Q}^2+16}+n^3 \left(1457-179 \tilde{Q}^2\right)\\
\nonumber&+n^2 \left(9425-2127 \tilde{Q}^2\right)+n \left(20518-6274 \tilde{Q}^2\right)-5385 \tilde{Q}^2+14535\biggr]\displaybreak\\
\nonumber&+\ell^3 \biggl[4 \left(120 n^3+705 n^2+1366 n+899\right) \sqrt{3 (\ell-1) (\ell+3) \tilde{Q}^2+16}+n^3 \left(988-676 \tilde{Q}^2\right)\\
\nonumber&+n^2 \left(6710-4454 \tilde{Q}^2\right)+n \left(14492-9372 \tilde{Q}^2\right)-6600 \tilde{Q}^2+10348\biggr]\\
\nonumber&+\ell^2 \biggl[2 \left(59 n^3+295 n^2+588 n+468\right) \sqrt{3 (\ell-1) (\ell+3) \tilde{Q}^2+16}\\
\nonumber&-(n+2) \left(n^2 \left(723 \tilde{Q}^2+97\right)+3 n \left(723 \tilde{Q}^2+97\right)+1956 \tilde{Q}^2-504\right)\biggr]\\
\nonumber&+\ell \biggl[2 (n+2) \left(n^2 \left(9 \tilde{Q}^2-565\right)+3 n \left(9 \tilde{Q}^2-565\right)-4 \left(9 \tilde{Q}^2+266\right)\right)\\
\nonumber&-4 \left(61 n^3+305 n^2+476 n+220\right) \sqrt{3 (\ell-1) (\ell+3) \tilde{Q}^2+16}\biggr]\\
&-120 (n+1) (n+2)^2 \sqrt{3 (\ell-1) (\ell+3) \tilde{Q}^2+16}+12 (n+1) (n+2)^2 \left(9 \tilde{Q}^2-41\right)\biggr\},
\end{align}}

The results for $\tilde{Q}=0$ are simple. For example we have
\begin{equation}\footnotesize
\begin{aligned}
\omega_3^+(\ell,n,0)&=\,-\frac{n+1}{\ell^2 (\ell+1)^5 (\ell+2)^5 \left(\ell^4+4 \ell^3-7 \ell^2-22 \ell+24\right)}\biggl\{15 \ell^{16} \left(25 n^3+75 n^2+86 n+36\right)\\
&+\ell^{15} \left(756 n^4+9024 n^3+23666 n^2+25924 n+10635\right)\\
&+3 \ell^{14} \left(154 n^5+4550 n^4+30365 n^3+70985 n^2+74362 n+29891\right)\\
&+\ell^{13} \left(6468 n^5+101412 n^4+496578 n^3+1049874 n^2+1052916 n+414787\right)\\
&+\ell^{12} \left(35196 n^5+385956 n^4+1522114 n^3+2925382 n^2+2804644 n+1082869\right)\\
&+\ell^{11} \left(86016 n^5+693552 n^4+2153328 n^3+3600596 n^2+3216904 n+1210625\right)\\
&+\ell^{10} \left(39760 n^5-87904 n^4-1511886 n^3-3906094 n^2-4101644 n-1592257\right)\\
&+\ell^9 \left(-289856 n^5-3115816 n^4-12385724 n^3-23675676 n^2-22375192 n-8462247\right)\\
&-3 \ell^8 \left(246260 n^5+2183260 n^4+7783923 n^3+14151321 n^2+13125202 n+4960511\right)\\
&-2 \ell^7 \left(363888 n^5+3001290 n^4+10332156 n^3+18697255 n^2+17577854 n+6802284\right)\\
&+\ell^6 \left(-132894 n^5-1063362 n^4-4100041 n^3-9041805 n^2-10390274 n-4784172\right)\\
&+6 \ell^5 \left(65346 n^5+520542 n^4+1624863 n^3+2471951 n^2+1820470 n+509992\right)\\
&+8 \ell^4 \left(53286 n^5+426288 n^4+1365032 n^3+2189278 n^2+1759685 n+567058\right)\\
&+88 \ell^3 \left(2442 n^5+19536 n^4+62751 n^3+101301 n^2+82270 n+26912\right)\\
&+16 \ell^2 \left(3623 n^5+28984 n^4+92786 n^3+148729 n^2+119432 n+38428\right)\\
&+32 \ell (n+2)^2 \left(241 n^3+964 n^2+1295 n+572\right)+384 (n+1)^2 (n+2)^3\biggr\},
\end{aligned}
\end{equation}

{\footnotesize\begin{align}
\nonumber\omega_3^-(\ell,n,0)&=\,-\frac{3 (n+1)}{\ell^2 (\ell+1)^5 (\ell+2)^5 \left(\ell^2+2 \ell-8\right) \left(\ell^2+2 \ell-3\right)^2}\biggl\{5 \ell^{18} \left(25 n^3+75 n^2+86 n+36\right)\\
\nonumber&+\ell^{17} \bigl(252 n^4+3258 n^3+9092 n^2+10408 n+4425\bigr)\\
\nonumber&+\ell^{16} \bigl(154 n^5+5054 n^4+37966 n^3+99224 n^2+113304 n+48897\bigr)\\
\nonumber&+\ell^{15} \bigl(2464 n^5+45188 n^4+260752 n^3+643190 n^2+731724 n+320220\bigr)\\
\nonumber&+\ell^{14} \bigl(17262 n^5+236610 n^4+1168195 n^3+2740075 n^2+3099718 n+1373280\bigr)\\
\nonumber&+\ell^{13} \bigl(69188 n^5+793636 n^4+3535114 n^3+7927998 n^2+8890460 n+3976230\bigr)\\
\nonumber&+2 \ell^{12} \left(84154 n^5+850862 n^4+3500778 n^3+7501860 n^2+8287284 n+3724179\right)\\
\nonumber&+4 \ell^{11} \left(53928 n^5+472806 n^4+1739794 n^3+3425097 n^2+3627146 n+1622127\right)\\
\nonumber&+\ell^{10} \left(-66724 n^5-1278476 n^4-7123933 n^3-17955107 n^2-21145974 n-9529848\right)\\
\nonumber&+\ell^9 \left(-985096 n^5-10656428 n^4-46400242 n^3-101836448 n^2-111953856 n-49214295\right)\\
\nonumber&+\ell^8 \left(-2469102 n^5-25776570 n^4-108825394 n^3-230903248 n^2-245888248 n-105114255\right)\\
\nonumber&-2 \ell^7 \left(1879384 n^5+19630910 n^4+81628616 n^3+169096401 n^2+175208146 n+72767760\right)\\
\nonumber&+\ell^6 \left(-3592058 n^5-37293334 n^4-153137219 n^3-313529871 n^2-321194878 n-131607588\right)\displaybreak\\
\nonumber&-2 \ell^5 \left(576222 n^5+6337866 n^4+28465433 n^3+64296769 n^2+71910306 n+31549320\right)\\
\nonumber&+24 \ell^4 \left(72936 n^5+583488 n^4+1719502 n^3+2252078 n^2+1174821 n+95490\right)\\
\nonumber&+72 \ell^3 \left(22354 n^5+178832 n^4+568471 n^3+897557 n^2+700422 n+212400\right)\\
\nonumber&-432 \ell^2 \left(1217 n^5+9736 n^4+28882 n^3+38531 n^2+21768 n+3636\right)\\
&-23328 \ell (n+2)^2 \left(41 n^3+164 n^2+199 n+76\right)-279936 (n+1)^2 (n+2)^3\biggr\}.
\end{align}}

\begin{equation}\footnotesize
\begin{aligned}
c_2^+(\ell,n)=&\,-\frac{n+1}{\ell^3 (\ell+1)^3 (\ell+2)^3}\times\\
&\bigl[10 \ell^8 \left(10 n^2+20 n+11\right)+\ell^7 \left(255 n^3+1565 n^2+2473 n+1243\right)\\
&+\ell^6 \left(210 n^4+2625 n^3+9495 n^2+12711 n+5909\right)\\
&+\ell^5 \left(1260 n^4+10200 n^3+29120 n^2+34849 n+15217\right)\\
&+\ell^4 \left(2820 n^4+19230 n^3+48822 n^2+54519 n+22769\right)\\
&+\ell^3 \left(2880 n^4+18495 n^3+44813 n^2+48454 n+19744\right)\\
&+\ell^2 \left(1362 n^4+8853 n^3+21515 n^2+23206 n+9384\right)\\
&+18 \ell \left(18 n^4+117 n^3+283 n^2+302 n+120\right)\\
&+12 (n+2)^2 \left(2 n^2+5 n+3\right)\bigr].
\end{aligned}
\end{equation}

\begin{equation}\footnotesize
\begin{aligned}
c_2^-(\ell,n)=&\,-\frac{n+1}{(\ell-1) \ell^3 (\ell+1)^3 (\ell+2)^3 (\ell+3)}\times\\
&\bigl[10 \ell^{10} \left(10 n^2+20 n+11\right)+\ell^9 \left(255 n^3+1765 n^2+3017 n+1607\right)\\
&+\ell^8 \left(210 n^4+3135 n^3+13189 n^2+20081 n+10337\right)\\
&+\ell^7 \left(1680 n^4+16701 n^3+56375 n^2+79108 n+39514\right)\\
&+\ell^6 \left(6150 n^4+51627 n^3+157937 n^2+211444 n+103060\right)\\
&+\ell^5 \left(13380 n^4+105779 n^3+315065 n^2+411681 n+195551\right)\\
&+\ell^4 \left(18822 n^4+151993 n^3+448163 n^2+568637 n+260789\right)\\
&+\ell^3 \left(17448 n^4+134151 n^3+375413 n^2+455782 n+202128\right)\\
&+\ell^2 \left(2250 n^4+13553 n^3+39135 n^2+57494 n+31736\right)\\
&-2 \ell \left(6318 n^4+40315 n^3+93509 n^2+92906 n+33208\right)\\
&-12 (n+2)^2 \left(486 n^2+1151 n+665\right)\bigr].
\end{aligned}
\end{equation}
\begin{equation}\footnotesize
\begin{aligned}
c_3^+&(\ell,n)=-\frac{n+1}{(\ell-2) (\ell-1) \ell^5 (\ell+1)^5 (\ell+2)^5 (\ell+3) (\ell+4)}\times\\
&\bigl[15 \ell^{17} \left(91 n^3+273 n^2+303 n+121\right)+\ell^{16} \left(5040 n^4+43365 n^3+104335 n^2+106385 n+40583\right)\\
&+\ell^{15} \left(7056 n^5+115920 n^4+571320 n^3+1160680 n^2+1092959 n+398063\right)\\
&+\ell^{14} \left(3696 n^6+128016 n^5+1131144 n^4+4134936 n^3+7325376 n^2+6398001 n+2223169\right)\\
&+\ell^{13} \left(51744 n^6+955136 n^5+6005776 n^4+17895274 n^3+28166734 n^2+22848697 n+7555941\right)\\
&+\ell^{12} \left(281568 n^6+3649184 n^5+17991064 n^4+45140146 n^3+63047590 n^2+46982515 n+14607239\right)\\
&+\ell^{11} \left(688128 n^6+6587840 n^5+24621856 n^4+47336904 n^3+51182552 n^2+29855341 n+7361149\right)\\
&+\ell^{10} \left(318080 n^6-767424 n^5-19740056 n^4-79113504 n^3-140832704 n^2-121424797 n-41453349\right)\\
&+\ell^9 \left(-2318848 n^6-29467424 n^5-147334992 n^4-381111663 n^3-546478093 n^2-414870374 n-130499224\right)\\
&+\ell^8 \left(-5910240 n^6-62116320 n^5-273254136 n^4-647699319 n^3-874604437 n^2-636568664 n-194296946\right)\\
&-8 \ell^7 \left(727776 n^6+7124306 n^5+29834722 n^4+68538983 n^3+90889405 n^2+65539020 n+19928374\right)\\
&-8 \ell^6 \left(132894 n^6+1262738 n^5+5557318 n^4+14238422 n^3+21585941 n^2+17741480 n+6050763\right)\\
&+8 \ell^5 \left(392076 n^6+3709140 n^5+14451756 n^4+29622189 n^3+33610805 n^2+19956976 n+4822764\right)\\
&+8 \ell^4 \left(426288 n^6+4049736 n^5+15980758 n^4+33537407 n^3+39480989 n^2+24717280 n+6426852\right)\\
&+64 \ell^3 \left(26862 n^6+255189 n^5+1007612 n^4+2117248 n^3+2497347 n^2+1567656 n+408996\right)\\
&+32 \ell^2 \left(14492 n^6+137674 n^5+542450 n^4+1134745 n^3+1329113 n^2+826216 n+212820\right)\\
&+128 \ell (n+2)^2 \left(482 n^4+2651 n^3+5428 n^2+4894 n+1635\right)+1536 (n+1)^2 (n+2)^3 (2 n+3)\bigr].
\end{aligned}
\end{equation}

\begin{equation}\footnotesize
\begin{aligned}
c_3^-&(\ell,n)=-\frac{n+1}{(\ell-2) (\ell-1)^2 \ell^5 (\ell+1)^5 (\ell+2)^5 (\ell+3)^2 (\ell+4)}\times\\
&\bigl[15 \ell^{19} \left(91 n^3+273 n^2+303 n+121\right)+\ell^{18} \left(5040 n^4+46095 n^3+115757 n^2+121939 n+47605\right)\\
&+3 \ell^{17} \left(2352 n^5+42000 n^4+225689 n^3+494859 n^2+495762 n+189152\right)\\
&+\ell^{16} \left(3696 n^6+142128 n^5+1412520 n^4+5799009 n^3+11424699 n^2+10912308 n+4066786\right)\\
&+2 \ell^{15} \left(29568 n^6+636312 n^5+4683960 n^4+16242717 n^3+29346615 n^2+26790223 n+9742869\right)\\
&+2 \ell^{14} \left(207144 n^6+3328344 n^5+20341608 n^4+62716827 n^3+105442501 n^2+92147897 n+32636935\right)\\
&+2 \ell^{13} \left(830256 n^6+11129152 n^5+59853152 n^4+168793783 n^3+266333757 n^2+222562652 n+76400934\right)\\
&+\ell^{12} \left(4039392 n^6+47469536 n^5+230529664 n^4+599814122 n^3+884578094 n^2+698202612 n+228547600\right)\\
&+\ell^{11} \left(5177088 n^6+52118560 n^5+217844000 n^4+485424457 n^3+601686747 n^2+388799865 n+100990311\right)\\
&+\ell^{10} \left(-1601376 n^6-37249760 n^5-261651952 n^4-881056901 n^3-1585967679 n^2-1466587949 n-544639571\right)\\
&+\ell^9 \left(-23642304 n^6-298886000 n^5-1569335248 n^4-4392315665 n^3-6886678995 n^2-5703560870 n-1936097508\right)\\
&+\ell^8 \left(-59258448 n^6-718739088 n^5-3644026392 n^4-9824708115 n^3-14784268801 n^2-11726896832 n-3812738890\right)\\
&-8 \ell^7 \left(11276304 n^6+136527410 n^5+682793314 n^4+1800593477 n^3+2636970447 n^2+2029697780 n+639473946\right)\\
&-8 \ell^6 \left(10776174 n^6+129579554 n^5+639764146 n^4+1663890578 n^3+2403039845 n^2+1822775504 n+565248175\right)\\
&-24 \ell^5 \left(1152444 n^6+14589612 n^5+77389468 n^4+217003577 n^3+335392665 n^2+268903168 n+86980532\right)\\
&+8 \ell^4 \left(5251392 n^6+49431600 n^5+183224790 n^4+338430639 n^3+319923477 n^2+137337136 n+16396436\right)\\
&+192 \ell^3 \left(201186 n^6+1885275 n^5+7338400 n^4+15170254 n^3+17523387 n^2+10683048 n+2672428\right)\\
&-32 \ell^2 \left(394308 n^6+3740566 n^5+14087966 n^4+26757055 n^3+26694575 n^2+12989272 n+2311628\right)\\
&-384 \ell (n+2)^2 \left(59778 n^4+324011 n^3+631772 n^2+528130 n+160591\right)-4608 (n+1)^2 (n+2)^3 (1458 n+2059)\bigr].
\end{aligned}
\end{equation}

\section{Singularity structure of several black hole spacetimes}\label{app:example}

We include a table presenting the singularity structure of several linear perturbations of different spacetime geometries, together with the corresponding instanton counting description. The geometries are asymptotically anti-de Sitter, but the singularity structure is the same for the corresponding asymptotically de Sitter spacetimes, the difference being in the imposed boundary conditions (and therefore in the relevant connection formulae to consider).

\setlength{\tabcolsep}{3pt}

\begin{longtable}{|>{\raggedright\arraybackslash}p{0.21\textwidth}|
                  >{\raggedright\arraybackslash}p{0.25\textwidth}|
                  >{\raggedright\arraybackslash}p{0.25\textwidth}|
                  >{\raggedright\arraybackslash}p{0.23\textwidth}|}
\hline
\textbf{SPACETIME GEOMETRY} & \textbf{TYPE OF PERTURBATION} & \textbf{SINGULARITY STRUCTURE} & \textbf{INSTANTON COUNTING} \\
\hline
\endfirsthead

\hline
\textbf{SPACETIME GEOMETRY} & \textbf{TYPE OF PERTURBATION} & \textbf{SINGULARITY STRUCTURE} & \textbf{INSTANTON COUNTING} \\
\hline
\endhead

Schwarzschild--AdS$_4$ &
Scalar field (generic mass) &
5 regular &
One instanton parameter is small, while the other remains finite in the $\mu\to 0$ limit. \\
\hline

Schwarzschild--AdS$_4$ &
Conformally coupled scalar field, electromagnetic perturbation, vector-type gravitational perturbation &
4 regular (compared to the previous case, radial infinity is a removable singularity) &
One small instanton parameter in the $R_h\to 0$ limit. \\
\hline

Schwarzschild--AdS$_4$ &
Scalar sector of gravitational perturbation &
5 regular (but not radial infinity) &
In the $R_h\to\infty$ limit, there is at least one instanton parameter with norm 1. \\
\hline

Schwarzschild--AdS$_5$ &
Scalar field (generic mass) &
4 regular &
One small instanton parameter in the $\mu\to 0$ limit. \\
\hline

Schwarzschild--AdS$_7$ &
Scalar field &
5 regular &
One instanton parameter is small, while the other remains finite in the $\mu\to 0$ limit. The difference with respect to the SAdS$_4$ case is that, in the large-$\ell$ regime, the quantization condition only involves the expansion in the small instanton parameter. The $K$-th correction in the $\omega$ expansion in $\mu$ can be found by considering the $4K$-th instanton expansion. \\
\hline

Kerr--AdS$_4$ &
Scalar field (generic mass) &
5 regular &
The rotation parameter $a$ is bounded by a function of the black-hole mass. In the small-$a$ regime, the instanton counting is as in the SAdS$_4$ case. \\
\hline

Kerr--AdS$_4$ &
Conformally coupled scalar field and ``spin-$s$ perturbations'' in terms of Newman--Penrose scalar quantities &
4 regular (compared to the previous case, radial infinity is a removable singularity) &
One small instanton parameter in the $R_h\to 0$ limit. \\
\hline

Kerr--AdS$_5$ &
Scalar field (generic mass) &
4 regular &
The rotation parameter $a$ is bounded by a function of the black-hole mass. In the small-$a$ regime, the instanton counting is as in the SAdS$_5$ case. \\
\hline

Kerr--AdS$_7$ with a single rotation parameter; small rotation parameter $a\sim 0$ &
Scalar field (generic mass) &
5 regular &
The same as for SAdS$_7$. \\
\hline

Kerr--AdS$_7$ with a single rotation parameter; large rotation parameter &
Scalar field (generic mass) &
5 regular &
The rotation parameter is not bounded; we take $a\sim\infty$. The instanton parameters behave as $1/a^2$ and $\mu/a^2$. The angular part is Heun, with the two-step connection procedure, but with integer parameters. \\
\hline

Kerr--AdS$_7$ with a single rotation parameter; black-brane limit &
Scalar field (generic mass) &
4 regular. One of the roots of the original $\Delta_r(r)$, in the variable $y=r^2$, goes to infinity, but infinity remains regular. &
One small instanton parameter, $t\sim \hat{\mu}+O(\hat{\mu}^2)$. The angular part becomes hypergeometric, and the only contribution of the separation constant is the term proportional to $a^2$. \\
\hline

Kerr--AdS$_9$ with a single rotation parameter; black-brane limit &
Scalar field (generic mass) &
5 regular &
The connection is the easy one with two nearby points, i.e. a hypergeometric-like connection, but the relevant instanton parameter is $t\sim 1/2$. The angular part is still hypergeometric, as above. \\
\hline

Kerr--AdS$_{2n+1}$ with a single rotation parameter; black-brane limit &
Scalar field (generic mass) &
$n+1$ regular: radial infinity, $r=0$, and $n-1$ of the $n$ roots of $\Delta_r(r)$ in the variable $y=r^2$ &
There are $n-2$ instanton parameters. \\
\hline

Reissner--Nordström--AdS$_4$ &
Scalar field (generic mass) &
5 regular (radial infinity and 4 roots of $r^2 f(r)$ in $r$; $r=0$ is not a singular point) &
The instanton counting is as in the SAdS$_4$ case. The confluence to the extremal limit would occur on the side of the small instanton parameter, while the other one would remain finite. \\
\hline

Reissner--Nordström--AdS$_4$ &
Electromagnetic perturbation and vector-type gravitational perturbation &
5 regular ($r=0$ and 4 roots of $r^2 f(r)$ in $r$; $r=\infty$ is not a singular point) &
The instanton counting is as in the SAdS$_7$ case: one instanton parameter is small, while the other is finite, but in the opposite way with respect to the scalar perturbation. \\
\hline

Reissner--Nordström--AdS$_5$ &
Scalar field (generic mass) &
4 regular (radial infinity and 3 roots of $r^4 f(r)$ in the variable $y=r^2$) &
In the small-$Q$ regime, the instanton counting is as in the SAdS$_5$ case. \\
\hline

Reissner--Nordström--AdS in $D=2n+1$ dimensions, $n>2$ &
Scalar field (generic mass) &
$2n$ regular (radial infinity and $2n-1$ roots of $r^{4n-2} f(r)$ in the variable $y=r^2$, where the relevant expression is $Q^2 + y^{n-2}\!\left(y^n + y^{n+1} - y\mu\right)$) &
There are $2n-3$ instanton parameters. \\
\hline

Reissner--Nordström--AdS in $D=2n+1$ dimensions, $n>2$; black-brane limit &
Scalar field (generic mass) &
It remains as above, since one still has $Q^2 + y^{n-2}\!\left(y^n + y^{n+1} - y\mu\right)$. &
There are $2n-3$ instanton parameters. \\
\hline

Reissner--Nordström--AdS in $D=2n$ dimensions, $n>2$ &
Scalar field (generic mass) &
$4n-3$ regular (radial infinity and $4n-4$ roots of $r^{4n-4} f(r)$ in the variable $r$) &
There are $4n-6$ instanton parameters. \\
\hline

C-metric &
Scalar field (generic mass) &
The radial equation has 4 regular singularities; the angular one has 5 regular singularities. &
In the small-$\alpha$ limit, where $\alpha$ is the acceleration parameter, the radial equation has one small instanton parameter. If $Q$ is also small, where $Q$ is the black-hole charge, the angular equation has one small and one large instanton parameter; the relevant connection is the hypergeometric-like one between $0$ and $t$, with $t$ small. \\
\hline

\end{longtable}

 \bibliographystyle{JHEP}
 \bibliography{biblio}

\providecommand{\href}[2]{#2}\begingroup\raggedright\begin{thebibliography}{100}

\bibitem{Marino:2015nla}
M.~Marino, \emph{{Spectral theory and mirror symmetry.}},
  \href{http://dx.doi.org/10.1090/pspum/098/01722}{\emph{Proc. Symp. Pure
  Math.} {\bf 98} (2018) 259}, [\href{http://arxiv.org/abs/1506.07757}{{\tt
  1506.07757}}].

\bibitem{ns}
N.~A. Nekrasov and S.~L. Shatashvili, \emph{{Quantization of Integrable Systems
  and Four Dimensional Gauge Theories}},
  \href{http://arxiv.org/abs/0908.4052}{{\tt 0908.4052}}.

\bibitem{mirmor}
A.~Mironov and A.~Morozov, \emph{{Nekrasov Functions and Exact Bohr-Sommerfeld
  Integrals}}, \href{http://dx.doi.org/10.1007/JHEP04(2010)040}{\emph{JHEP}
  {\bf 1004} (2010) 040}, [\href{http://arxiv.org/abs/0910.5679}{{\tt
  0910.5679}}].

\bibitem{mirmor2}
A.~Mironov and A.~Morozov, \emph{{Nekrasov Functions from Exact BS Periods: The
  Case of SU(N)}},
  \href{http://dx.doi.org/10.1088/1751-8113/43/19/195401}{\emph{J.Phys.} {\bf
  A43} (2010) 195401}, [\href{http://arxiv.org/abs/0911.2396}{{\tt
  0911.2396}}].

\bibitem{Zenkevich2011}
Y.~Zenkevich, \emph{Nekrasov prepotential with fundamental matter from the
  quantum spin chain},
  \href{http://dx.doi.org/https://doi.org/10.1016/j.physletb.2011.06.030}{\emph{Physics
  Letters B} {\bf 701} (2011) 630--639},
  [\href{http://arxiv.org/abs/1103.4843}{{\tt 1103.4843}}].

\bibitem{Nekrasov:2011bc}
N.~Nekrasov, A.~Rosly and S.~Shatashvili, \emph{{Darboux coordinates, Yang-Yang
  functional, and gauge theory}},
  \href{http://dx.doi.org/10.1016/j.nuclphysbps.2011.04.150}{\emph{Nucl. Phys.
  B Proc. Suppl.} {\bf 216} (2011) 69--93},
  [\href{http://arxiv.org/abs/1103.3919}{{\tt 1103.3919}}].

\bibitem{Kozlowski:2010tv}
K.~K. Kozlowski and J.~Teschner, \emph{{TBA for the Toda chain}},
  \href{http://arxiv.org/abs/1006.2906}{{\tt 1006.2906}}.

\bibitem{Alday:2010vg}
L.~F. Alday and Y.~Tachikawa, \emph{{Affine SL(2) conformal blocks from 4d
  gauge theories}},
  \href{http://dx.doi.org/10.1007/s11005-010-0422-4}{\emph{Lett. Math. Phys.}
  {\bf 94} (2010) 87--114}, [\href{http://arxiv.org/abs/1005.4469}{{\tt
  1005.4469}}].

\bibitem{Kanno:2011fw}
H.~Kanno and Y.~Tachikawa, \emph{{Instanton counting with a surface operator
  and the chain-saw quiver}},
  \href{http://dx.doi.org/10.1007/JHEP06(2011)119}{\emph{JHEP} {\bf 06} (2011)
  119}, [\href{http://arxiv.org/abs/1105.0357}{{\tt 1105.0357}}].

\bibitem{Jeong:2021rll}
S.~Jeong, N.~Lee and N.~Nekrasov, \emph{{Intersecting defects in gauge theory,
  quantum spin chains, and Knizhnik-Zamolodchikov equations}},
  \href{http://dx.doi.org/10.1007/JHEP10(2021)120}{\emph{JHEP} {\bf 10} (2021)
  120}, [\href{http://arxiv.org/abs/2103.17186}{{\tt 2103.17186}}].

\bibitem{Jeong:2023qdr}
S.~Jeong, N.~Lee and N.~Nekrasov, \emph{{Parallel surface defects, Hecke
  operators, and quantum Hitchin system}},
  \href{http://dx.doi.org/10.48550/arXiv.2304.04656}{\emph{arXiv e-prints: High
  Energy Physics - Theory} (4, 2023) },
  [\href{http://arxiv.org/abs/2304.04656}{{\tt 2304.04656}}].

\bibitem{Jeong:2018qpc}
S.~Jeong and N.~Nekrasov, \emph{{Opers, surface defects, and Yang-Yang
  functional}}, \href{http://dx.doi.org/10.4310/ATMP.2020.v24.n7.a4}{\emph{Adv.
  Theor. Math. Phys.} {\bf 24} (2020) 1789--1916},
  [\href{http://arxiv.org/abs/1806.08270}{{\tt 1806.08270}}].

\bibitem{Jeong:2017pai}
S.~Jeong, \emph{{Splitting of surface defect partition functions and integrable
  systems}},
  \href{http://dx.doi.org/10.1016/j.nuclphysb.2018.12.007}{\emph{Nucl. Phys. B}
  {\bf 938} (2019) 775--806}, [\href{http://arxiv.org/abs/1709.04926}{{\tt
  1709.04926}}].

\bibitem{Alday:2009fs}
L.~F. Alday, D.~Gaiotto, S.~Gukov, Y.~Tachikawa and H.~Verlinde, \emph{{Loop
  and surface operators in N=2 gauge theory and Liouville modular geometry}},
  \href{http://dx.doi.org/10.1007/JHEP01(2010)113}{\emph{JHEP} {\bf 1001}
  (2010) 113}, [\href{http://arxiv.org/abs/0909.0945}{{\tt 0909.0945}}].

\bibitem{Drukker:2009id}
N.~Drukker, J.~Gomis, T.~Okuda and J.~Teschner, \emph{{Gauge Theory Loop
  Operators and Liouville Theory}},
  \href{http://dx.doi.org/10.1007/JHEP02(2010)057}{\emph{JHEP} {\bf 1002}
  (2010) 057}, [\href{http://arxiv.org/abs/0909.1105}{{\tt 0909.1105}}].

\bibitem{Grassi:2019coc}
A.~Grassi, J.~Gu and M.~Mari\~no, \emph{{Non-perturbative approaches to the
  quantum Seiberg-Witten curve}},
  \href{http://dx.doi.org/10.1007/JHEP07(2020)106}{\emph{JHEP} {\bf 07} (2020)
  106}, [\href{http://arxiv.org/abs/1908.07065}{{\tt 1908.07065}}].

\bibitem{Grassi:2021wpw}
A.~Grassi, Q.~Hao and A.~Neitzke, \emph{{Exact WKB methods in SU(2) N$_{f}$ =
  1}}, \href{http://dx.doi.org/10.1007/JHEP01(2022)046}{\emph{JHEP} {\bf 01}
  (2022) 046}, [\href{http://arxiv.org/abs/2105.03777}{{\tt 2105.03777}}].

\bibitem{ghm}
A.~Grassi, Y.~Hatsuda and M.~Mari{\~n}o, \emph{Topological strings from quantum
  mechanics}, \href{http://dx.doi.org/10.1007/s00023-016-0479-4}{\emph{Annales
  Henri Poincar{\'e}} {\bf 17} (2016) 3177--3235}.

\bibitem{Bonelli:2022ten}
G.~Bonelli, C.~Iossa, D.~Panea~Lichtig and A.~Tanzini, \emph{{Irregular
  Liouville Correlators and Connection Formulae for Heun Functions}},
  \href{http://dx.doi.org/10.1007/s00220-022-04497-5}{\emph{Commun. Math.
  Phys.} {\bf 397} (2023) 635--727},
  [\href{http://arxiv.org/abs/2201.04491}{{\tt 2201.04491}}].

\bibitem{Wyllard:2009hg}
N.~Wyllard, \emph{{A(N-1) conformal Toda field theory correlation functions
  from conformal N = 2 SU(N) quiver gauge theories}},
  \href{http://dx.doi.org/10.1088/1126-6708/2009/11/002}{\emph{JHEP} {\bf 11}
  (2009) 002}, [\href{http://arxiv.org/abs/0907.2189}{{\tt 0907.2189}}].

\bibitem{Alday:2009aq}
L.~F. Alday, D.~Gaiotto and Y.~Tachikawa, \emph{{Liouville Correlation
  Functions from Four-dimensional Gauge Theories}},
  \href{http://dx.doi.org/10.1007/s11005-010-0369-5}{\emph{Lett. Math. Phys.}
  {\bf 91} (2010) 167--197}, [\href{http://arxiv.org/abs/0906.3219}{{\tt
  0906.3219}}].

\bibitem{Aminov:2020yma}
G.~Aminov, A.~Grassi and Y.~Hatsuda, \emph{{Black Hole Quasinormal Modes and
  Seiberg\textendash{}Witten Theory}},
  \href{http://dx.doi.org/10.1007/s00023-021-01137-x}{\emph{Annales Henri
  Poincare} {\bf 23} (2022) 1951--1977},
  [\href{http://arxiv.org/abs/2006.06111}{{\tt 2006.06111}}].

\bibitem{Bonelli:2021uvf}
G.~Bonelli, C.~Iossa, D.~P. Lichtig and A.~Tanzini, \emph{{Exact solution of
  Kerr black hole perturbations via CFT2 and instanton counting: Greybody
  factor, quasinormal modes, and Love numbers}},
  \href{http://dx.doi.org/10.1103/PhysRevD.105.044047}{\emph{Phys. Rev. D} {\bf
  105} (2022) 044047}, [\href{http://arxiv.org/abs/2105.04483}{{\tt
  2105.04483}}].

\bibitem{Casals:2021ugr}
M.~Casals and R.~T. da~Costa, \emph{{Hidden Spectral Symmetries and Mode
  Stability of Subextremal Kerr(-de Sitter) Black Holes}},
  \href{http://dx.doi.org/10.1007/s00220-022-04410-0}{\emph{Commun. Math.
  Phys.} {\bf 394} (2022) 797--832},
  [\href{http://arxiv.org/abs/2105.13329}{{\tt 2105.13329}}].

\bibitem{Bianchi:2021mft}
M.~Bianchi, D.~Consoli, A.~Grillo and J.~F. Morales, \emph{{More on the SW-QNM
  correspondence}},
  \href{http://dx.doi.org/10.1007/JHEP01(2022)024}{\emph{JHEP} {\bf 01} (2022)
  024}, [\href{http://arxiv.org/abs/2109.09804}{{\tt 2109.09804}}].

\bibitem{Consoli:2022eey}
D.~Consoli, F.~Fucito, J.~F. Morales and R.~Poghossian, \emph{{CFT description
  of BH's and ECO's: QNMs, superradiance, echoes and tidal responses}},
  \href{http://arxiv.org/abs/2206.09437}{{\tt 2206.09437}}.

\bibitem{Bianchi:2022wku}
M.~Bianchi and G.~Di~Russo, \emph{{Turning rotating D-branes and black holes
  inside out their photon-halo}},
  \href{http://dx.doi.org/10.1103/PhysRevD.106.086009}{\emph{Phys. Rev. D} {\bf
  106} (2022) 086009}, [\href{http://arxiv.org/abs/2203.14900}{{\tt
  2203.14900}}].

\bibitem{daCunha:2022ewy}
B.~C. da~Cunha and J.~a.~P. Cavalcante, \emph{{Expansions for semiclassical
  conformal blocks}},  \href{http://arxiv.org/abs/2211.03551}{{\tt
  2211.03551}}.

\bibitem{Bianchi:2023rlt}
M.~Bianchi, C.~Di~Benedetto, G.~Di~Russo and G.~Sudano, \emph{{Charge
  instability of JMaRT geometries}},
  \href{http://arxiv.org/abs/2305.00865}{{\tt 2305.00865}}.

\bibitem{Bautista:2023sdf}
Y.~F. Bautista, G.~Bonelli, C.~Iossa, A.~Tanzini and Z.~Zhou, \emph{{Black hole
  perturbation theory meets CFT2: Kerr-Compton amplitudes from
  Nekrasov-Shatashvili functions}},
  \href{http://dx.doi.org/10.1103/PhysRevD.109.084071}{\emph{Phys. Rev. D} {\bf
  109} (2024) 084071}, [\href{http://arxiv.org/abs/2312.05965}{{\tt
  2312.05965}}].

\bibitem{Bautista:2024agp}
Y.~F. Bautista, M.~Khalil, M.~Sergola, C.~Kavanagh and J.~Vines,
  \emph{{Post-Newtonian observables for aligned-spin binaries to sixth order in
  spin from gravitational self-force and Compton amplitudes}},
  \href{http://dx.doi.org/10.1103/PhysRevD.110.124005}{\emph{Phys. Rev. D} {\bf
  110} (2024) 124005}, [\href{http://arxiv.org/abs/2408.01871}{{\tt
  2408.01871}}].

\bibitem{Fucito:2024wlg}
F.~Fucito, J.~F. Morales and R.~Russo, \emph{{Gravitational wave forms for
  extreme mass ratio collisions from supersymmetric gauge theories}},
  \href{http://arxiv.org/abs/2408.07329}{{\tt 2408.07329}}.

\bibitem{Zhao:2024rrw}
H.~Zhao and R.-D. Zhu, \emph{{Connection formulae in the collision limit I:
  case studies in Lifshitz geometry}},
  \href{http://dx.doi.org/10.1088/1751-8121/ad7fa9}{\emph{J. Phys. A} {\bf 57}
  (2024) 455207}, [\href{http://arxiv.org/abs/2405.03564}{{\tt 2405.03564}}].

\bibitem{Cipriani:2024ygw}
A.~Cipriani, C.~Di~Benedetto, G.~Di~Russo, A.~Grillo and G.~Sudano,
  \emph{{Charge (in)stability and superradiance of Topological Stars}},
  \href{http://dx.doi.org/10.1007/JHEP07(2024)143}{\emph{JHEP} {\bf 07} (2024)
  143}, [\href{http://arxiv.org/abs/2405.06566}{{\tt 2405.06566}}].

\bibitem{Bianchi:2024mlq}
M.~Bianchi, G.~Dibitetto and J.~F. Morales, \emph{{Gauge theory meets
  cosmology}},
  \href{http://dx.doi.org/10.1088/1475-7516/2024/12/040}{\emph{JCAP} {\bf 12}
  (2024) 040}, [\href{http://arxiv.org/abs/2408.03243}{{\tt 2408.03243}}].

\bibitem{Arnaudo:2024bbd}
P.~Arnaudo, G.~Bonelli and A.~Tanzini, \emph{{One-loop corrections to near
  extremal Kerr thermodynamics from semiclassical Virasoro blocks}},
  \href{http://arxiv.org/abs/2412.16057}{{\tt 2412.16057}}.

\bibitem{Bianchi:2024rod}
M.~Bianchi, D.~Bini and G.~Di~Russo, \emph{{Scalar waves in a Topological Star
  spacetime: self-force and radiative losses}},
  \href{http://arxiv.org/abs/2411.19612}{{\tt 2411.19612}}.

\bibitem{Matone:2024ytm}
M.~Matone and N.~Dimakis, \emph{{Quantum Mechanics from General Relativity and
  the Quantum Friedmann Equation}},
  \href{http://arxiv.org/abs/2411.07961}{{\tt 2411.07961}}.

\bibitem{Arnaudo:2024rhv}
P.~Arnaudo, G.~Bonelli and A.~Tanzini, \emph{{One loop effective actions in
  Kerr-(A)dS black holes}},
  \href{http://dx.doi.org/10.1103/PhysRevD.110.106006}{\emph{Phys. Rev. D} {\bf
  110} (2024) 106006}, [\href{http://arxiv.org/abs/2405.13830}{{\tt
  2405.13830}}].

\bibitem{Gaiotto:2009we}
D.~Gaiotto, \emph{{N=2 dualities}},
  \href{http://dx.doi.org/10.1007/JHEP08(2012)034}{\emph{JHEP} {\bf 1208}
  (2012) 034}, [\href{http://arxiv.org/abs/0904.2715}{{\tt 0904.2715}}].

\bibitem{Gaiotto:2009hg}
D.~Gaiotto, G.~W. Moore and A.~Neitzke, \emph{{Wall-crossing, Hitchin systems,
  and the WKB approximation}},
  \href{http://dx.doi.org/10.1016/j.aim.2012.09.027}{\emph{Adv. Math.} {\bf
  234} (2013) 239--403}, [\href{http://arxiv.org/abs/0907.3987}{{\tt
  0907.3987}}].

\bibitem{Gaiotto:2014bza}
D.~Gaiotto, \emph{{Opers and TBA}},  \href{http://arxiv.org/abs/1403.6137}{{\tt
  1403.6137}}.

\bibitem{bpz}
A.~Belavin, A.~Polyakov and A.~Zamolodchikov, \emph{Infinite conformal symmetry
  in two-dimensional quantum field theory},
  \href{http://dx.doi.org/https://doi.org/10.1016/0550-3213(84)90052-X}{\emph{Nuclear
  Physics B} {\bf 241} (1984) 333--380}.

\bibitem{Dorn:1994xn}
H.~Dorn and H.~J. Otto, \emph{{Two and three point functions in Liouville
  theory}}, \href{http://dx.doi.org/10.1016/0550-3213(94)00352-1}{\emph{Nucl.
  Phys. B} {\bf 429} (1994) 375--388},
  [\href{http://arxiv.org/abs/hep-th/9403141}{{\tt hep-th/9403141}}].

\bibitem{ZAMOLODCHIKOV1996577}
A.~Zamolodchikov and A.~Zamolodchikov, \emph{Conformal bootstrap in liouville
  field theory},
  \href{http://dx.doi.org/https://doi.org/10.1016/0550-3213(96)00351-3}{\emph{Nuclear
  Physics B} {\bf 477} (1996) 577--605}.

\bibitem{Teschner:2001rv}
J.~Teschner, \emph{{Liouville theory revisited}},
  \href{http://dx.doi.org/10.1088/0264-9381/18/23/201}{\emph{Class. Quant.
  Grav.} {\bf 18} (2001) R153--R222},
  [\href{http://arxiv.org/abs/hep-th/0104158}{{\tt hep-th/0104158}}].

\bibitem{1986JETP631061Z}
A.~B. {Zamolodchikov}, \emph{{Two-dimensional conformal symmetry and critical
  four-spin correlation functions in the Ashkin-Teller model}}, {\emph{Soviet
  Journal of Experimental and Theoretical Physics} {\bf 63} (May, 1986) 1061}.

\bibitem{Matone:1995rx}
M.~Matone, \emph{{Instantons and recursion relations in N=2 SUSY gauge
  theory}}, \href{http://dx.doi.org/10.1016/0370-2693(95)00920-G}{\emph{Phys.
  Lett. B} {\bf 357} (1995) 342--348},
  [\href{http://arxiv.org/abs/hep-th/9506102}{{\tt hep-th/9506102}}].

\bibitem{Flume:2004rp}
R.~Flume, F.~Fucito, J.~F. Morales and R.~Poghossian, \emph{{Matone's relation
  in the presence of gravitational couplings}},
  \href{http://dx.doi.org/10.1088/1126-6708/2004/04/008}{\emph{JHEP} {\bf 04}
  (2004) 008}, [\href{http://arxiv.org/abs/hep-th/0403057}{{\tt
  hep-th/0403057}}].

\bibitem{lmn}
A.~S. Losev, A.~Marshakov and N.~A. Nekrasov, \emph{{Small instantons, little
  strings and free fermions}},  \href{http://arxiv.org/abs/hep-th/0302191}{{\tt
  hep-th/0302191}}.

\bibitem{Benini:2009mz}
F.~Benini, Y.~Tachikawa and B.~Wecht, \emph{{Sicilian gauge theories and N=1
  dualities}}, \href{http://dx.doi.org/10.1007/JHEP01(2010)088}{\emph{JHEP}
  {\bf 01} (2010) 088}, [\href{http://arxiv.org/abs/0909.1327}{{\tt
  0909.1327}}].

\bibitem{Hollands:2011zc}
L.~Hollands, C.~A. Keller and J.~Song, \emph{{Towards a 4d/2d correspondence
  for Sicilian quivers}},
  \href{http://dx.doi.org/10.1007/JHEP10(2011)100}{\emph{JHEP} {\bf 10} (2011)
  100}, [\href{http://arxiv.org/abs/1107.0973}{{\tt 1107.0973}}].

\bibitem{Coman:2019eex}
I.~Coman, E.~Pomoni and J.~Teschner, \emph{{Trinion Conformal Blocks from
  Topological strings}},
  \href{http://dx.doi.org/10.1007/JHEP09(2020)078}{\emph{JHEP} {\bf 09} (2020)
  078}, [\href{http://arxiv.org/abs/1906.06351}{{\tt 1906.06351}}].

\bibitem{Jia:2024zes}
H.~F. Jia and M.~Rangamani, \emph{{Holographic thermal correlators and
  quasinormal modes from semiclassical Virasoro blocks}},
  \href{http://dx.doi.org/10.1007/JHEP12(2024)047}{\emph{JHEP} {\bf 12} (2024)
  047}, [\href{http://arxiv.org/abs/2408.05208}{{\tt 2408.05208}}].

\bibitem{Mbook}
A.~Marshakov, \emph{Seiberg-Witten Theory and Integrable Systems}.
\newblock WORLD SCIENTIFIC, 1999,
  \href{http://dx.doi.org/10.1142/3936}{10.1142/3936}.

\bibitem{Moore:1997dj}
G.~W. Moore, N.~Nekrasov and S.~Shatashvili, \emph{{Integrating over Higgs
  branches}}, \href{http://dx.doi.org/10.1007/PL00005525}{\emph{Commun. Math.
  Phys.} {\bf 209} (2000) 97--121},
  [\href{http://arxiv.org/abs/hep-th/9712241}{{\tt hep-th/9712241}}].

\bibitem{Lossev:1997bz}
A.~Lossev, N.~Nekrasov and S.~L. Shatashvili, \emph{{Testing Seiberg-Witten
  solution}}, {\emph{NATO Sci. Ser. C} {\bf 520} (1999) 359--372},
  [\href{http://arxiv.org/abs/hep-th/9801061}{{\tt hep-th/9801061}}].

\bibitem{no2}
N.~Nekrasov and A.~Okounkov, \emph{{Seiberg-Witten theory and random
  partitions}}, \href{http://dx.doi.org/10.1007/0-8176-4467-9_15}{\emph{Prog.
  Math.} {\bf 244} (2006) 525--596},
  [\href{http://arxiv.org/abs/hep-th/0306238}{{\tt hep-th/0306238}}].

\bibitem{Jeong:2017mfh}
S.~Jeong and X.~Zhang, \emph{{BPZ equations for higher degenerate fields and
  nonperturbative Dyson-Schwinger equations}},
  \href{http://dx.doi.org/10.1103/PhysRevD.109.125006}{\emph{Phys. Rev. D} {\bf
  109} (2024) 125006}, [\href{http://arxiv.org/abs/1710.06970}{{\tt
  1710.06970}}].

\bibitem{Nekrasov:2002qd}
N.~A. Nekrasov, \emph{{Seiberg-Witten prepotential from instanton counting}},
  \href{http://dx.doi.org/10.4310/ATMP.2003.v7.n5.a4}{\emph{Adv.Theor.Math.Phys.}
  {\bf 7} (2004) 831--864}, [\href{http://arxiv.org/abs/hep-th/0206161}{{\tt
  hep-th/0206161}}].

\bibitem{Gaiotto:2014ina}
D.~Gaiotto and H.-C. Kim, \emph{{Surface defects and instanton partition
  functions}}, \href{http://dx.doi.org/10.1007/JHEP10(2016)012}{\emph{JHEP}
  {\bf 10} (2016) 012}, [\href{http://arxiv.org/abs/1412.2781}{{\tt
  1412.2781}}].

\bibitem{Piatek:2017fyn}
M.~Pik{a}tek and A.~R. Pietrykowski, \emph{{Solving Heun's equation using
  conformal blocks}},
  \href{http://dx.doi.org/10.1016/j.nuclphysb.2018.11.021}{\emph{Nucl. Phys. B}
  {\bf 938} (2019) 543--570}, [\href{http://arxiv.org/abs/1708.06135}{{\tt
  1708.06135}}].

\bibitem{LeFloch:2020uop}
B.~Le~Floch, \emph{{A slow review of the AGT correspondence}},
  \href{http://dx.doi.org/10.1088/1751-8121/ac5945}{\emph{J. Phys. A} {\bf 55}
  (2022) 353002}, [\href{http://arxiv.org/abs/2006.14025}{{\tt 2006.14025}}].

\bibitem{Lisovyy:2022flm}
O.~Lisovyy and A.~Naidiuk, \emph{{Perturbative connection formulas for Heun
  equations}}, \href{http://dx.doi.org/10.1088/1751-8121/ac9ba7}{\emph{J. Phys.
  A} {\bf 55} (2022) 434005}, [\href{http://arxiv.org/abs/2208.01604}{{\tt
  2208.01604}}].

\bibitem{Liu:2024eut}
P.~Liu and R.-D. Zhu, \emph{{Notes on Quasinormal Modes of charged de Sitter
  Blackholes from Quiver Gauge Theories}},
  \href{http://arxiv.org/abs/2412.18359}{{\tt 2412.18359}}.

\bibitem{Emparan:2003sy}
R.~Emparan and R.~C. Myers, \emph{{Instability of ultra-spinning black holes}},
  \href{http://dx.doi.org/10.1088/1126-6708/2003/09/025}{\emph{JHEP} {\bf 09}
  (2003) 025}, [\href{http://arxiv.org/abs/hep-th/0308056}{{\tt
  hep-th/0308056}}].

\bibitem{Cardoso:2004cj}
V.~Cardoso, G.~Siopsis and S.~Yoshida, \emph{{Scalar perturbations of higher
  dimensional rotating and ultra-spinning black holes}},
  \href{http://dx.doi.org/10.1103/PhysRevD.71.024019}{\emph{Phys. Rev. D} {\bf
  71} (2005) 024019}, [\href{http://arxiv.org/abs/hep-th/0412138}{{\tt
  hep-th/0412138}}].

\bibitem{Emparan:2008eg}
R.~Emparan and H.~S. Reall, \emph{{Black Holes in Higher Dimensions}},
  \href{http://dx.doi.org/10.12942/lrr-2008-6}{\emph{Living Rev. Rel.} {\bf 11}
  (2008) 6}, [\href{http://arxiv.org/abs/0801.3471}{{\tt 0801.3471}}].

\bibitem{Caldarelli:2008pz}
M.~M. Caldarelli, R.~Emparan and M.~J. Rodriguez, \emph{{Black Rings in
  (Anti)-deSitter space}},
  \href{http://dx.doi.org/10.1088/1126-6708/2008/11/011}{\emph{JHEP} {\bf 11}
  (2008) 011}, [\href{http://arxiv.org/abs/0806.1954}{{\tt 0806.1954}}].

\bibitem{Hennigar:2015gan}
R.~A. Hennigar, D.~Kubiz\v{n}\'ak, R.~B. Mann and N.~Musoke,
  \emph{{Ultraspinning limits and rotating hyperboloid membranes}},
  \href{http://dx.doi.org/10.1016/j.nuclphysb.2015.12.017}{\emph{Nucl. Phys. B}
  {\bf 903} (2016) 400--417}, [\href{http://arxiv.org/abs/1512.02293}{{\tt
  1512.02293}}].

\bibitem{Ponglertsakul:2020ufm}
S.~Ponglertsakul and B.~Gwak, \emph{{Massive scalar perturbations on
  Myers-Perry\textendash{}de Sitter black holes with a single rotation}},
  \href{http://dx.doi.org/10.1140/epjc/s10052-020-08616-1}{\emph{Eur. Phys. J.
  C} {\bf 80} (2020) 1023}, [\href{http://arxiv.org/abs/2007.16108}{{\tt
  2007.16108}}].

\bibitem{Aminov:2023jve}
G.~Aminov, P.~Arnaudo, G.~Bonelli, A.~Grassi and A.~Tanzini, \emph{{Black hole
  perturbation theory and multiple polylogarithms}},
  \href{http://dx.doi.org/10.1007/JHEP11(2023)059}{\emph{JHEP} {\bf 11} (2023)
  059}, [\href{http://arxiv.org/abs/2307.10141}{{\tt 2307.10141}}].

\bibitem{Lei:2023mqx}
Y.~Lei, H.~Shu, K.~Zhang and R.-D. Zhu, \emph{{Quasinormal modes of C-metric
  from SCFTs}}, \href{http://dx.doi.org/10.1007/JHEP02(2024)140}{\emph{JHEP}
  {\bf 02} (2024) 140}, [\href{http://arxiv.org/abs/2308.16677}{{\tt
  2308.16677}}].

\bibitem{Kodama:2003kk}
H.~Kodama and A.~Ishibashi, \emph{{Master equations for perturbations of
  generalized static black holes with charge in higher dimensions}},
  \href{http://dx.doi.org/10.1143/PTP.111.29}{\emph{Prog. Theor. Phys.} {\bf
  111} (2004) 29--73}, [\href{http://arxiv.org/abs/hep-th/0308128}{{\tt
  hep-th/0308128}}].

\bibitem{Loganayagam:2022teq}
R.~Loganayagam, M.~Rangamani and J.~Virrueta, \emph{{Holographic thermal
  correlators: a tale of Fuchsian ODEs and integration contours}},
  \href{http://dx.doi.org/10.1007/JHEP07(2023)008}{\emph{JHEP} {\bf 07} (2023)
  008}, [\href{http://arxiv.org/abs/2212.13940}{{\tt 2212.13940}}].

\bibitem{Son:2002sd}
D.~T. Son and A.~O. Starinets, \emph{{Minkowski space correlators in AdS / CFT
  correspondence: Recipe and applications}},
  \href{http://dx.doi.org/10.1088/1126-6708/2002/09/042}{\emph{JHEP} {\bf 09}
  (2002) 042}, [\href{http://arxiv.org/abs/hep-th/0205051}{{\tt
  hep-th/0205051}}].

\bibitem{Nunez:2003eq}
A.~Nunez and A.~O. Starinets, \emph{{AdS / CFT correspondence, quasinormal
  modes, and thermal correlators in N=4 SYM}},
  \href{http://dx.doi.org/10.1103/PhysRevD.67.124013}{\emph{Phys. Rev. D} {\bf
  67} (2003) 124013}, [\href{http://arxiv.org/abs/hep-th/0302026}{{\tt
  hep-th/0302026}}].

\bibitem{Policastro:2002se}
G.~Policastro, D.~T. Son and A.~O. Starinets, \emph{{From AdS / CFT
  correspondence to hydrodynamics}},
  \href{http://dx.doi.org/10.1088/1126-6708/2002/09/043}{\emph{JHEP} {\bf 09}
  (2002) 043}, [\href{http://arxiv.org/abs/hep-th/0205052}{{\tt
  hep-th/0205052}}].

\bibitem{Kovtun:2005ev}
P.~K. Kovtun and A.~O. Starinets, \emph{{Quasinormal modes and holography}},
  \href{http://dx.doi.org/10.1103/PhysRevD.72.086009}{\emph{Phys. Rev. D} {\bf
  72} (2005) 086009}, [\href{http://arxiv.org/abs/hep-th/0506184}{{\tt
  hep-th/0506184}}].

\bibitem{Lashkari:2016vgj}
N.~Lashkari, A.~Dymarsky and H.~Liu, \emph{{Eigenstate Thermalization
  Hypothesis in Conformal Field Theory}},
  \href{http://dx.doi.org/10.1088/1742-5468/aab020}{\emph{J. Stat. Mech.} {\bf
  1803} (2018) 033101}, [\href{http://arxiv.org/abs/1610.00302}{{\tt
  1610.00302}}].

\bibitem{Kulaxizi:2018dxo}
M.~Kulaxizi, G.~S. Ng and A.~Parnachev, \emph{{Black Holes, Heavy States, Phase
  Shift and Anomalous Dimensions}},
  \href{http://dx.doi.org/10.21468/SciPostPhys.6.6.065}{\emph{SciPost Phys.}
  {\bf 6} (2019) 065}, [\href{http://arxiv.org/abs/1812.03120}{{\tt
  1812.03120}}].

\bibitem{Karlsson:2019qfi}
R.~Karlsson, M.~Kulaxizi, A.~Parnachev and P.~Tadi\'c, \emph{{Black Holes and
  Conformal Regge Bootstrap}},
  \href{http://dx.doi.org/10.1007/JHEP10(2019)046}{\emph{JHEP} {\bf 10} (2019)
  046}, [\href{http://arxiv.org/abs/1904.00060}{{\tt 1904.00060}}].

\bibitem{Karlsson:2019dbd}
R.~Karlsson, M.~Kulaxizi, A.~Parnachev and P.~Tadi\'c, \emph{{Leading
  Multi-Stress Tensors and Conformal Bootstrap}},
  \href{http://dx.doi.org/10.1007/JHEP01(2020)076}{\emph{JHEP} {\bf 01} (2020)
  076}, [\href{http://arxiv.org/abs/1909.05775}{{\tt 1909.05775}}].

\bibitem{Karlsson:2021duj}
R.~Karlsson, A.~Parnachev and P.~Tadi\'c, \emph{{Thermalization in large-N
  CFTs}}, \href{http://dx.doi.org/10.1007/JHEP09(2021)205}{\emph{JHEP} {\bf 09}
  (2021) 205}, [\href{http://arxiv.org/abs/2102.04953}{{\tt 2102.04953}}].

\bibitem{Dodelson:2022eiz}
M.~Dodelson and A.~Zhiboedov, \emph{{Gravitational orbits, double-twist mirage,
  and many-body scars}},
  \href{http://dx.doi.org/10.1007/JHEP12(2022)163}{\emph{JHEP} {\bf 12} (2022)
  163}, [\href{http://arxiv.org/abs/2204.09749}{{\tt 2204.09749}}].

\bibitem{Dodelson:2022yvn}
M.~Dodelson, A.~Grassi, C.~Iossa, D.~Panea~Lichtig and A.~Zhiboedov,
  \emph{{Holographic thermal correlators from supersymmetric instantons}},
  \href{http://dx.doi.org/10.21468/SciPostPhys.14.5.116}{\emph{SciPost Phys.}
  {\bf 14} (2023) 116}, [\href{http://arxiv.org/abs/2206.07720}{{\tt
  2206.07720}}].

\bibitem{Bhatta:2022wga}
A.~Bhatta and T.~Mandal, \emph{{Exact thermal correlators of holographic
  CFTs}}, \href{http://dx.doi.org/10.1007/JHEP02(2023)222}{\emph{JHEP} {\bf 02}
  (2023) 222}, [\href{http://arxiv.org/abs/2211.02449}{{\tt 2211.02449}}].

\bibitem{Bhatta:2023qcl}
A.~Bhatta, S.~Chakrabortty, T.~Mandal and A.~Maurya, \emph{{Holographic thermal
  correlators for hyperbolic $CFT_s$}},
  \href{http://dx.doi.org/10.1007/JHEP11(2023)156}{\emph{JHEP} {\bf 11} (2023)
  156}, [\href{http://arxiv.org/abs/2308.14704}{{\tt 2308.14704}}].

\bibitem{He:2023wcs}
S.~He and Y.~Li, \emph{{Holographic Euclidean thermal correlator}},
  \href{http://dx.doi.org/10.1007/JHEP03(2024)024}{\emph{JHEP} {\bf 03} (2024)
  024}, [\href{http://arxiv.org/abs/2308.13518}{{\tt 2308.13518}}].

\bibitem{Ren:2024ifh}
J.~Ren and Z.~Yu, \emph{{Holographic thermal correlators from recursions}},
  \href{http://arxiv.org/abs/2412.02608}{{\tt 2412.02608}}.

\bibitem{BarraganAmado:2024tfu}
J.~Barrag\'an~Amado, S.~Chakrabortty and A.~Maurya, \emph{{The effect of
  resummation on retarded Green\textquoteright{}s function and greybody factor
  in AdS black holes}},
  \href{http://dx.doi.org/10.1007/JHEP11(2024)070}{\emph{JHEP} {\bf 11} (2024)
  070}, [\href{http://arxiv.org/abs/2409.07370}{{\tt 2409.07370}}].

\bibitem{Arnaudo:2024sen}
P.~Arnaudo and B.~Withers, \emph{{Exact low-temperature Green's functions in
  AdS/CFT: From Heun to confluent Heun}},
  \href{http://arxiv.org/abs/2412.01923}{{\tt 2412.01923}}.

\bibitem{review}
E.~Berti, V.~Cardoso and A.~O. Starinets, \emph{Quasinormal modes of black
  holes and black branes},
  \href{http://dx.doi.org/10.1088/0264-9381/26/16/163001}{\emph{Classical and
  Quantum Gravity} {\bf 26} (Jul, 2009) 163001}.

\bibitem{adscft}
J.~M. Maldacena, \emph{{The Large N limit of superconformal field theories and
  supergravity}},
  \href{http://dx.doi.org/10.1023/A:1026654312961}{\emph{Int.J.Theor.Phys.}
  {\bf 38} (1999) 1113--1133}, [\href{http://arxiv.org/abs/hep-th/9711200}{{\tt
  hep-th/9711200}}].

\bibitem{Witten:1998qj}
E.~Witten, \emph{{Anti-de Sitter space and holography}},
  \href{http://dx.doi.org/10.4310/ATMP.1998.v2.n2.a2}{\emph{Adv. Theor. Math.
  Phys.} {\bf 2} (1998) 253--291},
  [\href{http://arxiv.org/abs/hep-th/9802150}{{\tt hep-th/9802150}}].

\bibitem{Gubser:1998bc}
S.~S. Gubser, I.~R. Klebanov and A.~M. Polyakov, \emph{{Gauge theory
  correlators from noncritical string theory}},
  \href{http://dx.doi.org/10.1016/S0370-2693(98)00377-3}{\emph{Phys. Lett. B}
  {\bf 428} (1998) 105--114}, [\href{http://arxiv.org/abs/hep-th/9802109}{{\tt
  hep-th/9802109}}].

\bibitem{Li:2020dqm}
Y.-Z. Li and H.-Y. Zhang, \emph{{More on heavy-light bootstrap up to
  double-stress-tensor}},
  \href{http://dx.doi.org/10.1007/JHEP10(2020)055}{\emph{JHEP} {\bf 10} (2020)
  055}, [\href{http://arxiv.org/abs/2004.04758}{{\tt 2004.04758}}].

\bibitem{zamorecursion}
A.~B. Zamolodchikov, \emph{Conformal symmetry in two-dimensional space:
  Recursion representation of conformal block},
  \href{http://dx.doi.org/10.1007/BF01022967}{\emph{Theoretical and
  Mathematical Physics} {\bf 73} (1987) 1088--1093}.

\bibitem{Cho:2017oxl}
M.~Cho, S.~Collier and X.~Yin, \emph{{Recursive Representations of Arbitrary
  Virasoro Conformal Blocks}},
  \href{http://dx.doi.org/10.1007/JHEP04(2019)018}{\emph{JHEP} {\bf 04} (2019)
  018}, [\href{http://arxiv.org/abs/1703.09805}{{\tt 1703.09805}}].

\bibitem{Kodama:2007ph}
H.~Kodama, \emph{{Perturbations and Stability of Higher-Dimensional Black
  Holes}}, \href{http://dx.doi.org/10.1007/978-3-540-88460-6_11}{\emph{Lect.
  Notes Phys.} {\bf 769} (2009) 427--470},
  [\href{http://arxiv.org/abs/0712.2703}{{\tt 0712.2703}}].

\bibitem{unpGGDC}
G.~Bonelli, C.~Iossa, D.~P. Lichtig and A.~Tanzini, \emph{{Unpublished}}, .

\bibitem{Karlsson:2020ghx}
R.~Karlsson, M.~Kulaxizi, A.~Parnachev and P.~Tadi\'c, \emph{{Stress tensor
  sector of conformal correlators operators in the Regge limit}},
  \href{http://dx.doi.org/10.1007/JHEP07(2020)019}{\emph{JHEP} {\bf 07} (2020)
  019}, [\href{http://arxiv.org/abs/2002.12254}{{\tt 2002.12254}}].

\bibitem{Li:2019zba}
Y.-Z. Li, \emph{{Heavy-light Bootstrap from Lorentzian Inversion Formula}},
  \href{http://dx.doi.org/10.1007/JHEP07(2020)046}{\emph{JHEP} {\bf 07} (2020)
  046}, [\href{http://arxiv.org/abs/1910.06357}{{\tt 1910.06357}}].

\bibitem{Karlsson:2022osn}
R.~Karlsson, A.~Parnachev, V.~Prilepina and S.~Valach, \emph{{Thermal stress
  tensor correlators, OPE and holography}},
  \href{http://dx.doi.org/10.1007/JHEP09(2022)234}{\emph{JHEP} {\bf 09} (2022)
  234}, [\href{http://arxiv.org/abs/2206.05544}{{\tt 2206.05544}}].

\bibitem{kt2}
K.~Kozlowski and J.~Teschner, \emph{{TBA for the Toda chain}},
  \href{http://arxiv.org/abs/1006.2906}{{\tt 1006.2906}}.

\bibitem{Meneghelli:2013tia}
C.~Meneghelli and G.~Yang, \emph{{Mayer-Cluster Expansion of Instanton
  Partition Functions and Thermodynamic Bethe Ansatz}},
  \href{http://dx.doi.org/10.1007/JHEP05(2014)112}{\emph{JHEP} {\bf 05} (2014)
  112}, [\href{http://arxiv.org/abs/1312.4537}{{\tt 1312.4537}}].

\bibitem{post-zamo}
A.~B. Zamolodchikov, \emph{{Generalized Mathieu equations and Liouville TBA}},
  in \emph{Quantum Field Theories in Two Dimensions}, vol.~2.
\newblock World Scientific, 2012.

\bibitem{Fioravanti:2021dce}
D.~Fioravanti and D.~Gregori, \emph{{A new method for exact results on
  Quasinormal Modes of Black Holes}},
  \href{http://arxiv.org/abs/2112.11434}{{\tt 2112.11434}}.

\bibitem{Fioravanti:2019vxi}
D.~Fioravanti and D.~Gregori, \emph{{Integrability and cycles of deformed
  ${\cal N}=2$ gauge theory}},
  \href{http://dx.doi.org/10.1016/j.physletb.2020.135376}{\emph{Phys. Lett. B}
  {\bf 804} (2020) 135376}, [\href{http://arxiv.org/abs/1908.08030}{{\tt
  1908.08030}}].

\bibitem{Hollands:2017ahy}
L.~Hollands and O.~Kidwai, \emph{{Higher length-twist coordinates, generalized
  Heun\textquoteright{}s opers, and twisted superpotentials}},
  \href{http://dx.doi.org/10.4310/ATMP.2018.v22.n7.a2}{\emph{Adv. Theor. Math.
  Phys.} {\bf 22} (2018) 1713--1822},
  [\href{http://arxiv.org/abs/1710.04438}{{\tt 1710.04438}}].

\bibitem{Hollands:2013qza}
L.~Hollands and A.~Neitzke, \emph{{Spectral Networks and
  Fenchel\textendash{}Nielsen Coordinates}},
  \href{http://dx.doi.org/10.1007/s11005-016-0842-x}{\emph{Lett. Math. Phys.}
  {\bf 106} (2016) 811--877}, [\href{http://arxiv.org/abs/1312.2979}{{\tt
  1312.2979}}].

\bibitem{Hollands:2019wbr}
L.~Hollands and A.~Neitzke, \emph{{Exact WKB and abelianization for the $T_3$
  equation}}, \href{http://dx.doi.org/10.1007/s00220-020-03875-1}{\emph{Commun.
  Math. Phys.} {\bf 380} (2020) 131--186},
  [\href{http://arxiv.org/abs/1906.04271}{{\tt 1906.04271}}].

\bibitem{Imaizumi:2021cxf}
K.~Imaizumi, \emph{{Quantum periods and TBA equations for $\mathcal{N}=2\
  SU(2)\ N_f=2$ SQCD with flavor symmetry}},
  \href{http://dx.doi.org/10.1016/j.physletb.2021.136270}{\emph{Phys. Lett. B}
  {\bf 816} (2021) 136270}, [\href{http://arxiv.org/abs/2103.02248}{{\tt
  2103.02248}}].

\bibitem{seiberg1994}
N.~Seiberg and E.~Witten, \emph{Monopoles, duality and chiral symmetry breaking
  in {{N}} = 2 supersymmetric {{QCD}}}, {\emph{Nucl. Phys. B} {\bf 431} (Dec.,
  1994) 484--550}.

\bibitem{fb}
F.~Ferrari and A.~Bilal, \emph{{The Strong coupling spectrum of the
  Seiberg-Witten theory}},
  \href{http://dx.doi.org/10.1016/0550-3213(96)00150-2}{\emph{Nucl. Phys.} {\bf
  B469} (1996) 387--402}, [\href{http://arxiv.org/abs/hep-th/9602082}{{\tt
  hep-th/9602082}}].

\bibitem{selfdual}
A.~Klemm, W.~Lerche, P.~Mayr, C.~Vafa and N.~P. Warner, \emph{{Selfdual strings
  and N=2 supersymmetric field theory}},
  \href{http://dx.doi.org/10.1016/0550-3213(96)00353-7}{\emph{Nucl. Phys.} {\bf
  B477} (1996) 746--766}, [\href{http://arxiv.org/abs/hep-th/9604034}{{\tt
  hep-th/9604034}}].

\bibitem{Gaiotto:2012rg}
D.~Gaiotto, G.~W. Moore and A.~Neitzke, \emph{{Spectral networks}},
  \href{http://dx.doi.org/10.1007/s00023-013-0239-7}{\emph{Annales Henri
  Poincare} {\bf 14} (2013) 1643--1731},
  [\href{http://arxiv.org/abs/1204.4824}{{\tt 1204.4824}}].

\bibitem{Hollands:2021itj}
L.~Hollands, P.~R\"uter and R.~J. Szabo, \emph{{A geometric recipe for twisted
  superpotentials}},
  \href{http://dx.doi.org/10.1007/JHEP12(2021)164}{\emph{JHEP} {\bf 12} (2021)
  164}, [\href{http://arxiv.org/abs/2109.14699}{{\tt 2109.14699}}].

\bibitem{Hollands:2016kgm}
L.~Hollands and A.~Neitzke, \emph{{BPS states in the Minahan-Nemeschansky
  ${E_6}$ theory}},
  \href{http://dx.doi.org/10.1007/s00220-016-2798-1}{\emph{Commun. Math. Phys.}
  {\bf 353} (2017) 317--351}, [\href{http://arxiv.org/abs/1607.01743}{{\tt
  1607.01743}}].

\bibitem{Hao:2019ryd}
Q.~Hao, L.~Hollands and A.~Neitzke, \emph{{BPS states in the
  Minahan-Nemeschansky $E_7$ theory}},
  \href{http://dx.doi.org/10.1007/JHEP04(2020)039}{\emph{JHEP} {\bf 04} (2020)
  039}, [\href{http://arxiv.org/abs/1905.09879}{{\tt 1905.09879}}].

\bibitem{Jia:2026ryl}
H.~F. Jia and M.~Rangamani, \emph{{Exact holographic thermal spectral
  functions: OPE, non-perturbative corrections, and black hole singularity}},
  \href{http://arxiv.org/abs/2604.10803}{{\tt 2604.10803}}.

\end{thebibliography}\endgroup
 
\end{document}